\documentclass[nofootinbib,preprint,preprintnumbers,a4paper,11pt]{revtex4}
\usepackage{epsfig}
\usepackage{footnote}
\usepackage{ulem}
\usepackage{color}
\usepackage{array}
\usepackage{amssymb}
\usepackage{amsmath}
\usepackage{graphicx}
\usepackage{subfigure}
\usepackage{longtable}
\usepackage{verbatim}
\usepackage{amsfonts}
\usepackage{hyperref}
\usepackage{multirow}
\usepackage{cancel}

\begin{document}

\title{Generation of $SU(3)$ sum rule for charmed baryon decay}

\author{Di Wang$^{1}$}\email{wangdi@hunnu.edu.cn}

\address{%
$^1$Department of Physics, Hunan Normal University, and Key Laboratory of Low-Dimensional Quantum Structures and Quantum Control of Ministry of Education, Changsha 410081, China
}

\begin{abstract}
Flavor $SU(3)$ symmetry is a powerful tool to analyze charmed baryon decays.
In this work, we propose an approach to generate $SU(3)$ sum rules for the singly and doubly charmed baryon decays without writing the Wigner-Eckhart invariants explicitly.
The $SU(3)$ sum rules are computed routinely in several master formulas.
Hundreds of $SU(3)$ sum rules are found to serve as test of the flavor symmetry in the charmed baryon decays.
\end{abstract}

\maketitle


\section{Introduction}
Charmed baryon decays provide laboratories to study strong and weak interactions in heavy-to-light baryonic transitions.
In 2017, the LHCb collaboration observed the first doubly heavy baryon $\Xi_{cc}^{++}$ via $\Xi_{cc}^{++}\to \Lambda^+_cK^-\pi^+\pi^+$ decay \cite{Aaij:2017ueg} as proposed by \cite{Yu:2017zst}.
Subsequently, the measurements of lifetime and production rate of $\Xi_{cc}^{++}$ and the observations of other decay modes were performed \cite{LHCb:2019epo,LHCb:2019ybf,LHCb:2019qed,LHCb:2021eaf,LHCb:2022rpd,Aaij:2018wzf,LHCb:2018pcs}.
On the other hand, many new measurements of singly charmed baryon decays were performed by
Belle (II) \cite{Belle:2021btl,Berger:2018pli,Zupanc:2013iki,Yang:2015ytm,Pal:2017ypp},
BESIII \cite{BESIII:2022bkj,BESIII:2021fqx,BESIII:2020kap,BESIII:2020cpu,BESIII:2019odb,BESIII:2018qyg,Ablikim:2018jfs,Ablikim:2018woi,Ablikim:2018bir,Ablikim:2015prg,Ablikim:2015flg,Ablikim:2016tze,Ablikim:2016mcr,Ablikim:2016vqd,Ablikim:2017ors,Ablikim:2017iqd} and
LHCb \cite{Dudek:2019vxm,Aaij:2017nsd,Aaij:2017xva}.
Motivated by the experimental progress, great theoretical efforts are devoted to the study of doubly and singly charmed baryon weak decays.

For charmed baryon decays, the factorization approach could not give reliable predictions due to the lack of knowledge on low-energy inputs and complicated hard-scattering kernels.
Apart form model calculations, it is useful to study charmed baryon decays based on the flavor $SU(3)$ symmetry.
Among those $SU(3)$-inspired approaches, the $SU(3)$ irreducible representation amplitude (IRA) constructed by decomposing the effective Hamiltonian into irreducible representations \cite{Geng:2019awr,He:2021qnc,Li:2021rfj,Geng:2020zgr,Geng:2018rse,Geng:2019bfz,Geng:2019xbo,
Hsiao:2019yur,Jia:2019zxi,Wang:2019dls,Savage:1989qr,Sheikholeslami:1991ab,
Sharma:1996sc,Verma:1995dk,Wang:2017azm,Shi:2017dto,Wang:2018utj,Lu:2016ogy,
Geng:2017esc,Geng:2017mxn,Wang:2017gxe,Geng:2018plk,Geng:2018bow}, and the topological diagram amplitude (TDA) which is classified according to the topological structure of weak interaction \cite{Wang:2022wrb,Hsiao:2020iwc,Zhao:2018mov,Groote:2021pxt,Han:2021azw,
Chau:1995gk,Kohara:1991ug}, are widely used.
Both the IRA and TDA are the Wigner-Eckhart invariants \cite{Eckart30,Wigner59} and equivalent to each other \cite{He:2018joe,He:2018php,Wang:2020gmn}.

Based on the flavor $SU(3)$ symmetry, one can get some relations for several decay amplitudes, which help us to obtain the knowledge on the not measured channels and extract information of baryonic dynamics.
In most literatures, the $SU(3)$ sum rules are found by writing the decay amplitudes decomposed by the Wigner-Eckhart invariants and combining several modes to form a polygon in the complex plane. This method is laboriously and easy to miss.
To avoid writing the Wigner-Eckhart invariants, a direct approach to generate $SU(3)$ sum rules for charmed meson decays was proposed in Ref.~\cite{Grossman:2012ry}.
The key idea is that there exists two operators, $T_-$ and $S$, under which the Hamiltonian is invariant, $T_-\,H=0$ and $S\,H=0$. The invariance of Hamiltonian under these two operators is related to the generation of sum rule for the amplitudes, permitting us to obtain $SU(3)$ sum rules without computing the Wigner-Eckhart invariants.

In this work, we extend this approach to the charmed baryon decays.
The master formulas of $SU(3)$ sum rule for the two-body decays of singly and doubly charmed baryon are derived.
With these master formulas, abundant of $SU(3)$ sum rules are computed routinely without the Wigner-Eckhart invariants.
They may provide hints for exploration of new decay modes and serve as examination of the flavor symmetry.

The rest of this paper is structured as follows.
In Sec. \ref{sumrule}, we introduce the theoretical framework of computing $SU(3)$ sum rules.
In Sec. \ref{singly} and Sec. \ref{doubly}, we derive the master formulas of $SU(3)$ sum rules for the singly and doubly charmed baryon decays respectively and discuss their applications.
Sec. \ref{sum} is a brief summary. And the $SU(3)$ sum rules derived by the master formulas are listed in Appendices. \ref{res1} $\sim$ \ref{res4}.

\section{Theoretical framework}\label{sumrule}
In the Standard Model, the tree level effective Hamiltonian of charm decay can be written as \cite{Wang:2020gmn}
\begin{equation}\label{h}
 \mathcal H_{\rm eff}={G_F\over \sqrt 2}
 \sum_{q=d,s}V_{cq}^*V_{uq}\sum_{i^\prime=1}^2C_{i^\prime}(\mu)O_{i^\prime}(\mu)=\sum_{i,j,k=1}^3 H^{ij}_{k}O_{ij}^{k},
\end{equation}
where $O_{ij}^{k}$ denotes the four-quark operator, and $H$ is a $3\times 3\times 3$ complex matrix that can be obtained from the map $(\bar uq_1)(\bar q_2c)\rightarrow V^*_{cq_2}V_{uq_1}$ in current-current operators and $V_{cs}\simeq V_{ud} \approx 1$, $V_{cd}\simeq - V_{us} \approx -\lambda$ \cite{ParticleDataGroup:2020ssz}.
Operator $O_{ij}^k$, as the representation of $SU(3)$ group, can be decomposed into direct sum of irreducible representations, $3 \otimes\overline 3 \otimes 3 =  3_p\oplus3_t\oplus \overline 6 \oplus 15$.
The initial and final states, such as pseudoscalar mesons, can be written as
\begin{align}
  |M_\alpha\rangle = (M_\alpha)^{i}_{j}|M^{i}_{j} \rangle.
\end{align}
$|M^{i}_{j} \rangle$ is the quark composition of meson state, $|M^{i}_{j} \rangle = |q_i\bar q_j\rangle$. $(M_\alpha)$ is the coefficient matrix.
The decay amplitude of the two-body nonleptonic weak decay of charm hadron (taking $D_\gamma\to M_\alpha M_\beta$ as example) can be constructed to be
\begin{align}\label{amp}
\mathcal{A}(D_\gamma\to M_\alpha M_\beta)& =  \langle M_\alpha M_\beta |\mathcal{H}_{\rm eff}| D_\gamma\rangle \nonumber\\& ~= \sum_{\omega}\,(M_\alpha)^n_m\langle M^n_m|(M_\beta)^s_r\langle M^s_r||H^{jk}_lO^{jk}_l||(D_\gamma)_i|D_i\rangle\nonumber\\& ~~=\sum_\omega\,\langle M_{m}^{n} M^s_r |O^{jk}_{l} |D_{i}\rangle \times (M_\alpha)_{m}^{n}(M_\beta)_{r}^{s} H^{jk}_{l}(D_\gamma)_i\nonumber\\& ~~~= \sum_\omega X_{\omega}(C_\omega)_{\alpha\beta\gamma},
\end{align}
in which $\sum_{\omega}$ present summing over all the possible full contractions of tensor $\langle M_{m}^{n} M^s_r |O^{jk}_{l} |D_{i}\rangle$.
According to the Wigner-Eckhart theorem \cite{Eckart30,Wigner59}, tensor $X_\omega = \langle M_{m}^{n} M^s_r |O^{jk}_{l} |D_{i}\rangle$ is  the reduced matrix element which is independent of indices $\alpha$, $\beta$ and $\gamma$ in the limit of $SU(3)$ symmetry.
All the information of initial/final states is absorbed into the Clebsch-Gordan (CG) coefficient $(C_\omega)_{\alpha\beta\gamma}=(M_\alpha)_{m}^{n}(M_\beta)_{r}^{s} H^{jk}_{l}(D_\gamma)_i$.
If the effective Hamiltonian $\mathcal H_{\rm eff}$ is inserted into Eq.~\eqref{amp} without $SU(3)$ decomposition, it is the topological decomposition of decay amplitude.
If the effective Hamiltonian is decomposed into irreducible representation of $SU(3)$ group, Eq.~\eqref{amp} presents the $SU(3)$ irreducible representation amplitude. So Eq.~\eqref{amp} indicates equivalence of the TDA and IRA approach \cite{He:2018joe,He:2018php,Wang:2020gmn}.

To get $SU(3)$ sum rules, a widely used approach in literatures is to write the decay amplitude of each mode decomposed by the Wigner-Eckhart invariants, TDA or IRA, then combine several modes to form a polygon in the complex plane.
The detailed discussions about searching for $SU(3)$ sum rules in this way are shown in Refs.~\cite{Wang:2017azm,Savage:1989qr,Jia:2019zxi}, in which many amplitude sum rules for singly and doubly charmed baryon decays in the $SU(3)_F$ limit are found.
However, this method is laborious.
Ref.~\cite{Grossman:2012ry} provides an useful approach to generate $SU(3)$ sum rules for the charmed meson decays without computing the Wigner-Eckhart invariants.
The idea is that if there exists an operator $T$ under which $TH=0$, the invariance of Hamiltonian is related to the generation of sum rules for the amplitudes.
We can understand it as follows.
If $TH=0$, it follows that
\begin{equation}\label{rule}
  \langle M_\alpha M_\beta |T\mathcal{H}_{\rm eff}| D_\gamma\rangle = \sum_\omega\,\langle M_{m}^{n} M^s_r |O^{jk}_{l} |D_{i}\rangle \times (M_\alpha)_{m}^{n}(M_\beta)_{r}^{s} (TH)^{jk}_{l}(D_\gamma)_i = 0.
\end{equation}
One can apply the operator $T$ to the initial/final states rather than the effective Hamiltonian and expand the results with the initial/final states as bases.
Then the LHS of Eq.~\ref{rule} is turned into sum of several decay amplitudes. The RHS of Eq.~\ref{rule} is invariant.
Thereby, Eq.~\eqref{rule} is an abstract $SU(3)$ sum rule.

It is found in Ref.~\cite{Grossman:2012ry} that the effective Hamiltonian is invariant under the operators $T_-$ and $S$ in the $SU(3)_F$ limit, i.e., $T_- H=0$, $S H=0$.
$T_-$ and $S$ are expressed as \cite{Grossman:2012ry}
\begin{eqnarray}
 T_-=  \left( \begin{array}{ccc}
   0   & 0  & 0 \\
     1 &   0  & 0 \\
    \lambda & 0 & 0 \\
  \end{array}\right)\quad {\rm and} \quad S=  \left( \begin{array}{ccc}
   0   & 0  & 0 \\
     0 &   -\lambda  & 1 \\
    0 & -\lambda^2 & \lambda \\
  \end{array}\right).
\end{eqnarray}
$S$ is a linear combination of $U$-spin operators and $T_-$ is a linear combination of isospin and $V$-spin operators,
\begin{align}
  S = -\lambda U_3 -\lambda^2 U_- + U_+, \qquad T_- = I_- +\lambda V_-.
\end{align}
So the $SU(3)$ sum rules generated from $S$ are $U$-spin sum rules, and the $SU(3)$ sum rules generated from $T_-$ are combination of isospin sum rules and $V$-spin sum rules.

If we consider the linear flavor $SU(3)$ breaking, i.e., the NLO corrections of $SU(3)_F$ symmetry, the invariance of the effective Hamiltonian under $T_-$ and $S$ is broken, $T_-H_{\cancel{SU(3)_F}}\neq 0$ and $SH_{\cancel{SU(3)_F}}\neq 0$ \cite{Grossman:2012ry}.
And it seems that such operator under which $H_{\cancel{SU(3)_F}}$ being invariant is unavailable. So we only consider the flavor $SU(3)$ limit in this work.

\section{Singly charmed baryon decay}\label{singly}
According to above analysis, we need two steps to get concrete $SU(3)$ sum rules from Eq.~\eqref{rule}:
\begin{enumerate}
  \item Find one or several operators under which the effective Hamiltonian is invariant.
  \item Apply these operators to the initial/final states and compute coefficients of the obtained matrices expanded by the initial/final states.
\end{enumerate}
The master formula of $SU(3)$ sum rule for $D$ meson decays are derived in the appendix A of Ref.~\cite{Grossman:2012ry}.
In this work, we extend this approach to the charmed baryon decays. We shall derive the master formulas of $SU(3)$ sum rule for singly and doubly charmed baryon decays generated by $T_-$ and $S$.

The charmed baryon anti-triplet is
\begin{eqnarray}
 [\mathcal{B}_{c\overline 3}]=  \left( \begin{array}{ccc}
   0   & \Lambda_c^+  & \Xi_c^+ \\
    -\Lambda_c^+ &   0   & \Xi_c^0 \\
    -\Xi_c^+ & -\Xi_c^0 & 0 \\
  \end{array}\right).
\end{eqnarray}
With the Levi-Civita tensor, the charmed baryon anti-triplet can be written as
\begin{eqnarray}
[\mathcal{B}_{c\overline 3}]_{ij}=\epsilon_{ijk}[\mathcal{B}_{c\overline 3}]^{k}\qquad {\rm with}\qquad [\mathcal{B}_{c\overline 3}]^{k}=\left( \begin{array}{ccc}
     \Xi_c^0 \\
    -\Xi_c^+  \\
    \Lambda_c^+ \\
  \end{array}\right).
\end{eqnarray}
To obtain $SU(3)$ sum rules for the $\mathcal{B}_{\overline c3}\to M_8 \mathcal{B}_8$ decays, we apply operator $T$, where $T$ denotes $T_-$ or $S$, to the initial and final states.
A charmed baryon anti-triplet state can be expressed as
\begin{align}\label{cb3}
|[\mathcal{B}_{\overline c3}]_\alpha\rangle = ([\mathcal{B}_{\overline c3}]_\alpha)^i |[\mathcal{B}_{\overline c3}]^{i}\rangle,
\end{align}
in which $|[\mathcal{B}_{\overline c3}]^{i} \rangle$ is quark composition of baryon state, $([\mathcal{B}_{\overline c3}]_\alpha)^{i}$ is coefficient of quark composition.
Under the operator $T$, we have
\begin{align}\label{tc3}
&T|[\mathcal{B}_{\overline c3}]_\gamma\rangle = \sum_\alpha |[\mathcal{B}_{\overline c3}]_\alpha\rangle \langle [\mathcal{B}_{\overline c3}]_\alpha |T|[\mathcal{B}_{\overline c3}]_\gamma\rangle \nonumber\\&~~~= \sum_\alpha([\mathcal{B}_{\overline c3}]^\alpha)_j T^j_i([\mathcal{B}_{\overline c3}]_\gamma)^i|[\mathcal{B}_{\overline c3}]_\alpha\rangle = \sum_\alpha(T_{\mathcal{B}_{\overline c3}})^{\alpha}_{\gamma}|[\mathcal{B}_{\overline c3}]_\alpha\rangle.
\end{align}
The matrix element $(T_{\mathcal{B}_{\overline c3}})^{\alpha}_{\gamma}$ is coefficient of
$T|[\mathcal{B}_{\overline c3}]_\gamma\rangle$ with $|[\mathcal{B}_{\overline c3}]_\alpha\rangle$ as bases.
More specifically, $(T_{\mathcal{B}_{\overline c3}})_\gamma^\alpha$ is coefficient of matrix product $T\,[\mathcal{B}_{\overline c3}]_\gamma$ expanded by basis matrices $[\mathcal{B}_{\overline c3}]_\alpha$.
If we define the charmed baryon anti-triplet bases as
 $|[\mathcal{B}_{\overline c3}]_\alpha\rangle = ( |\Xi^0_c\rangle,\,\, |\Xi^+_c\rangle ,\,\, |\Lambda^+_c\rangle )$,
we get the coefficient matrices in the case of $T= S$ and $T_-$ as
\begin{eqnarray}
 [S]_{\mathcal{B}_{\overline c3}}= \left( \begin{array}{ccc}
   0   & 0  & 0 \\
     0 &   -\lambda  & -1 \\
    0 & \lambda^2 & \lambda \\
  \end{array}\right) \qquad {\rm and} \qquad
 [T_-]_{\mathcal{B}_{\overline c3}}= \left( \begin{array}{ccc}
   0   & 0  & 0 \\
     -1 &  0  & 0 \\
    \lambda & 0 & 0 \\
  \end{array}\right).
\end{eqnarray}

The pseudoscalar meson octet is
\begin{eqnarray}
 [M_8]=  \left( \begin{array}{ccc}
   \frac{1}{\sqrt 2} \pi^0+  \frac{1}{\sqrt 6} \eta_8    & \pi^+  & K^+ \\
    \pi^- &   - \frac{1}{\sqrt 2} \pi^0+ \frac{1}{\sqrt 6} \eta_8   & K^0 \\
    K^- & \overline K^0 & -\sqrt{2/3}\eta_8 \\
  \end{array}\right).
\end{eqnarray}
Two light quarks in the charmed baryon anti-triplet behave as a conjugate representation $\overline 3$.
While light quark pairs $q_i\overline q_j$ in meson octet behave as direct product of basic representation and conjugate representation, $3\times \overline 3$.
Operator $T^i_j$ acts on the basis of basic representation $q^j$ is written as $T^j_i q_j$ and on the basis of conjugate representation $\overline q^i$ is written as $\overline q^i T^j_i$. A conjugate representation in the initial state is equivalent to a basic representation in the final state. Thereby, operator $T$ acting on a pseudoscalar meson octet is not a simple left multiplication of the its coefficient matrix, but a commutator $[\,T,[M_8]_\alpha\,] = T[M_8]_\alpha - [M_8]_\alpha T$:
\begin{align}\label{m}
 [T, \langle [M_8]_\alpha|] = \sum_\beta{\rm Tr}\{[\,T,[M_8]_\alpha\,] [M_8]_\beta^T\}\langle [M_8]_\beta | = \sum_\beta[T_{M_8}]^\beta_\alpha \langle [M_8]_\beta |.
\end{align}
$[T_{M_8}]^\beta_\alpha$ is coefficient of commutator $[\,T,[M_8]_\alpha\,]$ expanded by basis matrices $[M_8]_\beta$.
If we define the pseudoscalar meson octet bases as
\begin{align}
 \langle [M_8]_\beta| = ( \langle \pi^+|,\,\,\langle \pi^0|,\,\,\langle \pi^-|,\,\,\langle K^+|,\,\,\langle K^0|,\,\,\langle \overline K^0|,\,\,\langle K^-|,\,\,\langle \eta_8|    ),
\end{align}
we get
\begin{eqnarray}
 [S]_{M_8}= \left( \begin{array}{cccccccc}
  \lambda & 0& 0& \lambda^2& 0& 0& 0& 0 \\
  0& 0& 0& 0& -\frac{1}{\sqrt{2}}\lambda^2& -\frac{1}{\sqrt{2}}& 0& 0 \\
 0& 0& -\lambda& 0& 0& 0& 1& 0 \\
  -1& 0& 0& -\lambda& 0& 0& 0& 0 \\
  0& \frac{1}{\sqrt{2}}& 0& 0& -2\lambda& 0& 0& -\frac{\sqrt{6}}{2}\\
 0& \frac{1}{\sqrt{2}}\lambda^2& 0& 0& 0& 2\lambda& 0& -\frac{\sqrt{6}}{2}\lambda^2\\
 0& 0& -\lambda^2& 0& 0& 0& \lambda& 0 \\
 0& 0& 0& 0& \frac{\sqrt{6}}{2}\lambda^2& \frac{\sqrt{6}}{2}& 0& 0 \\
  \end{array}\right),
\end{eqnarray}
and
\begin{eqnarray}
 [T_-]_{M_8}= \left( \begin{array}{cccccccc}
  0 & 0& 0& 0& 0& 0& 0& 0 \\
  -\sqrt{2}& 0& 0& -\frac{1}{\sqrt{2}}\lambda& 0& 0& 0& 0 \\
 0& \sqrt{2}& 0& 0& -\lambda& 0& 0& 0 \\
  0& 0& 0& 0& 0& 0& 0& 0 \\
  0& 0& 0& 1& 0& 0& 0& 0\\
 \lambda& 0& 0& 0& 0& 0& 0& 0\\
 0& \frac{1}{\sqrt{2}}\lambda& 0& 0& 0& -1& 0& \frac{\sqrt{6}}{2}\lambda \\
 0& 0& 0& -\frac{\sqrt{6}}{2}\lambda& 0&0& 0& 0 \\
  \end{array}\right).
\end{eqnarray}
The light baryon octet is
\begin{eqnarray}
 [\mathcal{B}_8]=  \left( \begin{array}{ccc}
   \frac{1}{\sqrt 2} \Sigma^0+  \frac{1}{\sqrt 6} \Lambda^0    & \Sigma^+  & p \\
    \Sigma^- &   - \frac{1}{\sqrt 2} \Sigma^0+ \frac{1}{\sqrt 6} \Lambda^0   & n \\
    \Xi^- & \Xi^0 & -\sqrt{2/3}\Lambda^0 \\
  \end{array}\right).
\end{eqnarray}
For the light baryon octet, matrices $[S]_{\mathcal{B}_8}$ and $ [T_-]_{\mathcal{B}_8}$ are the same with $[S]_{M_8}$ and $[T_-]_{M_8}$, respectively.
The light baryon octet bases are
\begin{align}
 \langle [\mathcal{B}_8]_\beta| = ( \langle \Sigma^+|,\,\,\langle \Sigma^0|,\,\,\langle \Sigma^-|,\,\,\langle p|,\,\,\langle n|,\,\,\langle \Xi^0|,\,\,\langle \Xi^-|,\,\,\langle \Lambda^0|).
\end{align}

With the matrix $T_{M_8}$, $T_{\mathcal{B}_8}$ and $T_{\mathcal{B}_{\overline c3}}$, the master formula of $SU(3)$ sum rule for the $[\mathcal{B}_{\overline c3}]_\gamma\to [M_8]_\alpha [\mathcal{B}_8]_\beta$ mode is derived to be
\begin{align}\label{rule1}
  \sum_\mu\left[(T_{M_8})_\alpha^\mu \mathcal{A}_{ \gamma \to \mu \beta} +  (T_{\mathcal{B}_8})_\beta^\mu \mathcal{A}_{\gamma\to \alpha\mu } + (T_{\mathcal{B}_{\overline c3}})_\gamma^\mu \mathcal{A}_{\mu\to \alpha \beta }\right] = 0.
\end{align}
This formula permits us to compute $SU(3)$ sum rules without computing the Wigner-Eckhart invariants.
A $SU(3)$ sum rule of $[\mathcal{B}_{\overline c3}]_\gamma\to [M_8]_\alpha [\mathcal{B}_8]_\beta$ decays is generated via Eq.~\eqref{rule1} if appropriate $\alpha$, $\beta$ and $\gamma$ are selected.
$T_-$ is a $\Delta Q = -1$ operator and $S$ is a QED charge preserving operator.
Thereby, operator $T_-$ produces a $SU(3)$ sum rule for each choice of $\alpha$, $\beta$ and $\gamma$ corresponding to a $\Delta Q = +1$ amplitude and operator $S$ produces a $SU(3)$ sum rule for each choice of $\alpha$, $\beta$ and $\gamma$ corresponding to a $\Delta Q = 0$ amplitude.
For example, the choice of $\{\gamma, \alpha, \beta\} = \{\Xi_c^0, K^0,p\}$ and $T=T_-$ generates a $SU(3)$ sum rule ${\rm SumT_-}\,[\Xi_c^0, K^0,p]$:
\begin{align}
\mathcal{A}(\Lambda_c^+\to K^0p)-\sqrt{\frac{3}{2}}\mathcal{A}(\Xi_c^0\to K^0\Lambda^0)-\frac{\mathcal{A}(\Xi_c^0\to K^0\Sigma^0)}{\sqrt{2}}-\mathcal{A}(\Xi_c^0\to \pi^-p)=0.
\end{align}
In the $[\mathcal{B}_{\overline c3}]_\beta\to \eta_1 [\mathcal{B}_8]_\alpha$ case, where $\eta_1$ is a $SU(3)$ singlet, the coefficient matrix of
$\eta_1$ state is commutative with $S$ and $T_-$, $[S, [\eta_1]]=[T_-, [\eta_1]]=0$.
Then Eq.~\eqref{rule1} is simplified to be
\begin{align}\label{rule2}
  \sum_\mu\left[ (T_{\mathcal{B}_8})_\alpha^\mu \mathcal{A}_{ \beta\to \eta_1\mu} + (T_{\mathcal{B}_{\overline c3}})_\beta^\mu \mathcal{A}_{\mu\to \eta_1 \alpha }\right] = 0.
\end{align}

The $SU(3)$ sum rules for the $\mathcal{B}_{\overline c3}\to M\mathcal{B}_{8}$ modes derived by Eqs.~\eqref{rule1} and \eqref{rule2} are listed in Appendix \ref{res1}.
Notice that $\Xi^0_c$ is an $U$-spin singlet.
In the $U$-spin sum rules generated from $S$  with the choice of $\gamma = \Xi^0_c$, the initial state is always $\Xi^0_c$.
The charmed baryons $\Xi^+_c$ and $\Lambda^+_c$ form an $U$-spin doublet. So the initial states in the $U$-spin sum rules with the choice of $\gamma = \Xi^+_c$ or $\Lambda^+_c$ are both $\Xi^+_c$ and $\Lambda^+_c$.

The light baryon decuplet is given by
{\small \begin{align}\label{b10}
[\mathcal{B}_{10}] = \left(\left( \begin{array}{ccc}
      \Delta^{++} &  \frac{1}{\sqrt{3}}\Delta^{+}  & \frac{1}{\sqrt{3}}\Sigma^{*+} \\
   \frac{1}{\sqrt{3}}\Delta^{+} &  \frac{1}{\sqrt{3}}\Delta^{0}  & \frac{1}{\sqrt{6}}\Sigma^{*0} \\
    \frac{1}{\sqrt{3}}\Sigma^{*+} & \frac{1}{\sqrt{6}}\Sigma^{*0} & \frac{1}{\sqrt{3}}\Xi^{*0} \\
  \end{array}\right)\left( \begin{array}{ccc}\frac{1}{\sqrt{3}}\Delta^{+} &  \frac{1}{\sqrt{3}}\Delta^{0}  & \frac{1}{\sqrt{6}}\Sigma^{*0} \\
   \frac{1}{\sqrt{3}}\Delta^{0} & \Delta^{-}  & \frac{1}{\sqrt{3}}\Sigma^{*-} \\
    \frac{1}{\sqrt{6}}\Sigma^{*0} & \frac{1}{\sqrt{3}}\Sigma^{*-} & \frac{1}{\sqrt{3}}\Xi^{*-}\\\end{array}\right)
    \left(\begin{array}{ccc}\frac{1}{\sqrt{3}}\Sigma^{*+} &  \frac{1}{\sqrt{6}}\Sigma^{*0}  & \frac{1}{\sqrt{3}}\Xi^{*0} \\
   \frac{1}{\sqrt{6}}\Sigma^{*0} & \frac{1}{\sqrt{3}}\Sigma^{*-}  & \frac{1}{\sqrt{3}}\Xi^{*-} \\
    \frac{1}{\sqrt{3}}\Xi^{*0} & \frac{1}{\sqrt{3}}\Xi^{*-} & \Omega^{-}\\\end{array}\right)\right).
\end{align}}
To get a master formula of $SU(3)$ sum rule for the $[\mathcal{B}_{c\overline 3}]_{\gamma}\to [M_8]_\alpha[\mathcal{B}_{10}]_\beta$ modes, we should derive the coefficient matrices $[S]_{\mathcal{B}_{10}}$ and $[T_-]_{\mathcal{B}_{10}}$.
There are three matrices in Eq.~\eqref{b10}. We label them as $D_1$, $D_2$ and $D_3$.
Similarly to Eq.~\eqref{cb3}, the light baryon decuplet can be written as
\begin{align}\label{cb10}
|[D_a]_\alpha\rangle = ([D_a]_\alpha)^{i}_j |[D_a]^{i}_j\rangle,
\end{align}
where subscript $a$ is used to distinguish $D_1$, $D_2$ and $D_3$.
For the meson decuplet, both the indices $i$ and $j$ in Eq.~\eqref{cb10} serve as the bases of basic representation of $SU(3)$ group.
We can compute coefficients of matrix product $T[D_a]_\alpha$ with $[D_a]_\beta$ as bases.
If we define the light baryon decuplet bases as
\begin{align}
 \langle [D_1]_\alpha| & = ( \langle \Delta^{++}|,\,\,\langle \Delta^0|,\,\,\langle \Xi^{*0}|,\,\,\langle \Delta^+|,\,\,\langle \Sigma^{*+}|,\,\,\langle \Sigma^{*0}| ),\\
  \langle [D_2]_\alpha| & = ( \langle \Delta^{+}|,\,\,\langle \Delta^-|,\,\,\langle \Xi^{*-}|,\,\,\langle \Delta^0|,\,\,\langle \Sigma^{*0}|,\,\,\langle \Sigma^{*-}| ),\\
  \langle [D_3]_\alpha| & = ( \langle \Sigma^{*+}|,\,\,\langle \Sigma^{*-}|,\,\,\langle \Omega^{-}|,\,\,\langle \Sigma^{*0}|,\,\,\langle \Xi^{*0}|,\,\,\langle \Xi^{*-}| ),
\end{align}
$[S]_{D_a}$ and $[T_-]_{D_a}$ are derived to be
\begin{eqnarray}
 [S]_{D_1}= \left( \begin{array}{cccccc}
  0 & 0& 0& 0& 0& 0 \\
  0 & -\frac{\lambda}{3}& 0& 0& 0& \frac{1}{3\sqrt{2}} \\
  0 & 0& \frac{\lambda}{3}& 0& 0& -\frac{\lambda^2}{3\sqrt{2}} \\    0 & 0& 0& -\frac{\lambda}{3}& \frac{1}{3}& 0 \\
  0 & 0& 0& -\frac{\lambda^2}{3}& \frac{\lambda}{3}& 0 \\
  0 & -\frac{\lambda^2}{3\sqrt{2}}& \frac{1}{3\sqrt{2}}& 0& 0& 0 \\
  \end{array}\right),
\end{eqnarray}
\begin{eqnarray}
 [S]_{D_2}= \left( \begin{array}{cccccc}
  0 & 0& 0& 0& 0& 0 \\
  0 & -\lambda& 0& 0& 0& \frac{1}{\sqrt{3}} \\
  0 & 0& \frac{\lambda}{3}& 0& 0& -\frac{\lambda^2}{3} \\
  0 & 0& 0& -\frac{\lambda}{3}& \frac{1}{3\sqrt{2}}& 0 \\
  0 & 0& 0& -\frac{\lambda^2}{3\sqrt{2}}& \frac{\lambda}{6}& 0 \\
  0 & -\frac{\lambda^2}{\sqrt{3}}& \frac{1}{3}& 0& 0& 0 \\
  \end{array}\right),
\end{eqnarray}
\begin{eqnarray}
 [S]_{D_3}= \left( \begin{array}{cccccc}
  0 & 0& 0& 0& 0& 0 \\
  0 & -\frac{\lambda}{3}& 0& 0& 0& \frac{1}{3} \\
  0 & 0& \lambda& 0& 0& -\frac{\lambda^2}{\sqrt{3}} \\
  0 & 0& 0& -\frac{\lambda}{6}& \frac{1}{3\sqrt{2}}& 0 \\
  0 & 0& 0& -\frac{\lambda^2}{3\sqrt{2}}& \frac{\lambda}{3}& 0 \\
  0 & -\frac{\lambda^2}{3}& \frac{1}{\sqrt{3}}& 0& 0& 0 \\
  \end{array}\right),
\end{eqnarray}
\begin{eqnarray}
 [T_-]_{D_1}= \left( \begin{array}{cccccc}
  0 & 0& 0& 0& 0& 0 \\
  0 & 0& 0& \frac{1}{3}& 0& 0 \\
  0 & 0& 0& 0& \frac{\lambda}{3}& 0 \\
  \frac{1}{\sqrt{3}} & 0& 0& 0& 0& 0 \\
  \frac{\lambda}{\sqrt{3}} & 0& 0& 0& 0& 0 \\
  0 & 0& 0& \frac{\lambda}{3\sqrt{2}}& \frac{1}{3\sqrt{2}}& 0 \\
  \end{array}\right),
\end{eqnarray}
\begin{eqnarray}
 [T_-]_{D_2}= \left( \begin{array}{cccccc}
  0 & 0& 0& 0& 0& 0 \\
  0 & 0& 0& \frac{1}{\sqrt{3}}& 0& 0 \\
  0 & 0& 0& 0& \frac{\lambda}{3\sqrt{2}}& 0 \\
  \frac{1}{3} & 0& 0& 0& 0& 0 \\
  \frac{\lambda}{3\sqrt{2}} & 0& 0& 0& 0& 0 \\
  0 & 0& 0& \frac{\lambda}{3}& \frac{1}{3\sqrt{2}}& 0 \\
  \end{array}\right),
\end{eqnarray}
\begin{eqnarray}
 [T_-]_{D_3}= \left( \begin{array}{cccccc}
  0 & 0& 0& 0& 0& 0 \\
  0 & 0& 0& \frac{1}{3\sqrt{2}}& 0& 0 \\
  0 & 0& 0& 0& \frac{\lambda}{\sqrt{3}}& 0 \\
  \frac{1}{3\sqrt{2}} & 0& 0& 0& 0& 0 \\
  \frac{\lambda}{3} & 0& 0& 0& 0& 0 \\
  0 & 0& 0& \frac{\lambda}{3\sqrt{2}}& \frac{1}{3}& 0 \\
  \end{array}\right).
\end{eqnarray}

The master formula of $SU(3)$ sum rule for the $[\mathcal{B}_{\overline c3}]_\gamma\to [M_8]_\alpha [\mathcal{B}_{10}]_\beta$ mode is
\begin{align}\label{rule3}
  \sum_\mu\left[(T_{M_8})_\alpha^\mu \mathcal{A}_{\mu \beta \gamma} +  3\sum_a(T_{D_a})_\beta^\mu \mathcal{A}_{\alpha\mu \gamma}  + (T_{\mathcal{B}_{\overline c3}})_\gamma^\mu \mathcal{A}_{\alpha \beta \mu}\right] = 0.
\end{align}
The factor $3$ in the second term is arisen from the fact that operator $T$ acts on the first, second and third quarks in the light baryon decuplet and get the same coefficient matrices due to the symmetric flavor indies of decuplet.
In the $[\mathcal{B}_{\overline c3}]_\beta\to \eta_1 [\mathcal{B}_{10}]_\alpha$ case, Eq.~\eqref{rule3} is simplified to be
\begin{align}\label{rule4}
  \sum_\mu\left[ 3\sum_a(T_{D_a})_\alpha^\mu \mathcal{A}_{\beta\to \eta_1\mu } + (T_{\mathcal{B}_{\overline c3}})_\beta^\mu \mathcal{A}_{\mu\to \eta_1 \alpha }\right] = 0.
\end{align}
The $SU(3)$ sum rules for the $\mathcal{B}_{\overline c3}\to M\mathcal{B}_{10}$ modes computed through Eq.~\eqref{rule3} and Eq.~\eqref{rule4} are listed in Appendix \ref{res2}.

The $SU(3)$ sum rules with pseudoscalar meson final state can be extended to the vector meson final state. All we need to do is replacing pseudoscalar meson with vector meson in Appendixes \ref{res1} and \ref{res2}. The pseudoscalar mesons $\eta_8$ and $\eta_1$ are not mass eigenstates.
The mass eigenstates $\eta$ and $\eta^\prime$ are mixing of $\eta_8$ and $\eta_1$,
\begin{eqnarray}
\left( \begin{array}{ccc}
\eta\\
\eta^\prime
\end{array}
\right)
=
\left(
\begin{array}{cc}
\cos\xi  &  -\sin\xi\\
\sin\xi  &  \cos\xi
\end{array}
\right)\left(
\begin{array}{c}
\eta_8\\
\eta_1
\end{array}\right).
\end{eqnarray}
Reversing the matrix equation, we have
\begin{eqnarray}\label{eta}
\left( \begin{array}{ccc}
\eta_8\\
\eta_1
\end{array}
\right)
=
\left(
\begin{array}{cc}
\cos\xi  &  \sin\xi\\
-\sin\xi  &  \cos\xi
\end{array}
\right)\left(
\begin{array}{c}
\eta\\
\eta^\prime
\end{array}\right).
\end{eqnarray}
Similarly, vector mesons $\omega_8$ and $\omega_1$ can be expressed as mixing of the mass eigenstates $\omega$ and $\phi$,
\begin{align}\label{phi}
\left(
\begin{array}{c}
\omega_8\\
\omega_1
\end{array}
\right)
=
\left(
\begin{array}{cc}
\cos\xi^\prime  &  \sin\xi^\prime\\
-\sin\xi^\prime  &  \cos\xi^\prime
\end{array}
\right)\left(
\begin{array}{c}
\phi\\
\omega
\end{array}
\right).
\end{align}
One can get the $SU(3)$ sum rules involving the mass eigenstates $\eta$, $\eta^\prime$, $\omega$ and $\phi$ from Appendices \ref{res1} and \ref{res2} according to Eqs.~\eqref{eta} and ~\eqref{phi}.

New $SU(3)$ sum rules can be found by combining the $SU(3)$ sum rules listed in Appendices \ref{res1} and \ref{res2}.
Among those $SU(3)$ sum rules, the ones that hold in isospin symmetry are more accurate than others.
One can search for the isospin sum rules by combining the $SU(3)$ sum rules generated from operator $T_-$.
In our previous work \cite{Jia:2019zxi}, we list the isospin sum rules for singly charmed baryon decays. Please see Eqs.~(20)-(22) and Appendix B of Ref.~\cite{Jia:2019zxi} for details.
In addition, if a $SU(3)$ sum rule involves only two decay modes, it also an equation of decay rate.
The rate equations for singly charmed baryon decays can be found in Ref.~\cite{Savage:1989qr}.
Among those rate equations, the following equations hold under isospin symmetry,
\begin{equation}
\mathcal{B}r(\Lambda_c^+\rightarrow\Sigma^+\pi^0)=\mathcal{B}r(\Lambda_c^+\rightarrow\Sigma^0\pi^+),
\end{equation}
\begin{equation}
\mathcal{B}r(\Lambda_c^+\rightarrow\Sigma^{*+}\pi^0)=\mathcal{B}r(\Lambda_c^+\rightarrow\Sigma^{*0}\pi^+),
\end{equation}
\begin{equation}
\mathcal{B}r(\Lambda_c^+\rightarrow\Delta^{++}K^-)=3\mathcal{B}r(\Lambda_c^+\rightarrow\Delta^+\overline{K}^0),
\end{equation}
\begin{equation}
\mathcal{B}r(\Lambda_c^+\rightarrow\Delta^+K^0)=\mathcal{B}r(\Lambda_c^+\rightarrow\Delta^0K^+),
\end{equation}
which could be used to test the isospin symmetry in the charmed baryon decay modes by experimental measurements.


\section{Doubly charmed baryon decay}\label{doubly}

The doubly charmed baryon triplet can be written as
$|[\mathcal{B}_{cc}]_\alpha\rangle = ([\mathcal{B}_{cc}]_\alpha)_i |[\mathcal{B}_{cc}]_{i}\rangle$,
in which the light quarks serve bases of basic representation of $SU(3)$ group.
Since a basic representation in the final state is equivalent to a conjugate representation in the initial state, we compute coefficients of matrix product $[\mathcal{B}_{cc}]_\alpha\,T$ with $[\mathcal{B}_{cc}]_\beta$ as bases.
If we define the doubly charmed baryon triplet bases as
 $|[\mathcal{B}_{cc}]_\alpha\rangle = ( |\Xi_{cc}^{++}\rangle,\,\, |\Xi_{cc}^{+}\rangle ,\,\, |\Omega_{cc}^{+}\rangle )$,
the coefficient matrices in the case of $T= S$ and $T_-$ are
\begin{eqnarray}
 [S]_{\mathcal{B}_{cc}}= \left( \begin{array}{ccc}
   0   & 0  & 0 \\
     0 &   -\lambda  & -\lambda^2 \\
    0 & 1 & \lambda \\
  \end{array}\right) \qquad {\rm and} \qquad
 [T_-]_{\mathcal{B}_{cc}}= \left( \begin{array}{ccc}
   0   & 1  & \lambda \\
     0 &  0  & 0 \\
    0 & 0 & 0 \\
  \end{array}\right).
\end{eqnarray}

The master formula of $SU(3)$ sum rule \eqref{rule} for the $[\mathcal{B}_{cc}]_\gamma\to [M_8]_\alpha [\mathcal{B}_{\overline c3}]_\beta$ mode is
\begin{align}\label{rule5}
  \sum_\mu\left[(T_{M_8})_\alpha^\mu \mathcal{A}_{\gamma\to \mu \beta } -  (T^T_{\mathcal{B}_{\overline c3}})_\beta^\mu \mathcal{A}_{\gamma\to \alpha\mu } -(T_{\mathcal{B}_{cc}})_\gamma^\mu \mathcal{A}_{\mu\to\alpha \beta }\right] = 0.
\end{align}
The superscript $T$ in the second term denotes the transposition of matrices $[S]_{\mathcal{B}_{\overline c3}}$ and $[T_-]_{\mathcal{B}_{\overline c3}}$, which is arisen from the initial-final transformation for the charmed baryon anti-triplet.
The minus signs in the last two terms are used to match the minus sign of commutator in Eq.~\eqref{m}.
In the $[\mathcal{B}_{cc}]_\beta\to \eta_1 [\mathcal{B}_{\overline c3}]_\alpha$ case, Eq.~\eqref{rule5} is simplified to be
\begin{align}\label{rule6}
  \sum_\mu\left[ (T^T_{\mathcal{B}_{\overline c3}})_\alpha^\mu \mathcal{A}_{ \beta\to \eta_1\mu} + (T_{\mathcal{B}_{cc}})_\beta^\mu \mathcal{A}_{\mu\to\eta_1 \alpha }\right] = 0.
\end{align}
The $SU(3)$ sum rules for the $\mathcal{B}_{cc}\to M\mathcal{B}_{\overline c3}$ modes computed by Eq.~\eqref{rule5} and \eqref{rule6} are listed in Appendix \ref{res3}.

The charmed baryon sextet is
\begin{eqnarray}
 [\mathcal{B}_{c6}]=  \left( \begin{array}{ccc}
   \Sigma_c^{++}   &  \frac{1}{\sqrt{2}}\Sigma_c^{+}  & \frac{1}{\sqrt{2}}\Xi_c^{*+} \\
   \frac{1}{\sqrt{2}}\Sigma_c^{+} &   \Sigma_c^{0}   & \frac{1}{\sqrt{2}}\Xi_c^{*0} \\
    \frac{1}{\sqrt{2}}\Xi_c^{*+} & \frac{1}{\sqrt{2}}\Xi_c^{*0} & \Omega_c^0 \\
  \end{array}\right).
\end{eqnarray}
If we define the charmed baryon sextet bases as
 $|[\mathcal{B}_{cc}]_\alpha\rangle = ( |\Sigma_{c}^{++}\rangle,\,\, |\Sigma_{c}^{0}\rangle,\,\, |\Omega_{c}^{0}\rangle, \,\, |\Sigma_{c}^{+}\rangle,\,\, |\Xi_{c}^{*+}\rangle,\,\, |\Xi_{c}^{*0}\rangle )$,
the coefficient matrices in the case of $T= S$ and $T_-$ are
\begin{eqnarray}
 [S]_{\mathcal{B}_{c6}}= \left( \begin{array}{cccccc}
  0 & 0& 0& 0& 0& 0 \\
  0 & -\lambda& 0& 0& 0& \frac{1}{\sqrt{2}} \\
  0 & 0& \lambda& 0& 0& -\frac{\lambda^2}{\sqrt{2}} \\
  0 & 0& 0& -\frac{\lambda}{2}& \frac{1}{2}& 0 \\
  0 & 0& 0& -\frac{\lambda^2}{2}& \frac{\lambda}{2}& 0 \\
  0 & -\frac{\lambda^2}{\sqrt{2}}& \frac{1}{\sqrt{2}}& 0& 0& 0 \\
  \end{array}\right),\qquad
 [T_-]_{\mathcal{B}_{c6}}= \left( \begin{array}{cccccc}
  0 & 0& 0& 0& 0& 0 \\
  0 & 0& 0& \frac{1}{\sqrt{2}}& 0& 0 \\
  0 & 0& 0& 0& \frac{\lambda}{\sqrt{2}}& 0 \\
  \frac{1}{\sqrt{2}} & 0& 0& 0& 0& 0 \\
  \frac{\lambda}{\sqrt{2}} & 0& 0& 0& 0& 0 \\
  0 & 0& 0& \frac{\lambda}{2}& \frac{1}{2}& 0 \\
  \end{array}\right).
\end{eqnarray}
The master formula of $SU(3)$ sum rule for the $[\mathcal{B}_{cc}]_\gamma\to [M_8]_\alpha [\mathcal{B}_{c6}]_\beta$ mode is
\begin{align}\label{rule7}
  \sum_\mu\left[(T_{M_8})_\alpha^\mu \mathcal{A}_{\gamma\to \mu \beta } +  2(T_{\mathcal{B}_{c6}})_\beta^\mu \mathcal{A}_{\gamma\to \alpha\mu } - (T_{\mathcal{B}_{cc}})_\gamma^\mu \mathcal{A}_{ \mu\to \alpha \beta}\right] = 0.
\end{align}
The factor $2$ in the second term is arisen from the fact that operator $T$ acts on the first and second quarks in the charmed baryon sextet getting the same coefficient matrices.
In the $[\mathcal{B}_{ cc}]_\beta\to \eta_1 [\mathcal{B}_{c6}]_\alpha$ case, Eq.~\eqref{rule7} is simplified to be
\begin{align}\label{rule8}
  \sum_\mu\left[ 2(T_{\mathcal{B}_{c6}})_\alpha^\mu \mathcal{A}_{\beta\to \eta_1\mu } - (T_{\mathcal{B}_{cc}})_\beta^\mu \mathcal{A}_{ \mu\to\eta_1 \alpha}\right] = 0.
\end{align}
The $SU(3)$ sum rules for the $\mathcal{B}_{cc}\to M\mathcal{B}_{c6}$ modes derived from Eq.~\eqref{rule7} and Eq.~\eqref{rule8} are listed  in Appendix \ref{res4}.

The $SU(3)$ sum rules that involve two decay modes in Appendices \ref{res3} and \ref{res4} are related to equations of decay rate.
By combining the $SU(3)$ sum rules generated from operator $T_-$, we find six isospin sum rules for doubly charmed baryon decays:
\begin{align}
-\sqrt{2}\mathcal{A}(\Xi_{cc}^{+}\to  \pi^0\Xi_c^{+})-\mathcal{A}(\Xi_{cc}^{++}\to \pi^+\Xi_c^+)+\mathcal{A}(\Xi_{cc}^{+}\to \pi^+\Xi_c^{0})=0,
\end{align}
\begin{align}
&\sqrt{2}\mathcal{A}(\Xi_{cc}^{++}\to  \pi^+\Sigma_c^{+})+2\mathcal{A}(\Xi_{cc}^{+}\to \pi^0\Sigma_c^{+})-\mathcal{A}(\Xi_{cc}^{+}\to \pi^+\Sigma_c^{0})\nonumber\\&~~~-\sqrt{2}\mathcal{A}(\Xi_{cc}^{++}\to \pi^0 \Sigma_c^{++})+\mathcal{A}(\Xi_{cc}^{+}\to \pi^-\Sigma_c^{++})=0,
\end{align}
\begin{align}
\mathcal{A}(\Xi_{cc}^{++}\to  \overline K^0\Sigma_c^{++})+\mathcal{A}(\Xi_{cc}^{+}\to K^-\Sigma_c^{++})-\sqrt{2}\mathcal{A}(\Xi_{cc}^{+}\to \overline K^0\Sigma_c^{+})=0,
\end{align}
\begin{align}
\mathcal{A}(\Xi_{cc}^{++}\to   K^0\Sigma_c^{++})+\sqrt{2}\mathcal{A}(\Xi_{cc}^{++}\to K^+\Sigma_c^{+})-\sqrt{2}\mathcal{A}(\Xi_{cc}^{+}\to  K^0\Sigma_c^{+})-\mathcal{A}(\Xi_{cc}^{+}\to  K^+\Sigma_c^{0})=0,
\end{align}
\begin{align}
-\sqrt{2}\mathcal{A}(\Xi_{cc}^{+}\to  \pi^0\Xi_c^{*+})-\mathcal{A}(\Xi_{cc}^{++}\to \pi^+\Xi_c^{*+})+\sqrt{2}\mathcal{A}(\Xi_{cc}^{+}\to \pi^+\Xi_c^{*0})=0,
\end{align}
\begin{align}
\mathcal{A}(\Omega_{cc}^{+}\to  \pi^-\Sigma_c^{++})+2\mathcal{A}(\Omega_{cc}^{++}\to \pi^0\Sigma_c^{+})-\mathcal{A}(\Omega_{cc}^{+}\to \pi^+\Sigma_c^{0})=0.
\end{align}

\section{Summary}\label{sum}

Flavor $SU(3)$ symmetry is a powerful tool to analyze the charmed baryon decays since the factorization scheme based on the expansion of $1/m_c$ is not always valid.
In this work, we propose a method to compute amplitude sum rules for the singly and doubly charmed baryon decays in the $SU(3)_F$ limit without writing the Wigner-Eckhart invariants.
Hundreds of $SU(3)$ sum rules are derived systematically which can be tested by experiments.

\begin{acknowledgements}

This work was supported in part by the National Natural Science Foundation of China under Grants No. 12105099.

\end{acknowledgements}

\appendix

\section{SU(3) sum rules of $\mathcal{B}_{\overline c3}\to M\mathcal{B}_{8}$ modes}\label{res1}

\begin{align}
&{\rm SumS}\,[\Xi_c^0,\pi^0,\Sigma^0]:\nonumber\\&\,\mathcal{A}(\Xi_c^0\to K^0\Sigma^0)+\mathcal{A}(\Xi_c^0\to \pi^0n)+\lambda^2\mathcal{A}(\Xi_c^0\to \overline K^0\Sigma^0)+\lambda^2\mathcal{A}(\Xi_c^0\to \pi^0\Xi^0)=0.
\end{align}
\begin{align}
&{\rm SumS}\,[\Xi_c^0,\pi^0,\Lambda^0]:\nonumber\\&\,\frac{\mathcal{A}(\Xi_c^0\to K^0\Lambda^0)}{\sqrt{2}}-\sqrt{\frac{3}{2}}\mathcal{A}(\Xi_c^0\to \pi^0n)+\frac{\lambda^2\mathcal{A}(\Xi_c^0\to \overline K^0\Lambda^0)}{\sqrt{2}}-\sqrt{\frac{3}{2}}\lambda^2\mathcal{A}(\Xi_c^0\to \pi^0\Xi^0)=0.
\end{align}
\begin{align}
&{\rm SumS}\,[\Xi_c^0,\pi^0,n]:\nonumber\\&\,-2\lambda\mathcal{A}(\Xi_c^0\to \pi^0n)+\frac{\lambda^2\mathcal{A}(\Xi_c^0\to \overline K^0n)}{\sqrt{2}}+\sqrt{\frac{3}{2}}\lambda^2\mathcal{A}(\Xi_c^0\to \pi^0\Lambda^0)-\frac{\lambda^2\mathcal{A}(\Xi_c^0\to \pi^0\Sigma^0)}{\sqrt{2}}=0.
\end{align}
\begin{align}
&{\rm SumS}\,[\Xi_c^0,\pi^0,\Xi^0]:\nonumber\\&\,\frac{\mathcal{A}(\Xi_c^0\to K^0\Xi^0)}{\sqrt{2}}+\sqrt{\frac{3}{2}}\mathcal{A}(\Xi_c^0\to \pi^0\Lambda^0)-\frac{\mathcal{A}(\Xi_c^0\to \pi^0\Sigma^0)}{\sqrt{2}}+2\lambda\mathcal{A}(\Xi_c^0\to \pi^0\Xi^0)=0.
\end{align}
\begin{align}
&{\rm SumS}\,[\Xi_c^0,\eta_8,\Sigma^0]:\nonumber\\&\,\frac{\mathcal{A}(\Xi_c^0\to \eta_8n)}{\sqrt{2}}-\sqrt{\frac{3}{2}}\mathcal{A}(\Xi_c^0\to K^0\Sigma^0)+\frac{\lambda^2\mathcal{A}(\Xi_c^0\to \eta_8\Xi^0)}{\sqrt{2}}-\sqrt{\frac{3}{2}}\lambda^2\mathcal{A}(\Xi_c^0\to \overline K^0\Sigma^0)=0.
\end{align}
\begin{align}
&{\rm SumS}\,[\Xi_c^0,\eta_8,\Lambda^0]:\nonumber\\\,&\mathcal{A}(\Xi_c^0\to \eta_8n)+\mathcal{A}(\Xi_c^0\to K^0\Lambda^0)+\lambda^2\mathcal{A}(\Xi_c^0\to \eta_8\Xi^0)+\lambda^2\mathcal{A}(\Xi_c^0\to \overline K^0\Lambda^0)=0.
\end{align}
\begin{align}
&{\rm SumS}\,[\Xi_c^0,\eta_8,n]:\nonumber\\&\,-2\lambda\mathcal{A}(\Xi_c^0\to \eta_8n)+\sqrt{\frac{3}{2}}\lambda^2\mathcal{A}(\Xi_c^0\to \eta_8\Lambda^0)-\frac{\lambda^2\mathcal{A}(\Xi_c^0\to \eta_8\Sigma^0)}{\sqrt{2}}-\sqrt{\frac{3}{2}}\lambda^2\mathcal{A}(\Xi_c^0\to \overline K^0n)=0.
\end{align}
\begin{align}
&{\rm SumS}\,[\Xi_c^0,\eta_8,\Xi^0]:\nonumber\\&\,\sqrt{\frac{3}{2}}\lambda\mathcal{A}(\Xi_c^0\to \eta_8\Lambda^0)-\frac{\mathcal{A}(\Xi_c^0\to \eta_8\Sigma^0)}{\sqrt{2}}-\sqrt{\frac{3}{2}}\mathcal{A}(\Xi_c^0\to K^0\Xi^0)+2\lambda\mathcal{A}(\Xi_c^0\to \eta_8\Xi^0)=0.
\end{align}
\begin{align}
&{\rm SumS}\,[\Xi_c^0,K^0,\Sigma^0]:\nonumber\\\,&-2\lambda\mathcal{A}(\Xi_c^0\to K^0\Sigma^0)+\sqrt{\frac{3}{2}}\lambda^2\mathcal{A}(\Xi_c^0\to \eta_8\Sigma^0)+\frac{\lambda^2\mathcal{A}(\Xi_c^0\to K^0\Xi^0)}{\sqrt{2}}-\frac{\lambda^2\mathcal{A}(\Xi_c^0\to \pi^0\Sigma^0)}{\sqrt{2}}=0.
\end{align}
\begin{align}
{\rm SumS}\,[\Xi_c^0,K^0,\Lambda^0]:&\,-2\lambda\mathcal{A}(\Xi_c^0\to K^0\Lambda^0)+\sqrt{\frac{3}{2}}\lambda^2\mathcal{A}(\Xi_c^0\to \eta_8\Lambda^0)\nonumber\\&~~~-\sqrt{\frac{3}{2}}\lambda^2\mathcal{A}(\Xi_c^0\to K^0\Xi^0)-\frac{\lambda^2\mathcal{A}(\Xi_c^0\to \pi^0\Lambda^0)}{\sqrt{2}}=0.
\end{align}
\begin{align}
&{\rm SumS}\,[\Xi_c^0,K^0,n]:\nonumber\\&\,\sqrt{\frac{3}{2}}\mathcal{A}(\Xi_c^0\to \eta_8n)+\sqrt{\frac{3}{2}}\mathcal{A}(\Xi_c^0\to K^0\Lambda^0)-\frac{\mathcal{A}(\Xi_c^0\to K^0\Sigma^0)}{\sqrt{2}}-\frac{\mathcal{A}(\Xi_c^0\to \pi^0n)}{\sqrt{2}}=0.
\end{align}
\begin{align}
&{\rm SumS}\,[\Xi_c^0,K^0,\Xi^0]:\nonumber\\&\,\sqrt{\frac{3}{2}}\mathcal{A}(\Xi_c^0\to K^0\Lambda^0)-\frac{\mathcal{A}(\Xi_c^0\to K^0\Sigma^0)}{\sqrt{2}}+\sqrt{\frac{3}{2}}\lambda^2\mathcal{A}(\Xi_c^0\to \eta_8\Xi^0)-\frac{\lambda^2\mathcal{A}(\Xi_c^0\to \pi^0\Xi^0)}{\sqrt{2}}=0.
\end{align}
\begin{align}
&{\rm SumS}\,[\Xi_c^0,\overline K^0,\Sigma^0]:\nonumber\\&\,\sqrt{\frac{3}{2}}\mathcal{A}(\Xi_c^0\to \eta_8\Sigma^0)+\frac{\mathcal{A}(\Xi_c^0\to \overline K^0n)}{\sqrt{2}}-\frac{\mathcal{A}(\Xi_c^0\to \pi^0\Sigma^0)}{\sqrt{2}}+2\lambda\mathcal{A}(\Xi_c^0\to \overline K^0\Sigma^0)=0.
\end{align}
\begin{align}
&{\rm SumS}\,[\Xi_c^0,\overline K^0,\Lambda^0]:\nonumber\\&\,\sqrt{\frac{3}{2}}\mathcal{A}(\Xi_c^0\to \eta_8\Lambda^0)-\sqrt{\frac{3}{2}}\mathcal{A}(\Xi_c^0\to \overline K^0n)-\frac{\mathcal{A}(\Xi_c^0\to \pi^0\Lambda^0)}{\sqrt{2}}+2\lambda\mathcal{A}(\Xi_c^0\to \overline K^0\Lambda^0)=0.
\end{align}
\begin{align}
&{\rm SumS}\,[\Xi_c^0,\overline K^0,n]:\nonumber\\&\,\sqrt{\frac{3}{2}}\mathcal{A}(\Xi_c^0\to \eta_8n)-\frac{\mathcal{A}(\Xi_c^0\to  \pi^0n)}{\sqrt{2}}+\sqrt{\frac{3}{2}}\lambda^2\mathcal{A}(\Xi_c^0\to \overline K^0\Lambda^0)-\frac{\lambda^2\mathcal{A}(\Xi_c^0\to \overline K^0\Sigma^0)}{\sqrt{2}}=0.
\end{align}
\begin{align}
&{\rm SumS}\,[\Xi_c^0,\overline K^0,\Xi^0]:\nonumber\\&\,\sqrt{\frac{3}{2}}\mathcal{A}(\Xi_c^0\to \eta_8\Xi^0)+\sqrt{\frac{3}{2}}\mathcal{A}(\Xi_c^0\to  \overline K^0\Lambda^0)-\frac{\mathcal{A}(\Xi_c^0\to \overline K^0\Sigma^0)}{\sqrt{2}}-\frac{\mathcal{A}(\Xi_c^0\to \pi^0\Xi^0)}{\sqrt{2}}=0.
\end{align}
\begin{align}
{\rm SumS}\,[\Xi_c^0, K^+,\Xi^-]:\,\mathcal{A}(\Xi_c^0\to K^+\Sigma^-)+\lambda^2\mathcal{A}(\Xi_c^0\to \pi^+\Xi^-)=0.
\end{align}
\begin{align}
{\rm SumS}\,[\Xi_c^0, K^+,\Sigma^-]:\,-2\lambda\mathcal{A}(\Xi_c^0\to K^+\Sigma^-)-\lambda^2\mathcal{A}(\Xi_c^0\to K^+\Xi^-)+\lambda^2\mathcal{A}(\Xi_c^0\to \pi^+\Sigma^-)=0.
\end{align}
\begin{align}
{\rm SumS}\,[\Xi_c^0, \pi^+,\Xi^-]:\,-\mathcal{A}(\Xi_c^0\to K^+\Xi^-)+\mathcal{A}(\Xi_c^0\to \pi^+\Sigma^-)+2\lambda\mathcal{A}(\Xi_c^0\to \pi^+\Xi^-)=0.
\end{align}
\begin{align}
{\rm SumS}\,[\Xi_c^0, \pi^+,\Sigma^-]:\,\mathcal{A}(\Xi_c^0\to K^+\Sigma^-)+\lambda^2\mathcal{A}(\Xi_c^0\to \pi^+\Xi^-)=0.
\end{align}
\begin{align}
{\rm SumS}\,[\Xi_c^0, K^-,p]:\,\mathcal{A}(\Xi_c^0\to \pi^-p)+\lambda^2\mathcal{A}(\Xi_c^0\to K^-\Sigma^+)=0.
\end{align}
\begin{align}
{\rm SumS}\,[\Xi_c^0, \pi^-,p]:\,-2\lambda\mathcal{A}(\Xi_c^0\to \pi^-p)-\lambda^2\mathcal{A}(\Xi_c^0\to K^-p)+\lambda^2\mathcal{A}(\Xi_c^0\to \pi^-\Sigma^+)=0.
\end{align}
\begin{align}
{\rm SumS}\,[\Xi_c^0, K^-,\Sigma^+]:\,-\mathcal{A}(\Xi_c^0\to K^-p)+\mathcal{A}(\Xi_c^0\to \pi^-\Sigma^+)+2\lambda\mathcal{A}(\Xi_c^0\to K^-\Sigma^+)=0.
\end{align}
\begin{align}
{\rm SumS}\,[\Xi_c^0, \pi^-,\Sigma^+]:\,\mathcal{A}(\Xi_c^0\to \pi^-p)+\lambda^2\mathcal{A}(\Xi_c^0\to K^-\Sigma^+)=0.
\end{align}
\begin{align}
&{\rm SumS}\,[\Xi_c^+, K^+,\Sigma^0]:\nonumber\\\,&-2\lambda\mathcal{A}(\Xi_c^+\to K^+\Sigma^0)+\lambda^2\mathcal{A}(\Lambda_c^+\to K^+\Sigma^0)+\frac{\lambda^2\mathcal{A}(\Xi_c^+\to K^+\Xi^0)}{\sqrt{2}}+\lambda^2\mathcal{A}(\Xi_c^+\to \pi^+\Sigma^0)=0.
\end{align}
\begin{align}
{\rm SumS}\,[\Xi_c^+, K^+,\Lambda^0]:&-2\lambda\mathcal{A}(\Xi_c^+\to K^+\Lambda^0)+\lambda^2\mathcal{A}(\Lambda_c^+\to K^+\Lambda^0)\nonumber\\&~~~-\sqrt{\frac{3}{2}}\lambda^2\mathcal{A}(\Xi_c^+\to K^+\Xi^0)+\lambda^2\mathcal{A}(\Xi_c^+\to \pi^+\Lambda^0)=0.
\end{align}
\begin{align}
&{\rm SumS}\,[\Xi_c^+, K^+,n]:\nonumber\\&\,\mathcal{A}(\Lambda_c^+\to K^+n)+\sqrt{\frac{3}{2}}\mathcal{A}(\Xi_c^+\to K^+\Lambda^0)-\frac{\mathcal{A}(\Xi_c^+\to K^+\Sigma^0)}{\sqrt{2}}+\mathcal{A}(\Xi_c^+\to \pi^+n)=0.
\end{align}
\begin{align}
&{\rm SumS}\,[\Xi_c^+, K^+,\Xi^0]:\nonumber\\&\,\sqrt{\frac{3}{2}}\mathcal{A}(\Xi_c^+\to K^+\Lambda^0)-\frac{\mathcal{A}(\Xi_c^+\to K^+\Sigma^0)}{\sqrt{2}}+\lambda^2\mathcal{A}(\Lambda_c^+\to K^+\Xi^0)+\lambda^2\mathcal{A}(\Xi_c^+\to \pi^+\Xi^0)=0.
\end{align}
\begin{align}
&{\rm SumS}\,[\Xi_c^+, \pi^+,\Sigma^0]:\nonumber\\&\,-\mathcal{A}(\Xi_c^+\to K^+\Sigma^0)+\frac{\mathcal{A}(\Xi_c^+\to \pi^+n)}{\sqrt{2}}+\lambda^2\mathcal{A}(\Lambda_c^+\to \pi^+\Sigma^0)+\frac{\lambda^2\mathcal{A}(\Xi_c^+\to \pi^+\Xi^0)}{\sqrt{2}}=0.
\end{align}
\begin{align}
&{\rm SumS}\,[\Xi_c^+, \pi^+,\Lambda^0]:\nonumber\\&\,-\mathcal{A}(\Xi_c^+\to K^+\Lambda^0)-\sqrt{\frac{3}{2}}\mathcal{A}(\Xi_c^+\to \pi^+n)+\lambda^2\mathcal{A}(\Lambda_c^+\to \pi^+\Lambda^0)-\sqrt{\frac{3}{2}}\lambda^2\mathcal{A}(\Xi_c^+\to \pi^+\Xi^0)=0.
\end{align}
\begin{align}
&{\rm SumS}\,[\Xi_c^+, \pi^+,n]:\nonumber\\&\,-2\lambda\mathcal{A}(\Xi_c^+\to \pi^+n)+\lambda^2\mathcal{A}(\Lambda_c^+\to \pi^+n)+\sqrt{\frac{3}{2}}\lambda^2\mathcal{A}(\Xi_c^+\to \pi^+\Lambda^0)-\frac{\lambda^2\mathcal{A}(\Xi_c^+\to \pi^+\Sigma^0)}{\sqrt{2}}=0.
\end{align}
\begin{align}
&{\rm SumS}\,[\Xi_c^+, \pi^+,\Xi^0]:\nonumber\\&\,-\mathcal{A}(\Xi_c^+\to K^+\Xi^0)+\sqrt{\frac{3}{2}}\mathcal{A}(\Xi_c^+\to \pi^+\Lambda^0)-\frac{\mathcal{A}(\Xi_c^+\to \pi^+\Sigma^0)}{\sqrt{2}}+2\lambda\mathcal{A}(\Xi_c^+\to \pi^+\Xi^0)=0.
\end{align}
\begin{align}
&{\rm SumS}\,[\Xi_c^+, \pi^0,p]:\nonumber\\&\,-2\lambda\mathcal{A}(\Xi_c^+\to \pi^0p)+\lambda^2\mathcal{A}(\Lambda_c^+\to \pi^0p)+\frac{\lambda^2\mathcal{A}(\Xi_c^+\to \overline K^0p)}{\sqrt{2}}+\lambda\mathcal{A}(\Xi_c^+\to \pi^0\Sigma^+)=0.
\end{align}
\begin{align}
&{\rm SumS}\,[\Xi_c^+, \eta_8,p]:\nonumber\\&\,-2\lambda\mathcal{A}(\Xi_c^+\to \eta_8p)+\lambda^2\mathcal{A}(\Lambda_c^+\to \eta_8p)+\lambda^2\mathcal{A}(\Xi_c^+\to \eta_8\Sigma^+)-\sqrt{\frac{3}{2}}\lambda^2\mathcal{A}(\Xi_c^+\to \overline K^0p)=0.
\end{align}
\begin{align}
&{\rm SumS}\,[\Xi_c^+, K^0,p]:\nonumber\\&\,\mathcal{A}(\Lambda_c^+\to K^0p)+\sqrt{\frac{3}{2}}\mathcal{A}(\Xi_c^+\to \eta_8p)+\mathcal{A}(\Xi_c^+\to K^0\Sigma^+)-\frac{\mathcal{A}(\Xi_c^+\to \pi^0p)}{\sqrt{2}}=0.
\end{align}
\begin{align}
&{\rm SumS}\,[\Xi_c^+, \overline K^0,p]:\nonumber\\&\,\sqrt{\frac{3}{2}}\mathcal{A}(\Xi_c^+\to \eta_8p)-\frac{\mathcal{A}(\Xi_c^+\to \pi^0p)}{\sqrt{2}}+\lambda^2\mathcal{A}(\Lambda_c^+\to \overline K^0p)+\lambda^2\mathcal{A}(\Xi_c^+\to \overline K^0\Sigma^+)=0.
\end{align}
\begin{align}
&{\rm SumS}\,[\Xi_c^+, \pi^0,\Sigma^+]:\nonumber\\&\,\frac{\mathcal{A}(\Xi_c^+\to K^0\Sigma^+)}{\sqrt{2}}-\mathcal{A}(\Xi_c^+\to \pi^0p)+\lambda^2\mathcal{A}(\Lambda_c^+\to \pi^0\Sigma^+)+\frac{\lambda^2\mathcal{A}(\Xi_c^+\to \overline K^0\Sigma^+)}{\sqrt{2}}=0.
\end{align}
\begin{align}
&{\rm SumS}\,[\Xi_c^+, \eta_8,\Sigma^+]:\nonumber\\&\,-\mathcal{A}(\Xi_c^+\to \eta_8p)-\sqrt{\frac{3}{2}}\mathcal{A}(\Xi_c^+\to K^0\Sigma^+)+\lambda^2\mathcal{A}(\Lambda_c^+\to \eta_8\Sigma^+)-\sqrt{\frac{3}{2}}\lambda^2\mathcal{A}(\Xi_c^+\to \overline K^0\Sigma^+)=0.
\end{align}
\begin{align}
{\rm SumS}\,[\Xi_c^+, K^0,\Sigma^+]:&\,-2\lambda\mathcal{A}(\Xi_c^+\to K^0\Sigma^+)+\lambda^2\mathcal{A}(\Lambda_c^+\to K^0\Sigma^+)\nonumber\\&~~~+\sqrt{\frac{3}{2}}\lambda^2\mathcal{A}(\Lambda_c^+\to \eta_8\Sigma^+)-\frac{\lambda^2\mathcal{A}(\Xi_c^+\to \pi^0\Sigma^+)}{\sqrt{2}}=0.
\end{align}
\begin{align}
&{\rm SumS}\,[\Xi_c^+, \overline K^0,\Sigma^+]:\nonumber\\&\,\sqrt{\frac{3}{2}}\mathcal{A}(\Xi_c^+\to \eta_8\Sigma^+)-\mathcal{A}(\Xi_c^+\to \overline K^0p)-\frac{\mathcal{A}(\Xi_c^+\to \pi^0\Sigma^+)}{\sqrt{2}}+2\lambda\mathcal{A}(\Xi_c^+\to \overline K^0\Sigma^+)=0.
\end{align}
\begin{align}
&{\rm SumS}\,[\Lambda_c^+, K^+,\Sigma^0]:\nonumber\\&\,\frac{\mathcal{A}(\Lambda_c^+\to K^+n)}{\sqrt{2}}-\mathcal{A}(\Xi_c^+\to K^+\Sigma^0)+\frac{\lambda^2\mathcal{A}(\Lambda_c^+\to K^+\Xi^0)}{\sqrt{2}}+\lambda^2\mathcal{A}(\Lambda_c^+\to \pi^+\Sigma^0)=0.
\end{align}
\begin{align}
{\rm SumS}\,[\Lambda_c^+, K^+,\Lambda^0]:&\,-\sqrt{\frac{3}{2}}\mathcal{A}(\Lambda_c^+\to K^+n)-\mathcal{A}(\Xi_c^+\to K^+\Lambda^0)\nonumber\\&~~~-\sqrt{\frac{3}{2}}\lambda^2\mathcal{A}(\Lambda_c^+\to K^+\Xi^0)+\lambda^2\mathcal{A}(\Lambda_c^+\to \pi^+\Lambda^0)=0.
\end{align}
\begin{align}
{\rm SumS}\,[\Lambda_c^+, K^+,n]:&\,-2\lambda\mathcal{A}(\Lambda_c^+\to K^+n)+\sqrt{\frac{3}{2}}\lambda^2\mathcal{A}(\Lambda_c^+\to K^+\Lambda^0)\nonumber\\&~~~-\frac{\lambda^2\mathcal{A}(\Lambda_c^+\to K^+\Sigma^0)}{\sqrt{2}}+\lambda^2\mathcal{A}(\Lambda_c^+\to \pi^+n)=0.
\end{align}
\begin{align}
&{\rm SumS}\,[\Lambda_c^+, K^+,\Xi^0]:\nonumber\\&\,\sqrt{\frac{3}{2}}\mathcal{A}(\Lambda_c^+\to K^+\Lambda^0)-\frac{\mathcal{A}(\Lambda_c^+\to K^+\Sigma^0)}{\sqrt{2}}-\mathcal{A}(\Xi_c^+\to K^+\Xi^0)+2\lambda\mathcal{A}(\Lambda_c^+\to K^+\Xi^0)=0.
\end{align}
\begin{align}
&{\rm SumS}\,[\Lambda_c^+, \pi^+,\Sigma^0]:\nonumber\\&\,-\mathcal{A}(\Lambda_c^+\to K^+\Sigma^0)+\frac{\mathcal{A}(\Lambda_c^+\to \pi^+n)}{\sqrt{2}}-\mathcal{A}(\Xi_c^+\to \pi^+\Sigma^0)+2\lambda\mathcal{A}(\Lambda_c^+\to \pi^+\Sigma^0)=0.
\end{align}
\begin{align}
&{\rm SumS}\,[\Lambda_c^+, \pi^+,\Lambda^0]:\nonumber\\&\,-\mathcal{A}(\Lambda_c^+\to K^+\Lambda^0)-\sqrt{\frac{3}{2}}\mathcal{A}(\Lambda_c^+\to \pi^+n)-\mathcal{A}(\Xi_c^+\to \pi^+\Lambda^0)+2\lambda\mathcal{A}(\Lambda_c^+\to \pi^+\Lambda^0)=0.
\end{align}
\begin{align}
&{\rm SumS}\,[\Lambda_c^+, \pi^+,n]:\nonumber\\&\,-\mathcal{A}(\Lambda_c^+\to K^+n)-\mathcal{A}(\Lambda_c^+\to \pi^+n)+\sqrt{\frac{3}{2}}\lambda^2\mathcal{A}(\Lambda_c^+\to \pi^+\Lambda^0)-\frac{\lambda^2\mathcal{A}(\Lambda_c^+\to \pi^+\Sigma^0)}{\sqrt{2}}=0.
\end{align}
\begin{align}
&{\rm SumS}\,[\Lambda_c^+, \pi^+,\Xi^0]:\nonumber\\&\,-\mathcal{A}(\Lambda_c^+\to K^+\Xi^0)+\sqrt{\frac{3}{2}}\mathcal{A}(\Lambda_c^+\to \pi^+\Lambda^0)-\frac{\mathcal{A}(\Lambda_c^+\to \pi^+\Sigma^0)}{\sqrt{2}}-\mathcal{A}(\Xi_c^+\to \pi^+\Xi^0)=0.
\end{align}
\begin{align}
&{\rm SumS}\,[\Lambda_c^+, \pi^0,p]:\nonumber\\&\,\frac{\mathcal{A}(\Lambda_c^+\to K^0p)}{\sqrt{2}}-\mathcal{A}(\Xi_c^+\to \pi^0p)+\frac{\lambda^2\mathcal{A}(\Lambda_c^+\to \overline K^0p)}{\sqrt{2}}+\lambda^2\mathcal{A}(\Lambda_c^+\to \pi^0\Sigma^+)=0.
\end{align}
\begin{align}
&{\rm SumS}\,[\Lambda_c^+, \eta_8,p]:\nonumber\\&\,-\sqrt{\frac{3}{2}}\mathcal{A}(\Lambda_c^+\to K^0p)-\mathcal{A}(\Xi_c^+\to \eta_8p)+\lambda^2\mathcal{A}(\Lambda_c^+\to  \eta_8\Sigma^+)+\sqrt{\frac{3}{2}}\lambda^2\mathcal{A}(\Lambda_c^+\to \overline K^0p)=0.
\end{align}
\begin{align}
&{\rm SumS}\,[\Lambda_c^+, K^0,p]:\nonumber\\&\,-2\lambda\mathcal{A}(\Lambda_c^+\to K^0p)+\sqrt{\frac{3}{2}}\lambda^2\mathcal{A}(\Lambda_c^+\to \eta_8p)+\lambda^2\mathcal{A}(\Lambda_c^+\to  K^0\Sigma^+)-\frac{\lambda^2\mathcal{A}(\Lambda_c^+\to \pi^0p)}{\sqrt{2}}=0.
\end{align}
\begin{align}
&{\rm SumS}\,[\Lambda_c^+, \overline K^0,p]:\nonumber\\&\,\sqrt{\frac{3}{2}}\mathcal{A}(\Lambda_c^+\to \eta_8p)-\frac{\mathcal{A}(\Lambda_c^+\to \pi^0p)}{\sqrt{2}}-\mathcal{A}(\Xi_c^+\to  \overline K^0p)+2\lambda\mathcal{A}(\Lambda_c^+\to \overline K^0p)=0.
\end{align}
\begin{align}
&{\rm SumS}\,[\Lambda_c^+, \pi^0,\Sigma^+]:\nonumber\\&\,\frac{\mathcal{A}(\Lambda_c^+\to K^0\Sigma^+)}{\sqrt{2}}-\mathcal{A}(\Lambda_c^+\to \pi^0p)-\mathcal{A}(\Xi_c^+\to \pi^0\Sigma^+)+2\lambda\mathcal{A}(\Lambda_c^+\to \pi^0\Sigma^+)=0.
\end{align}
\begin{align}
&{\rm SumS}\,[\Lambda_c^+, \eta_8,\Sigma^+]:\nonumber\\&\,-\mathcal{A}(\Lambda_c^+\to \eta_8p)-\sqrt{\frac{3}{2}}\mathcal{A}(\Lambda_c^+\to K^0\Sigma^+)-\mathcal{A}(\Xi_c^+\to \eta_8\Sigma^+)+2\lambda\mathcal{A}(\Lambda_c^+\to \eta_8\Sigma^+)=0.
\end{align}
\begin{align}
&{\rm SumS}\,[\Lambda_c^+, K^0,\Sigma^+]:\nonumber\\&\,-\mathcal{A}(\Lambda_c^+\to K^0p)-\mathcal{A}(\Xi_c^+\to K^0\Sigma^+)+\sqrt{\frac{3}{2}}\lambda^2\mathcal{A}(\Lambda_c^+\to \eta_8\Sigma^+)-\frac{\lambda^2\mathcal{A}(\Lambda_c^+\to \pi^0\Sigma^+)}{\sqrt{2}}=0.
\end{align}
\begin{align}
&{\rm SumS}\,[\Lambda_c^+, \overline K^0,\Sigma^+]:\nonumber\\&\,\sqrt{\frac{3}{2}}\mathcal{A}(\Lambda_c^+\to \eta_8\Sigma^+)-\mathcal{A}(\Lambda_c^+\to \overline K^0p)-\frac{\mathcal{A}(\Lambda_c^+\to \pi^0\Sigma^+)}{\sqrt{2}}-\mathcal{A}(\Xi_c^+\to \overline K^0\Sigma^+)=0.
\end{align}
\begin{align}
{\rm SumT_-}\,[\Xi_c^0,  K^+,\Sigma^0]:&\,\mathcal{A}(\Xi_c^0\to K^0\Sigma^0)+\sqrt{2}\mathcal{A}(\Xi_c^0\to K^+\Sigma^-)-\mathcal{A}(\Xi_c^+\to K^+\Sigma^0)\nonumber\\&~~~-\lambda\mathcal{A}(\Lambda_c^+\to K^+\Sigma^0)-\sqrt{\frac{3}{2}}\lambda\mathcal{A}(\Xi_c^0\to \eta_8\Sigma^0)\nonumber\\&~~~~~+\frac{\lambda\mathcal{A}(\Xi_c^0\to K^+\Xi^-)}{\sqrt{2}}-\frac{\lambda\mathcal{A}(\Xi_c^0\to \pi^0\Sigma^0)}{\sqrt{2}}=0.
\end{align}
\begin{align}
{\rm SumT_-}\,[\Xi_c^0,  K^+,\Lambda^0]:&\,\mathcal{A}(\Xi_c^0\to K^0\Lambda^0)-\mathcal{A}(\Xi_c^+\to K^+\Lambda^0)\nonumber\\&~~~+\lambda\mathcal{A}(\Lambda_c^+\to K^+\Lambda^0)-\sqrt{\frac{3}{2}}\lambda\mathcal{A}(\Xi_c^0\to \eta_8\Lambda^0)\nonumber\\&~~~~~+\sqrt{\frac{3}{2}}\lambda\mathcal{A}(\Xi_c^0\to K^+\Xi^-)-\frac{\lambda\mathcal{A}(\Xi_c^0\to \pi^0\Lambda^0)}{\sqrt{2}}=0.
\end{align}
\begin{align}
&{\rm SumT_-}\,[\Xi_c^0,  K^+,n]:\nonumber\\&\,\mathcal{A}(\Xi_c^0\to K^+n)-\sqrt{\frac{3}{2}}\mathcal{A}(\Xi_c^0\to \eta_8n)-\mathcal{A}(\Xi_c^0\to K^+\Sigma^-)-\frac{\mathcal{A}(\Xi_c^0\to \pi^0n)}{\sqrt{2}}=0.
\end{align}
\begin{align}
{\rm SumT_-}\,[\Xi_c^0,  K^+,\Xi^0]:&\,\mathcal{A}(\Xi_c^0\to K^0\Xi^0)-\mathcal{A}(\Xi_c^0\to K^+\Xi^-)-\mathcal{A}(\Xi_c^+\to K^+\Xi^0)\nonumber\\&~~~+\lambda\mathcal{A}(\Lambda_c^+\to K^+\Xi^0)-\sqrt{\frac{3}{2}}\lambda\mathcal{A}(\Xi_c^0\to \eta_8\Xi^0)-\frac{\lambda\mathcal{A}(\Xi_c^0\to \pi^0\Xi^0)}{\sqrt{2}}=0.
\end{align}
\begin{align}
{\rm SumT_-}\,[\Xi_c^0,  \pi^+,\Sigma^0]:&\,-\sqrt{2}\mathcal{A}(\Xi_c^0\to \pi^0\Sigma^0)+\sqrt{2}\mathcal{A}(\Xi_c^0\to \pi^+\Sigma^-)-\mathcal{A}(\Xi_c^+\to \pi^+\Sigma^0)\nonumber\\&~~~+\lambda\mathcal{A}(\Lambda_c^+\to \pi^+\Sigma^0)+\lambda\mathcal{A}(\Xi_c^0\to \overline K^0\Sigma^0)+\frac{\lambda\mathcal{A}(\Xi_c^0\to \pi^+\Xi^-)}{\sqrt{2}}=0.
\end{align}
\begin{align}
{\rm SumT_-}\,[\Xi_c^0,  \pi^+,\Lambda^0]:&\,-\sqrt{2}\mathcal{A}(\Xi_c^0\to \pi^0\Lambda^0)-\mathcal{A}(\Xi_c^+\to \pi^+\Lambda^0)+\lambda\mathcal{A}(\Lambda_c^+\to \pi^+\Lambda^0)\nonumber\\&~~~+\lambda\mathcal{A}(\Xi_c^0\to \overline K^0\Lambda^0)+\sqrt{\frac{3}{2}}\lambda\mathcal{A}(\Xi_c^0\to \pi^+\Xi^-)=0.
\end{align}
\begin{align}
{\rm SumT_-}\,[\Xi_c^0,  \pi^+,n]:&\,-\sqrt{2}\mathcal{A}(\Xi_c^0\to \pi^0n)-\mathcal{A}(\Xi_c^+\to \pi^+n)+\lambda\mathcal{A}(\Lambda_c^+\to \pi^+n)\nonumber\\&+\lambda\mathcal{A}(\Xi_c^0\to \overline K^0n)-\lambda\mathcal{A}(\Xi_c^0\to \pi^+\Sigma^-)=0.
\end{align}
\begin{align}
{\rm SumT_-}\,[\Xi_c^0,  \pi^+,\Xi^0]:\,\sqrt{2}\mathcal{A}(\Xi_c^0\to \pi^0\Xi^0)+\mathcal{A}(\Xi_c^0\to \pi^+\Xi^-)+\mathcal{A}(\Xi_c^+\to \pi^+\Xi^0)=0.
\end{align}
\begin{align}
{\rm SumT_-}\,[\Xi_c^0,  \pi^0,p]:&\,\mathcal{A}(\Xi_c^0\to \pi^0n)+\sqrt{2}\mathcal{A}(\Xi_c^0\to \pi^-p)-\mathcal{A}(\Xi_c^+\to \pi^0p)+\lambda\mathcal{A}(\Lambda_c^+\to \pi^0p)\nonumber\\&~~~+\frac{\lambda\mathcal{A}(\Xi_c^0\to K^-p)}{\sqrt{2}}-\sqrt{\frac{3}{2}}\lambda\mathcal{A}(\Xi_c^0\to \pi^0\Lambda^0)-\frac{\lambda\mathcal{A}(\Xi_c^0\to \pi^0\Sigma^0)}{\sqrt{2}}=0.
\end{align}
\begin{align}
{\rm SumT_-}\,[\Xi_c^0,  \eta_8,p]:&\,\mathcal{A}(\Xi_c^0\to \eta_8n)-\mathcal{A}(\Xi_c^+\to \eta_8p)+\lambda\mathcal{A}(\Lambda_c^+\to \eta_8p)-\sqrt{\frac{3}{2}}\lambda\mathcal{A}(\Xi_c^0\to \eta_8\Lambda^0)\nonumber\\&~~~-\frac{\lambda\mathcal{A}(\Xi_c^0\to \eta_8\Sigma^0)}{\sqrt{2}}+\sqrt{\frac{3}{2}}\lambda\mathcal{A}(\Xi_c^0\to K^-p)=0.
\end{align}
\begin{align}
&{\rm SumT_-}\,[\Xi_c^0, K^0,p]:\nonumber\\&\,\mathcal{A}(\Lambda_c^+\to K^0p)-\sqrt{\frac{3}{2}}\mathcal{A}(\Xi_c^0\to K^0\Lambda^0)-\frac{\mathcal{A}(\Xi_c^0\to K^0\Sigma^0)}{\sqrt{2}}-\mathcal{A}(\Xi_c^0\to \pi^-p)=0.
\end{align}
\begin{align}
{\rm SumT_-}\,[\Xi_c^0, \overline K^0,p]:&\,-\mathcal{A}(\Xi_c^0\to K^-p)+\mathcal{A}(\Xi_c^0\to \overline K^0n)-\mathcal{A}(\Lambda_c^+\to \overline K^0p)+\lambda\mathcal{A}(\Lambda_c^+\to \overline K^0p)\nonumber\\&~~~-\sqrt{\frac{3}{2}}\lambda\mathcal{A}(\Xi_c^0\to \overline K^0\Lambda^0)-\frac{\lambda\mathcal{A}(\Xi_c^0\to \overline K^0\Sigma^0)}{\sqrt{2}}=0.
\end{align}
\begin{align}
{\rm SumT_-}\,[\Xi_c^0,  \pi^0,\Sigma^+]:&\,-\sqrt{2}\mathcal{A}(\Xi_c^0\to \pi^0\Sigma^0)+\sqrt{2}\mathcal{A}(\Xi_c^0\to \pi^-\Sigma^+)-\mathcal{A}(\Lambda_c^+\to \pi^0\Sigma^+)\nonumber\\&~~~+\lambda\mathcal{A}(\Lambda_c^+\to \pi^0\Sigma^+)+\frac{\lambda\mathcal{A}(\Xi_c^0\to K^-\Sigma^+)}{\sqrt{2}}+\lambda\mathcal{A}(\Xi_c^0\to \pi^0\Xi^0)=0.
\end{align}
\begin{align}
{\rm SumT_-}\,[\Xi_c^0,  \eta_8,\Sigma^+]:&\,-\sqrt{2}\mathcal{A}(\Xi_c^0\to \eta_8\Sigma^0)-\mathcal{A}(\Xi_c^+\to \eta_8\Sigma^+)+\lambda\mathcal{A}(\Lambda_c^+\to \eta_8\Sigma^+)\nonumber\\&~~~+\lambda\mathcal{A}(\Xi_c^0\to \eta_8\Xi^0)+\sqrt{\frac{3}{2}}\lambda\mathcal{A}(\Xi_c^0\to K^-\Sigma^+)=0.
\end{align}
\begin{align}
{\rm SumT_-}\,[\Xi_c^0, K^0,\Sigma^+]:&\,-\sqrt{2}\mathcal{A}(\Xi_c^0\to K^0\Sigma^0)-\mathcal{A}(\Xi_c^+\to K^0\Sigma^+)+\lambda\mathcal{A}(\Lambda_c^+\to K^0\Sigma^+)\nonumber\\&~~~+\lambda\mathcal{A}(\Xi_c^0\to K^0\Xi^0)-\lambda\mathcal{A}(\Xi_c^0\to \pi^-\Sigma^+)=0.
\end{align}
\begin{align}
{\rm SumT_-}\,[\Xi_c^0, \overline K^0,\Sigma^+]:\,-\mathcal{A}(\Xi_c^0\to K^-\Sigma^+)-\sqrt{2}\mathcal{A}(\Xi_c^0\to \overline K^0\Sigma^0)+\mathcal{A}(\Xi_c^+\to \overline K^0\Sigma^+)=0.
\end{align}
\begin{align}
{\rm SumT_-}\,[\Xi_c^+, K^+,\Sigma^+]:&\,\mathcal{A}(\Xi_c^+\to K^0\Sigma^+)-\sqrt{2}\mathcal{A}(\Xi_c^+\to K^+\Sigma^0)+\sqrt{\frac{3}{2}}\lambda\mathcal{A}(\Lambda_c^+\to \eta_8\Sigma^+)\nonumber\\&~~~+\lambda\mathcal{A}(\Xi_c^+\to K^+\Xi^0)-\frac{\lambda\mathcal{A}(\Xi_c^+\to \pi^0\Sigma^+)}{\sqrt{2}}=0.
\end{align}
\begin{align}
{\rm SumT_-}\,[\Lambda_c^+, K^+,\Sigma^+]:&\,\mathcal{A}(\Lambda_c^+\to K^0\Sigma^+)-\sqrt{2}\mathcal{A}(\Lambda_c^+\to K^+\Sigma^0)-\sqrt{\frac{3}{2}}\lambda\mathcal{A}(\Lambda_c^+\to \eta_8\Sigma^+)\nonumber\\&~~~+\lambda\mathcal{A}(\Lambda_c^+\to K^+\Xi^0)-\frac{\lambda\mathcal{A}(\Lambda_c^+\to \pi^0\Sigma^+)}{\sqrt{2}}=0.
\end{align}
\begin{align}
{\rm SumT_-}\,[\Xi_c^+, K^+,p]:&\,-\sqrt{2}\mathcal{A}(\Xi_c^+\to \pi^0p)+\mathcal{A}(\Xi_c^+\to \pi^+n)+\lambda\mathcal{A}(\Xi_c^+\to \overline K^0p)\nonumber\\&~~~-\sqrt{\frac{3}{2}}\lambda\mathcal{A}(\Xi_c^+\to \pi^+\Lambda^0)-\frac{\lambda\mathcal{A}(\Xi_c^+\to \pi^+\Sigma^0)}{\sqrt{2}}=0.
\end{align}
\begin{align}
{\rm SumT_-}\,[\Lambda_c^+, K^+,p]:&\,-\sqrt{2}\mathcal{A}(\Lambda_c^+\to \pi^0p)+\mathcal{A}(\Lambda_c^+\to \pi^+n)+\lambda\mathcal{A}(\Lambda_c^+\to \overline K^0p)\nonumber\\&~~~-\sqrt{\frac{3}{2}}\lambda\mathcal{A}(\Lambda_c^+\to \pi^+\Lambda^0)-\frac{\lambda\mathcal{A}(\Lambda_c^+\to \pi^+\Sigma^0)}{\sqrt{2}}=0.
\end{align}
\begin{align}
&{\rm SumS}\,[\Xi_c^0, \eta_1,\Sigma^0]:\,\mathcal{A}(\Xi_c^0\to \eta_1n)+\lambda^2\mathcal{A}(\Xi_c^0\to \eta_1\Xi^0)=0.
\end{align}
\begin{align}
&{\rm SumS}\,[\Xi_c^0, \eta_1,\Lambda^0]:\,\mathcal{A}(\Xi_c^0\to \eta_1n)+\lambda^2\mathcal{A}(\Xi_c^0\to \eta_1\Xi^0)=0.
\end{align}
\begin{align}
&{\rm SumS}\,[\Xi_c^0, \eta_1,n]:\,-2\lambda\mathcal{A}(\Xi_c^0\to \eta_1n)+\sqrt{\frac{3}{2}}\lambda^2\mathcal{A}(\Xi_c^0\to \eta_1\Lambda^0)-\frac{\lambda^2\mathcal{A}(\Xi_c^0\to \eta_1\Sigma^0)}{\sqrt{2}}=0.
\end{align}
\begin{align}
&{\rm SumS}\,[\Xi_c^0, \eta_1,\Xi^0]:\,\sqrt{\frac{3}{2}}\mathcal{A}(\Xi_c^0\to \eta_1\Lambda^0)-\frac{\mathcal{A}(\Xi_c^0\to \eta_1\Sigma^0)}{\sqrt{2}}+2\lambda\mathcal{A}(\Xi_c^0\to \eta_1\Xi^0)=0.
\end{align}
\begin{align}
&{\rm SumS}\,[\Xi_c^+, \eta_1,p]:\,-2\lambda\mathcal{A}(\Xi_c^+\to \eta_1p)+\lambda^2\mathcal{A}(\Lambda_c^+\to \eta_1p)+\lambda^2\mathcal{A}(\Xi_c^+\to \eta_1\Sigma^+)=0.
\end{align}
\begin{align}
&{\rm SumS}\,[\Xi_c^+, \eta_1,\Sigma^+]:\,-\mathcal{A}(\Xi_c^+\to \eta_1p)+\lambda^2\mathcal{A}(\Lambda_c^+\to \eta_1\Sigma^+)=0.
\end{align}
\begin{align}
&{\rm SumS}\,[\Lambda_c^+, \eta_1,p]:\,-\mathcal{A}(\Xi_c^+\to \eta_1p)+\lambda^2\mathcal{A}(\Lambda_c^+\to \eta_1\Sigma^+)=0.
\end{align}
\begin{align}
&{\rm SumS}\,[\Lambda_c^+, \eta_1,\Sigma^+]:\,-\mathcal{A}(\Lambda_c^+\to \eta_1p)-\mathcal{A}(\Xi_c^+\to \eta_1\Sigma^+)+2\lambda\mathcal{A}(\Lambda_c^+\to \eta_1\Sigma^+)=0.
\end{align}
\begin{align}
{\rm SumT_-}\,[\Xi_c^0, \eta_1,p]:&\,\mathcal{A}(\Xi_c^0\to \eta_1n)-\mathcal{A}(\Xi_c^+\to \eta_1p)+\lambda\mathcal{A}(\Lambda_c^+\to \eta_1p)\nonumber\\&~~~-\sqrt{\frac{3}{2}}\lambda\mathcal{A}(\Xi_c^0\to \eta_1\Lambda^0)-\frac{\lambda\mathcal{A}(\Xi_c^0\to \eta_1\Sigma^0)}{\sqrt{2}}=0.
\end{align}
\begin{align}
&{\rm SumT_-}\,[\Xi_c^0, \eta_1,\Sigma^+]:\nonumber\\&\,-\sqrt{2}\mathcal{A}(\Xi_c^0\to \eta_1\Sigma^0)-\mathcal{A}(\Xi_c^+\to \eta_1\Sigma^+)+\lambda\mathcal{A}(\Lambda_c^+\to \eta_1\Sigma^+)+\lambda\mathcal{A}(\Xi_c^0\to \eta_1\Xi^0)=0.
\end{align}

\section{SU(3) sum rules of $\mathcal{B}_{\overline c3}\to M\mathcal{B}_{10}$ modes}\label{res2}

\begin{align}
&{\rm SumS}\,[\Xi_c^0, \pi^+,\Delta^-]:\,2\lambda\mathcal{A}(\Xi_c^0\to \pi^+\Delta^-)+\sqrt{3}\lambda^2\mathcal{A}(\Xi_c^0\to \pi^+\Sigma^{*-})=0.
\end{align}
\begin{align}
&{\rm SumS}\,[\Xi_c^0, \pi^+,\Sigma^{*-}]:\,-\mathcal{A}(\Xi_c^0\to K^+\Sigma^{*-})+\sqrt{3}\mathcal{A}(\Xi_c^0\to \pi^+\Delta^{-})-2\lambda^2\mathcal{A}(\Xi_c^0\to \pi^+\Xi^{*-})=0.
\end{align}
\begin{align}
&{\rm SumS}\,[\Xi_c^0, \pi^+,\Xi^{*-}]:\,-\mathcal{A}(\Xi_c^0\to K^+\Xi^{*-})+2\mathcal{A}(\Xi_c^0\to \pi^+\Sigma^{*-})+2\lambda\mathcal{A}(\Xi_c^0\to \pi^+\Xi^{*-})=0.
\end{align}
\begin{align}
&{\rm SumS}\,[\Xi_c^0, \pi^+,\Omega^{-}]:\,-\mathcal{A}(\Xi_c^0\to K^+\Omega^{-})+\sqrt{3}\mathcal{A}(\Xi_c^0\to \pi^+\Xi^{*-})=0.
\end{align}
\begin{align}
&{\rm SumS}\,[\Xi_c^0, K^+,\Delta^-]:\,-\sqrt{3}\mathcal{A}(\Xi_c^0\to K^+\Sigma^{*-})+\mathcal{A}(\Xi_c^0\to \pi^+\Delta^{-})=0.
\end{align}
\begin{align}
&{\rm SumS}\,[\Xi_c^0, K^+,\Sigma^{*-}]:\nonumber\\&\,-2\lambda\mathcal{A}(\Xi_c^0\to K^+\Sigma^{*-})-2\lambda^2\mathcal{A}(\Xi_c^0\to K^+\Xi^{*-})+\lambda^2\mathcal{A}(\Xi_c^0\to \pi^+\Sigma^{*-})=0.
\end{align}
\begin{align}
&{\rm SumS}\,[\Xi_c^0, K^+,\Xi^{*-}]:\nonumber\\&\,2\mathcal{A}(\Xi_c^0\to K^+\Sigma^{*-})-\sqrt{3}\lambda^2\mathcal{A}(\Xi_c^0\to K^+\Omega^{-})+\lambda^2\mathcal{A}(\Xi_c^0\to \pi^+\Xi^{*-})=0.
\end{align}
\begin{align}
&{\rm SumS}\,[\Xi_c^0, K^+,\Omega^{-}]:\,\sqrt{3}\mathcal{A}(\Xi_c^0\to K^+\Xi^{*-})+2\lambda\mathcal{A}(\Xi_c^0\to K^+\Omega^{-})=0.
\end{align}
\begin{align}
&{\rm SumS}\,[\Xi_c^0, \pi^0,\Delta^{0}]:\,-2\lambda\mathcal{A}(\Xi_c^0\to \pi^0\Delta^{0})+\frac{\lambda^2\mathcal{A}(\Xi_c^0\to \overline K^0\Delta^{0})}{\sqrt{2}}-\sqrt{2}\lambda^2\mathcal{A}(\Xi_c^0\to \pi^0\Sigma^{*0})=0.
\end{align}
\begin{align}
&{\rm SumS}\,[\Xi_c^0, \pi^0,\Sigma^{*0}]:\nonumber\\&\,\mathcal{A}(\Xi_c^0\to K^0\Sigma^{*0})+2\mathcal{A}(\Xi_c^0\to \pi^0\Delta^{0})+\lambda^2\mathcal{A}(\Xi_c^0\to \overline K^0\Sigma^{*0})-2\lambda^2\mathcal{A}(\Xi_c^0\to \pi^0\Xi^{*0})=0.
\end{align}
\begin{align}
&{\rm SumS}\,[\Xi_c^0, \pi^0,\Xi^{*0}]:\,\frac{\mathcal{A}(\Xi_c^0\to K^0\Xi^{*0})}{\sqrt{2}}+\sqrt{2}\mathcal{A}(\Xi_c^0\to \pi^0\Sigma^{*0})+2\lambda\mathcal{A}(\Xi_c^0\to \pi^0\Xi^{*0})=0.
\end{align}
\begin{align}
&{\rm SumS}\,[\Xi_c^0, \eta_8,\Delta^{0}]:\,2\lambda\mathcal{A}(\Xi_c^0\to \eta_8\Delta^{0})+\sqrt{2}\lambda^2\mathcal{A}(\Xi_c^0\to \eta_8\Sigma^{*0})+\frac{\sqrt{6}}{2}\lambda^2\mathcal{A}(\Xi_c^0\to \overline K^0\Delta^{0})=0.
\end{align}
\begin{align}
&{\rm SumS}\,[\Xi_c^0, \eta_8,\Sigma^{*0}]:\nonumber\\&\,2\mathcal{A}(\Xi_c^0\to \eta_8\Delta^{0})-\sqrt{3}\mathcal{A}(\Xi_c^0\to K^0\Sigma^{*0})-2\lambda^2\mathcal{A}(\Xi_c^0\to \eta_8\Xi^{*0})-\sqrt{3}\lambda^2\mathcal{A}(\Xi_c^0\to \overline K^0\Sigma^{*0})=0.
\end{align}
\begin{align}
&{\rm SumS}\,[\Xi_c^0, \eta_8,\Xi^{*0}]:\,\sqrt{2}\mathcal{A}(\Xi_c^0\to \eta_8\Sigma^{*0})-\sqrt{\frac{3}{2}}\mathcal{A}(\Xi_c^0\to K^0\Xi^{*0})+2\lambda\mathcal{A}(\Xi_c^0\to \eta_8\Xi^{*0})=0.
\end{align}
\begin{align}
&{\rm SumS}\,[\Xi_c^0, K^0,\Delta^{0}]:\,\sqrt{\frac{3}{2}}\mathcal{A}(\Xi_c^0\to \eta_8\Delta^{0})-\sqrt{2}\mathcal{A}(\Xi_c^0\to K^0\Sigma^{*0})-\frac{1}{\sqrt{2}}\mathcal{A}(\Xi_c^0\to \pi^0\Delta^{0})=0.
\end{align}
\begin{align}
{\rm SumS}\,[\Xi_c^0, K^0,\Sigma^{*0}]:&\,-2\lambda\mathcal{A}(\Xi_c^0\to K^0\Sigma^{*0})+\sqrt{\frac{3}{2}}\lambda^2\mathcal{A}(\Xi_c^0\to \eta_8\Sigma^{*0})\nonumber\\&~~~-\sqrt{2}\lambda^2\mathcal{A}(\Xi_c^0\to K^0\Xi^{*0})-\frac{1}{\sqrt{2}}\lambda^2\mathcal{A}(\Xi_c^0\to \pi^0\Sigma^{*0})=0.
\end{align}
\begin{align}
&{\rm SumS}\,[\Xi_c^0, K^0,\Xi^{*0}]:\nonumber\\&\,\sqrt{2}\mathcal{A}(\Xi_c^0\to K^0\Sigma^{*0})+\sqrt{\frac{3}{2}}\lambda^2\mathcal{A}(\Xi_c^0\to \eta_8\Xi^{*0})-\frac{1}{\sqrt{2}}\lambda^2\mathcal{A}(\Xi_c^0\to \pi^0\Xi^{*0})=0.
\end{align}
\begin{align}
&{\rm SumS}\,[\Xi_c^0, \overline K^0,\Delta^{0}]:\,\sqrt{\frac{3}{2}}\mathcal{A}(\Xi_c^0\to \eta_8\Delta^{0})-\frac{1}{\sqrt{2}}\mathcal{A}(\Xi_c^0\to \pi^0\Delta^{0})-\sqrt{2}\lambda^2\mathcal{A}(\Xi_c^0\to \overline K^0\Sigma^{*0})=0.
\end{align}
\begin{align}
&{\rm SumS}\,[\Xi_c^0, \overline K^0,\Sigma^{*0}]:\nonumber\\&\,\sqrt{\frac{3}{2}}\mathcal{A}(\Xi_c^0\to \eta_8\Sigma^{*0})+\sqrt{2}\mathcal{A}(\Xi_c^0\to \overline K^0\Delta^{0})-\frac{1}{\sqrt{2}}\mathcal{A}(\Xi_c^0\to \pi^0\Sigma^{*0})+2\lambda\mathcal{A}(\Xi_c^0\to \overline K^0\Sigma^{*0})=0.
\end{align}
\begin{align}
&{\rm SumS}\,[\Xi_c^0, \overline K^0,\Xi^{*0}]:\,\sqrt{\frac{3}{2}}\mathcal{A}(\Xi_c^0\to \eta_8\Xi^{*0})+\sqrt{2}\mathcal{A}(\Xi_c^0\to \overline K^0\Sigma^{*0})-\frac{\mathcal{A}(\Xi_c^0\to \pi^0\Xi^{*0})}{\sqrt{2}}=0.
\end{align}
\begin{align}
&{\rm SumS}\,[\Xi_c^0, \pi^-,\Delta^{+}]:\,2\lambda\mathcal{A}(\Xi_c^0\to \pi^-\Delta^{+})+\lambda^2\mathcal{A}(\Xi_c^0\to K^-\Delta^{+})+\lambda^2\mathcal{A}(\Xi_c^0\to \pi^-\Sigma^{*+})=0.
\end{align}
\begin{align}
&{\rm SumS}\,[\Xi_c^0, \pi^-,\Sigma^{*+}]:\,\mathcal{A}(\Xi_c^0\to \pi^-\Delta^{+})-\lambda^2\mathcal{A}(\Xi_c^0\to K^-\Sigma^{*+})=0.
\end{align}
\begin{align}
&{\rm SumS}\,[\Xi_c^0, K^-,\Delta^{+}]:\,\mathcal{A}(\Xi_c^0\to \pi^-\Delta^{+})-\lambda^2\mathcal{A}(\Xi_c^0\to K^-\Sigma^{*+})=0.
\end{align}
\begin{align}
&{\rm SumS}\,[\Xi_c^0, K^-,\Sigma^{*+}]:\,\mathcal{A}(\Xi_c^0\to K^-\Delta^{+})+\mathcal{A}(\Xi_c^0\to \pi^-\Sigma^{*+})+2\lambda\mathcal{A}(\Xi_c^0\to K^-\Sigma^{*+})=0.
\end{align}
\begin{align}
&{\rm SumS}\,[\Xi_c^+, \pi^+,\Delta^{0}]:\nonumber\\&\,-2\lambda\mathcal{A}(\Xi_c^+\to \pi^+\Delta^{0})+\lambda^2\mathcal{A}(\Lambda_c^+\to \pi^+\Delta^{0})-\sqrt{2}\lambda^2\mathcal{A}(\Xi_c^+\to \pi^+\Sigma^{*0})=0.
\end{align}
\begin{align}
{\rm SumS}\,[\Xi_c^+, \pi^+,\Sigma^{*0}]:&\,-\mathcal{A}(\Xi_c^+\to K^+\Sigma^{*0})+\sqrt{2}\mathcal{A}(\Xi_c^+\to \pi^+\Delta^{0})\nonumber\\&~~~+\lambda^2\mathcal{A}(\Lambda_c^+\to \pi^+\Sigma^{*0})-\sqrt{2}\lambda^2\mathcal{A}(\Xi_c^+\to \pi^+\Xi^{*0})=0.
\end{align}
\begin{align}
&{\rm SumS}\,[\Xi_c^+, \pi^+,\Xi^{*0}]:\,-\mathcal{A}(\Xi_c^+\to K^+\Xi^{*0})+\sqrt{2}\mathcal{A}(\Xi_c^+\to \pi^+\Sigma^{*0})+2\lambda\mathcal{A}(\Xi_c^+\to \pi^+\Xi^{*0})=0.
\end{align}
\begin{align}
&{\rm SumS}\,[\Xi_c^+, K^+,\Delta^{0}]:\,\mathcal{A}(\Lambda_c^+\to K^+\Delta^{0})-\sqrt{2}\mathcal{A}(\Xi_c^+\to K^+\Sigma^{*0})+\mathcal{A}(\Xi_c^+\to \pi^+\Delta^{0})=0.
\end{align}
\begin{align}
{\rm SumS}\,[\Xi_c^+, K^+,\Sigma^{*0}]:&\,-2\lambda\mathcal{A}(\Xi_c^+\to K^+\Sigma^{*0})+\lambda^2\mathcal{A}(\Lambda_c^+\to K^+\Sigma^{*0})\nonumber\\&~~~-\sqrt{2}\lambda^2\mathcal{A}(\Xi_c^+\to K^+\Xi^{*0})+\lambda^2\mathcal{A}(\Xi_c^+\to \pi^+\Sigma^{*0})=0.
\end{align}
\begin{align}
&{\rm SumS}\,[\Xi_c^+, K^+,\Xi^{*0}]:\nonumber\\&\,\sqrt{2}\mathcal{A}(\Xi_c^+\to K^+\Sigma^{*0})+\lambda^2\mathcal{A}(\Lambda_c^+\to K^+\Xi^{*0})+\lambda^2\mathcal{A}(\Xi_c^+\to \pi^+\Xi^{*0})=0.
\end{align}
\begin{align}
{\rm SumS}\,[\Xi_c^+, \pi^0,\Delta^{+}]:&\,-2\lambda\mathcal{A}(\Xi_c^+\to \pi^0\Delta^{+})+\lambda^2\mathcal{A}(\Lambda_c^+\to \pi^0\Delta^{+})\nonumber\\&~~~+\frac{\lambda^2\mathcal{A}(\Xi_c^+\to \overline K^0\Delta^{+})}{\sqrt{2}}-\lambda^2\mathcal{A}(\Xi_c^+\to \pi^0\Sigma^{*+})=0.
\end{align}
\begin{align}
&{\rm SumS}\,[\Xi_c^+, \pi^0,\Sigma^{*+}]:\nonumber\\&\,\frac{\mathcal{A}(\Xi_c^+\to K^0\Sigma^{*+})}{\sqrt{2}}+\mathcal{A}(\Xi_c^+\to \pi^0\Delta^{+})+\lambda^2\mathcal{A}(\Lambda_c^+\to \pi^0\Sigma^{*+})+\frac{\lambda^2\mathcal{A}(\Xi_c^+\to \overline K^0\Sigma^{*+})}{\sqrt{2}}=0.
\end{align}
\begin{align}
{\rm SumS}\,[\Xi_c^+, \eta_8,\Delta^{+}]:&\,-2\lambda\mathcal{A}(\Xi_c^+\to \eta_8\Delta^{+})+\lambda^2\mathcal{A}(\Lambda_c^+\to \eta_8\Delta^{+})\nonumber\\&~~~-\lambda^2\mathcal{A}(\Xi_c^+\to \eta_8\Sigma^{*+})-\frac{\sqrt{6}}{2}\lambda^2\mathcal{A}(\Xi_c^+\to \overline K^0\Delta^{+})=0.
\end{align}
\begin{align}
{\rm SumS}\,[\Xi_c^+, \eta_8,\Sigma^{*+}]:&\,\mathcal{A}(\Xi_c^+\to \eta_8\Delta^{+})-\sqrt{\frac{3}{2}}\mathcal{A}(\Xi_c^+\to K^0\Sigma^{*+})\nonumber\\&~~~+\lambda^2\mathcal{A}(\Lambda_c^+\to \eta_8\Sigma^{*+})-\sqrt{\frac{3}{2}}\lambda^2\mathcal{A}(\Xi_c^+\to \overline K^0\Sigma^{*+})=0.
\end{align}
\begin{align}
&{\rm SumS}\,[\Xi_c^+, K^0,\Delta^{+}]:\nonumber\\&\,\mathcal{A}(\Lambda_c^+\to K^0\Delta^{+})+\frac{\sqrt{6}}{2}\mathcal{A}(\Xi_c^+\to \eta_8\Delta^{+})-\mathcal{A}(\Xi_c^+\to K^0\Sigma^{*+})-\frac{\mathcal{A}(\Xi_c^+\to \pi^0\Delta^{+})}{\sqrt{2}}=0.
\end{align}
\begin{align}
{\rm SumS}\,[\Xi_c^+, K^0,\Sigma^{*+}]:&\,-2\lambda\mathcal{A}(\Xi_c^+\to K^0\Sigma^{*+})+\lambda^2\mathcal{A}(\Lambda_c^+\to K^0\Sigma^{*+})\nonumber\\&~~~+\sqrt{\frac{3}{2}}\lambda^2\mathcal{A}(\Xi_c^+\to \eta_8\Sigma^{*+})-\frac{\lambda^2\mathcal{A}(\Xi_c^+\to \pi^0\Sigma^{*+})}{\sqrt{2}}=0.
\end{align}
\begin{align}
&{\rm SumS}\,[\Xi_c^+, \overline K^0,\Delta^{+}]:\nonumber\\&\,\sqrt{\frac{3}{2}}\mathcal{A}(\Xi_c^+\to \eta_8\Delta^{+})-\frac{\mathcal{A}(\Xi_c^+\to \pi^0\Delta^{+})}{\sqrt{2}}+\lambda^2\mathcal{A}(\Lambda_c^+\to \overline K^0\Delta^{+})-\lambda^2\mathcal{A}(\Xi_c^+\to \overline K^0\Sigma^{*+})=0.
\end{align}
\begin{align}
&{\rm SumS}\,[\Xi_c^+, \overline K^0,\Sigma^{*+}]:\nonumber\\&\,\sqrt{\frac{3}{2}}\mathcal{A}(\Xi_c^+\to \eta_8\Sigma^{*+})+\mathcal{A}(\Xi_c^+\to \overline K^0\Delta^{+})-\frac{\mathcal{A}(\Xi_c^+\to \pi^0\Sigma^{*+})}{\sqrt{2}}+2\lambda\mathcal{A}(\Xi_c^+\to \overline K^0\Sigma^{*+})=0.
\end{align}
\begin{align}
&{\rm SumS}\,[\Lambda_c^+, \pi^+,\Delta^{0}]:\,\mathcal{A}(\Lambda_c^+\to K^+\Delta^{+})+\mathcal{A}(\Xi_c^+\to \pi^+\Delta^{0})+\sqrt{2}\lambda^2\mathcal{A}(\Lambda_c^+\to \pi^+\Sigma^{*0})=0.
\end{align}
\begin{align}
&{\rm SumS}\,[\Lambda_c^+, \pi^+,\Sigma^{*0}]:\nonumber\\&\,-\mathcal{A}(\Lambda_c^+\to K^+\Sigma^{*0})+\sqrt{2}\mathcal{A}(\Lambda_c^+\to \pi^+\Delta^{0})-\mathcal{A}(\Xi_c^+\to \pi^+\Sigma^{*0})+2\lambda\mathcal{A}(\Lambda_c^+\to \pi^+\Sigma^{*0})=0.
\end{align}
\begin{align}
&{\rm SumS}\,[\Lambda_c^+, \pi^+,\Xi^{*0}]:\,-\mathcal{A}(\Lambda_c^+\to K^+\Xi^{*0})+\sqrt{2}\mathcal{A}(\Lambda_c^+\to \pi^+\Sigma^{*0})-\mathcal{A}(\Xi_c^+\to \pi^+\Xi^{*0})=0.
\end{align}
\begin{align}
&{\rm SumS}\,[\Lambda_c^+, K^+,\Delta^{0}]:\nonumber\\&\,-2\lambda\mathcal{A}(\Lambda_c^+\to K^+\Delta^{0})-\sqrt{2}\lambda^2\mathcal{A}(\Lambda_c^+\to K^+\Sigma^{*0})+\lambda^2\mathcal{A}(\Lambda_c^+\to \pi^+\Delta^{0})=0.
\end{align}
\begin{align}
{\rm SumS}\,[\Lambda_c^+, K^+,\Sigma^{*0}]:&\,-\mathcal{A}(\Xi_c^+\to K^+\Sigma^{*0})+\lambda^2\mathcal{A}(\Lambda_c^+\to \pi^+\Sigma^{*0})\nonumber\\&~~~+\sqrt{2}\mathcal{A}(\Lambda_c^+\to K^+\Delta^{0})-\sqrt{2}\lambda^2\mathcal{A}(\Lambda_c^+\to K^+\Xi^{*0})=0.
\end{align}
\begin{align}
&{\rm SumS}\,[\Lambda_c^+, K^+,\Xi^{*0}]:\,\sqrt{2}\mathcal{A}(\Lambda_c^+\to K^+\Sigma^{*0})-\mathcal{A}(\Xi_c^+\to K^+\Xi^{*0})+2\lambda\mathcal{A}(\Lambda_c^+\to K^+\Xi^{*0})=0.
\end{align}
\begin{align}
&{\rm SumS}\,[\Lambda_c^+, \pi^0,\Delta^{+}]:\nonumber\\&\,-\mathcal{A}(\Xi_c^+\to \pi^0\Delta^{+})-\lambda^2\mathcal{A}(\Lambda_c^+\to \pi^0\Sigma^{*+})+\frac{\mathcal{A}(\Lambda_c^+\to K^0\Delta^{+})}{\sqrt{2}}+\frac{\lambda^2\mathcal{A}(\Lambda_c^+\to \overline K^0\Delta^{+})}{\sqrt{2}}=0.
\end{align}
\begin{align}
&{\rm SumS}\,[\Lambda_c^+, \pi^0,\Sigma^{*+}]:\nonumber\\&\,\frac{\mathcal{A}(\Lambda_c^+\to K^0\Sigma^{*+})}{\sqrt{2}}+\mathcal{A}(\Lambda_c^+\to \pi^0\Delta^{+})-\mathcal{A}(\Xi_c^+\to \pi^0\Sigma^{*+})+2\lambda\mathcal{A}(\Lambda_c^+\to \pi^0\Sigma^{*+})=0.
\end{align}
\begin{align}
{\rm SumS}\,[\Lambda_c^+, \eta_8,\Delta^{+}]:&\,-\frac{\sqrt{6}}{2}\mathcal{A}(\Lambda_c^+\to K^0\Delta^{+})-\mathcal{A}(\Xi_c^+\to \eta_8\Delta^{+})\nonumber\\&~~~-\lambda^2\mathcal{A}(\Lambda_c^+\to \eta_8\Sigma^{*+})+\frac{\sqrt{6}}{2}\lambda^2\mathcal{A}(\Lambda_c^+\to \overline K^0\Delta^{+})=0.
\end{align}
\begin{align}
&{\rm SumS}\,[\Lambda_c^+, \eta_8,\Sigma^{*+}]:\nonumber\\&\,\mathcal{A}(\Lambda_c^+\to \eta_8\Delta^{+})-\sqrt{\frac{3}{2}}\mathcal{A}(\Lambda_c^+\to K^0\Sigma^{*+})-\mathcal{A}(\Lambda_c^+\to \eta_8\Sigma^{*+})+2\lambda\mathcal{A}(\Lambda_c^+\to \eta_8\Sigma^{*+})=0.
\end{align}
\begin{align}
{\rm SumS}\,[\Lambda_c^+, K^0,\Delta^{+}]:&\,-2\lambda\mathcal{A}(\Lambda_c^+\to K^0\Delta^{+})+\frac{\sqrt{6}}{2}\lambda^2\mathcal{A}(\Lambda_c^+\to \eta_8\Delta^{+})\nonumber\\&~~~-\lambda^2\mathcal{A}(\Lambda_c^+\to K^0\Sigma^{*+})-\frac{\sqrt{2}}{2}\lambda^2\mathcal{A}(\Lambda_c^+\to \pi^0\Delta^{+})=0.
\end{align}
\begin{align}
{\rm SumS}\,[\Lambda_c^+, K^0,\Sigma^{*+}]:&\,\mathcal{A}(\Lambda_c^+\to K^0\Delta^{+})-\mathcal{A}(\Xi_c^+\to K^0\Sigma^{*+})\nonumber\\&~~~+\sqrt{\frac{3}{2}}\lambda^2\mathcal{A}(\Lambda_c^+\to \eta_8\Sigma^{*+})-\frac{\sqrt{2}}{2}\lambda^2\mathcal{A}(\Lambda_c^+\to \pi^0\Sigma^{*+})=0.
\end{align}
\begin{align}
&{\rm SumS}\,[\Lambda_c^+, \overline K^0,\Delta^{+}]:\nonumber\\&\,\sqrt{\frac{3}{2}}\mathcal{A}(\Lambda_c^+\to \eta_8\Delta^{+})-\frac{\mathcal{A}(\Lambda_c^+\to \pi^0\Delta^{+})}{\sqrt{2}}-\mathcal{A}(\Xi_c^+\to \overline K^0\Delta^{+})+2\lambda\mathcal{A}(\Lambda_c^+\to \overline K^0\Delta^{+})=0.
\end{align}
\begin{align}
&{\rm SumS}\,[\Lambda_c^+, \overline K^0,\Sigma^{*+}]:\nonumber\\&\,\sqrt{\frac{3}{2}}\mathcal{A}(\Lambda_c^+\to \eta_8\Sigma^{*+})+\mathcal{A}(\Lambda_c^+\to \overline K^0\Delta^{+})-\frac{\mathcal{A}(\Lambda_c^+\to  \pi^0\Sigma^{*+})}{\sqrt{2}}-\mathcal{A}(\Xi_c^+\to \overline K^0\Sigma^{*+})=0.
\end{align}
\begin{align}
&{\rm SumS}\,[\Xi_c^+, K^-,\Delta^{++}]:\nonumber\\&\,-2\lambda\mathcal{A}(\Xi_c^+\to  \pi^-\Delta^{++})+\lambda^2\mathcal{A}(\Lambda_c^+\to \pi^-\Delta^{++})-2\lambda^2\mathcal{A}(\Xi_c^+\to K^-\Delta^{++})=0.
\end{align}
\begin{align}
&{\rm SumS}\,[\Xi_c^+, \pi^-,\Delta^{++}]:\,\mathcal{A}(\Xi_c^+\to  \pi^-\Delta^{++})+\mathcal{A}(\Lambda_c^+\to K^-\Delta^{++})=0.
\end{align}
\begin{align}
&{\rm SumS}\,[\Lambda_c^+, \pi^-,\Delta^{++}]:\,\mathcal{A}(\Xi_c^+\to  \pi^-\Delta^{++})+\mathcal{A}(\Lambda_c^+\to K^-\Delta^{++})=0.
\end{align}
\begin{align}
&{\rm SumS}\,[\Lambda_c^+, K^-,\Delta^{++}]:\nonumber\\&\,\mathcal{A}(\Lambda_c^+\to  \pi^-\Delta^{++})-\mathcal{A}(\Xi_c^+\to K^-\Delta^{++})+2\lambda\mathcal{A}(\Lambda_c^+\to K^-\Delta^{++})=0.
\end{align}
\begin{align}
{\rm SumT_-}\,[\Xi_c^0,\pi^+,\Delta^{0}]:&\,-\sqrt{2}\mathcal{A}(\Xi_c^0\to  \pi^0\Delta^{0})+\sqrt{3}\mathcal{A}(\Xi_c^0\to \pi^+\Delta^{-})-\mathcal{A}(\Xi_c^+\to \pi^+\Delta^{0})\nonumber\\&~~~+\lambda\mathcal{A}(\Lambda_c^+\to \pi^+\Delta^{0})+\lambda\mathcal{A}(\Xi_c^0\to \overline K^0\Delta^{0})+\lambda\mathcal{A}(\Xi_c^0\to  \pi^+\Sigma^{*-})=0.
\end{align}
\begin{align}
{\rm SumT_-}\,[\Xi_c^0,\pi^+,\Sigma^{*0}]:&\,-\sqrt{2}\mathcal{A}(\Xi_c^0\to  \pi^0\Sigma^{*0})+\sqrt{2}\mathcal{A}(\Xi_c^0\to \pi^+\Sigma^{*-})\nonumber\\&~~~-\mathcal{A}(\Xi_c^+\to \pi^+\Sigma^{*0})+\lambda\mathcal{A}(\Lambda_c^+\to \pi^+\Sigma^{*0})\nonumber\\&~~~~~+\lambda\mathcal{A}(\Xi_c^0\to \overline K^0\Sigma^{*0})+\sqrt{2}\lambda\mathcal{A}(\Xi_c^0\to  \pi^+\Xi^{*-})=0.
\end{align}
\begin{align}
&{\rm SumT_-}\,[\Xi_c^0,\pi^+,\Xi^{*0}]:\,-\sqrt{2}\mathcal{A}(\Xi_c^0\to  \pi^0\Xi^{*0})+\mathcal{A}(\Xi_c^0\to \pi^+\Xi^{*-})-\mathcal{A}(\Xi_c^+\to \pi^+\Xi^{*0})=0.
\end{align}
\begin{align}
&{\rm SumT_-}\,[\Xi_c^0,K^+,\Delta^{0}]:\nonumber\\&\,\mathcal{A}(\Lambda_c^+\to  K^+\Delta^{0})-\sqrt{\frac{3}{2}}\mathcal{A}(\Xi_c^0\to \eta_8\Delta^{0})+\mathcal{A}(\Xi_c^0\to K^+\Sigma^{*-})-\frac{\mathcal{A}(\Xi_c^0\to \pi^0\Delta^{0})}{\sqrt{2}}=0.
\end{align}
\begin{align}
{\rm SumT_-}\,[\Xi_c^0,K^+,\Sigma^{*0}]:&\,\mathcal{A}(\Xi_c^0\to  K^0\Sigma^{*0})+\sqrt{2}\mathcal{A}(\Xi_c^0\to K^+\Sigma^{*-})-\mathcal{A}(\Xi_c^+\to K^+\Sigma^{*0})\nonumber\\&~~~+\lambda\mathcal{A}(\Lambda_c^+\to K^+\Sigma^{*0})-\sqrt{\frac{3}{2}}\lambda\mathcal{A}(\Xi_c^0\to \eta_8\Sigma^{*0})\nonumber\\&~~~~~+\sqrt{2}\lambda\mathcal{A}(\Xi_c^0\to K^+\Xi^{*-})-\frac{\lambda\mathcal{A}(\Xi_c^0\to \pi^0\Sigma^{*0})}{\sqrt{2}}=0.
\end{align}
\begin{align}
{\rm SumT_-}\,[\Xi_c^0,K^+,\Xi^{*0}]:&\,\mathcal{A}(\Xi_c^0\to  K^0\Xi^{*0})+\mathcal{A}(\Xi_c^0\to K^+\Xi^{*-})-\mathcal{A}(\Xi_c^+\to K^+\Xi^{*0})\nonumber\\&~~~+\lambda\mathcal{A}(\Lambda_c^+\to K^+\Xi^{*0})-\sqrt{\frac{3}{2}}\lambda\mathcal{A}(\Xi_c^0\to \eta_8\Xi^{*0})\nonumber\\&~~~~~+\sqrt{3}\lambda\mathcal{A}(\Xi_c^0\to K^+\Omega^{-})-\frac{\lambda\mathcal{A}(\Xi_c^0\to \pi^0\Xi^{*0})}{\sqrt{2}}=0.
\end{align}
\begin{align}
{\rm SumT_-}\,[\Xi_c^0,\pi^0,\Delta^{+}]:&\,2\mathcal{A}(\Xi_c^0\to  \pi^0\Delta^{0})+\sqrt{2}\mathcal{A}(\Xi_c^0\to \pi^-\Delta^{+})-\mathcal{A}(\Xi_c^+\to \pi^0\Delta^{+})\nonumber\\&~~~+\lambda\mathcal{A}(\Lambda_c^+\to \pi^0\Delta^{+})+\frac{\lambda\mathcal{A}(\Xi_c^0\to K^-\Delta^{+})}{\sqrt{2}}\nonumber\\&~~~~~+\sqrt{2}\lambda\mathcal{A}(\Xi_c^0\to \pi^0\Sigma^{*0})=0.
\end{align}
\begin{align}
{\rm SumT_-}\,[\Xi_c^0,\pi^0,\Sigma^{*+}]:&\,\sqrt{2}\mathcal{A}(\Xi_c^0\to  \pi^0\Sigma^{*0})+\sqrt{2}\mathcal{A}(\Xi_c^0\to \pi^-\Sigma^{*+})-\mathcal{A}(\Xi_c^+\to \pi^0\Sigma^{*+})\nonumber\\&~~~+\lambda\mathcal{A}(\Lambda_c^+\to \pi^0\Sigma^{*+})+\frac{\lambda\mathcal{A}(\Xi_c^0\to K^-\Sigma^{*+})}{\sqrt{2}}\nonumber\\&~~~~~+2\lambda\mathcal{A}(\Xi_c^0\to \pi^0\Xi^{*0})=0.
\end{align}
\begin{align}
{\rm SumT_-}\,[\Xi_c^0,\eta_8,\Delta^{+}]:&\,2\mathcal{A}(\Xi_c^0\to  \eta_8\Delta^{0})-\mathcal{A}(\Xi_c^+\to \eta_8\Delta^{+})+\lambda\mathcal{A}(\Lambda_c^+\to \eta_8\Delta^{+})\nonumber\\&~~~+\sqrt{2}\lambda\mathcal{A}(\Xi_c^0\to \eta_8\Sigma^{*0})+\sqrt{\frac{3}{2}}\lambda\mathcal{A}(\Xi_c^0\to K^-\Delta^{+})=0.
\end{align}
\begin{align}
{\rm SumT_-}\,[\Xi_c^0,\eta_8,\Sigma^{*+}]:&\,\sqrt{2}\mathcal{A}(\Xi_c^0\to  \eta_8\Sigma^{*0})-\mathcal{A}(\Xi_c^+\to \eta_8\Sigma^{*+})+\lambda\mathcal{A}(\Lambda_c^+\to \eta_8\Sigma^{*+})\nonumber\\&~~~+2\lambda\mathcal{A}(\Xi_c^0\to \eta_8\Xi^{*0})+\sqrt{\frac{3}{2}}\lambda\mathcal{A}(\Xi_c^0\to K^-\Sigma^{*+})=0.
\end{align}
\begin{align}
&{\rm SumT_-}\,[\Xi_c^0,K^0,\Delta^{+}]:\,\mathcal{A}(\Lambda_c^+\to  K^0\Delta^{+})+\sqrt{2}\mathcal{A}(\Xi_c^0\to K^0\Sigma^{*0})-\mathcal{A}(\Xi_c^0\to \pi^-\Delta^{+})=0.
\end{align}
\begin{align}
{\rm SumT_-}\,[\Xi_c^0,K^0,\Sigma^{*+}]:&\,\sqrt{2}\mathcal{A}(\Xi_c^0\to  K^0\Sigma^{*0})-\mathcal{A}(\Xi_c^+\to K^0\Sigma^{*+})+\lambda\mathcal{A}(\Lambda_c^+\to K^0\Sigma^{*+})\nonumber\\&~~~+2\lambda\mathcal{A}(\Xi_c^0\to K^0\Xi^{*0})-\lambda\mathcal{A}(\Xi_c^0\to \pi^-\Sigma^{*+})=0.
\end{align}
\begin{align}
{\rm SumT_-}\,[\Xi_c^0,\overline K^0,\Delta^{+}]:&\,-\mathcal{A}(\Xi_c^0\to  K^-\Delta^{+})+2\mathcal{A}(\Xi_c^0\to \overline K^0\Delta^{0})-\mathcal{A}(\Xi_c^+\to \overline K^0\Delta^{+})\nonumber\\&~~~+\lambda\mathcal{A}(\Lambda_c^+\to \overline K^0\Delta^{+})+\sqrt{2}\lambda\mathcal{A}(\Xi_c^0\to \overline K^0\Sigma^{*0})=0.
\end{align}
\begin{align}
&{\rm SumT_-}\,[\Xi_c^0,\overline K^0,\Sigma^{*+}]:\,-\mathcal{A}(\Xi_c^0\to  K^-\Sigma^{*+})+\sqrt{2}\mathcal{A}(\Xi_c^0\to \overline K^0\Sigma^{*0})-\mathcal{A}(\Xi_c^+\to \overline K^0\Sigma^{*+})=0.
\end{align}
\begin{align}
{\rm SumT_-}\,[\Xi_c^0,\pi^-,\Delta^{++}]:&\,\sqrt{3}\mathcal{A}(\Xi_c^0\to  \pi^-\Delta^{+})-\mathcal{A}(\Xi_c^+\to \pi^-\Delta^{++})\nonumber\\&~~~+\lambda\mathcal{A}(\Lambda_c^+\to \pi^-\Delta^{++})+\sqrt{3}\lambda\mathcal{A}(\Xi_c^0\to \pi^-\Sigma^{*+})=0.
\end{align}
\begin{align}
{\rm SumT_-}\,[\Xi_c^0,K^-,\Delta^{++}]:&\,\sqrt{3}\mathcal{A}(\Xi_c^0\to  K^-\Delta^{+})-\mathcal{A}(\Xi_c^+\to K^-\Delta^{++})\nonumber\\&~~~+\lambda\mathcal{A}(\Lambda_c^+\to K^-\Delta^{++})+\sqrt{3}\lambda\mathcal{A}(\Xi_c^0\to K^-\Sigma^{*+})=0.
\end{align}
\begin{align}
{\rm SumT_-}\,[\Xi_c^+,\pi^+,\Delta^{+}]:&\,-\sqrt{2}\mathcal{A}(\Xi_c^+\to  \pi^0\Delta^{+})+2\mathcal{A}(\Xi_c^+\to \pi^+\Delta^{0})\nonumber\\&~~~+\lambda\mathcal{A}(\Lambda_c^+\to \overline K^0\Delta^{+})+\sqrt{2}\lambda\mathcal{A}(\Xi_c^+\to \pi^+\Sigma^{*0})=0.
\end{align}
\begin{align}
{\rm SumT_-}\,[\Xi_c^+,\pi^+,\Sigma^{*+}]:&\,-\sqrt{2}\mathcal{A}(\Xi_c^+\to  \pi^0\Sigma^{*+})+\sqrt{2}\mathcal{A}(\Xi_c^+\to \pi^+\Sigma^{*0})\nonumber\\&~~~+\lambda\mathcal{A}(\Xi_c^+\to \overline K^0\Sigma^{*+})+2\lambda\mathcal{A}(\Xi_c^+\to \pi^+\Xi^{*0})=0.
\end{align}
\begin{align}
&{\rm SumT_-}\,[\Xi_c^+,K^+,\Delta^{+}]:\,\sqrt{3}\mathcal{A}(\Xi_c^+\to  \eta_8\Delta^{+})-2\mathcal{A}(\Xi_c^+\to K^+\Sigma^{*0})+\mathcal{A}(\Xi_c^+\to \pi^0\Delta^{+})=0.
\end{align}
\begin{align}
{\rm SumT_-}\,[\Xi_c^+,K^+,\Sigma^{*+}]:&\,\mathcal{A}(\Xi_c^+\to  K^0\Sigma^{*+})+\sqrt{2}\mathcal{A}(\Xi_c^+\to K^+\Sigma^{*0})-\frac{\sqrt{6}}{2}\lambda\mathcal{A}(\Xi_c^+\to \eta_8\Sigma^{*+})\nonumber\\&~~~+2\lambda\mathcal{A}(\Xi_c^+\to K^+\Xi^{*0})-\frac{\lambda\mathcal{A}(\Xi_c^+\to \pi^0\Sigma^{*+})}{\sqrt{2}}=0.
\end{align}
\begin{align}
{\rm SumT_-}\,[\Xi_c^+,\pi^0,\Delta^{++}]:&\,\sqrt{3}\mathcal{A}(\Xi_c^+\to  \pi^0\Delta^{+})+\sqrt{2}\mathcal{A}(\Xi_c^+\to \pi^-\Delta^{++})\nonumber\\&~~~+\frac{\lambda\mathcal{A}(\Xi_c^+\to K^-\Delta^{++})}{\sqrt{2}}+\sqrt{3}\lambda\mathcal{A}(\Xi_c^+\to \pi^0\Sigma^{*+})=0.
\end{align}
\begin{align}
&{\rm SumT_-}\,[\Xi_c^+,\eta_8,\Delta^{++}]:\nonumber\\&\,\sqrt{3}\mathcal{A}(\Xi_c^+\to  \eta_8\Delta^{+})+\sqrt{3}\lambda\mathcal{A}(\Xi_c^+\to \eta_8\Sigma^{*+})+\frac{\sqrt{6}}{2}\lambda\mathcal{A}(\Xi_c^+\to K^-\Delta^{++})=0.
\end{align}
\begin{align}
&{\rm SumT_-}\,[\Xi_c^+,K^0,\Delta^{++}]:\,\sqrt{3}\mathcal{A}(\Xi_c^+\to  K^0\Sigma^{*+})-\mathcal{A}(\Xi_c^+\to \pi^-\Delta^{++})=0.
\end{align}
\begin{align}
&{\rm SumT_-}\,[\Xi_c^+,\overline K^0,\Delta^{++}]:\nonumber\\&~~~\,-\mathcal{A}(\Xi_c^+\to  K^-\Delta^{++})+\sqrt{3}\mathcal{A}(\Xi_c^+\to \overline K^0\Delta^{+})+\sqrt{3}\lambda\mathcal{A}(\Xi_c^+\to \overline K^0\Sigma^{*+})=0.
\end{align}
\begin{align}
{\rm SumT_-}\,[\Lambda_c^+,\pi^+,\Delta^{+}]:&\,-\sqrt{2}\mathcal{A}(\Lambda_c^+\to  \pi^0\Delta^{+})+2\mathcal{A}(\Lambda_c^+\to \pi^+\Delta^{0})\nonumber\\&~~~+\lambda\mathcal{A}(\Lambda_c^+\to \overline K^0\Delta^{+})+\sqrt{2}\lambda\mathcal{A}(\Lambda_c^+\to \pi^+\Sigma^{*0})=0.
\end{align}
\begin{align}
&{\rm SumT_-}\,[\Lambda_c^+,\pi^+,\Sigma^{*+}]:\,-\mathcal{A}(\Lambda_c^+\to  \pi^0\Sigma^{*+})+\mathcal{A}(\Lambda_c^+\to \pi^+\Sigma^{*0})=0.
\end{align}
\begin{align}
{\rm SumT_-}\,[\Lambda_c^+,K^+,\Delta^{+}]:&\,\mathcal{A}(\Lambda_c^+\to  K^0\Delta^{+})+2\mathcal{A}(\Lambda_c^+\to K^+\Delta^{0})-\sqrt{\frac{3}{2}}\lambda\mathcal{A}(\Lambda_c^+\to \eta_8\Delta^{+})\nonumber\\&~~~+\sqrt{2}\lambda\mathcal{A}(\Lambda_c^+\to K^+\Sigma^{*0})-\frac{\lambda\mathcal{A}(\Lambda_c^+\to \pi^0\Delta^{+})}{\sqrt{2}}=0.
\end{align}
\begin{align}
{\rm SumT_-}\,[\Lambda_c^+,K^+,\Sigma^{*+}]:&\,\mathcal{A}(\Lambda_c^+\to  K^0\Sigma^{*+})+\sqrt{2}\mathcal{A}(\Lambda_c^+\to K^+\Sigma^{*0})-\frac{\sqrt{6}}{2}\lambda\mathcal{A}(\Lambda_c^+\to \eta_8\Sigma^{*+})\nonumber\\&~~~+2\lambda\mathcal{A}(\Lambda_c^+\to K^+\Xi^{*0})-\frac{\lambda\mathcal{A}(\Lambda_c^+\to \pi^0\Sigma^{*+})}{\sqrt{2}}=0.
\end{align}
\begin{align}
{\rm SumT_-}\,[\Lambda_c^+,\pi^0,\Delta^{++}]:&\,\sqrt{3}\mathcal{A}(\Lambda_c^+\to  \pi^0\Lambda^{++})+\sqrt{2}\mathcal{A}(\Lambda_c^+\to \pi^-\Delta^{++})\nonumber\\&~~~+\frac{\lambda\mathcal{A}(\Lambda_c^+\to K^-\Delta^{++})}{\sqrt{2}}+\sqrt{3}\lambda\mathcal{A}(\Lambda_c^+\to \pi^0\Sigma^{*+})=0.
\end{align}
\begin{align}
&{\rm SumT_-}\,[\Lambda_c^+,\eta_8,\Delta^{++}]:\nonumber\\&~~~\,\sqrt{3}\mathcal{A}(\Lambda_c^+\to  \eta_8\Lambda^{+})+\sqrt{3}\lambda\mathcal{A}(\Lambda_c^+\to \eta_8\Sigma^{*+})+\frac{\sqrt{6}}{2}\lambda\mathcal{A}(\Lambda_c^+\to K^-\Delta^{++})=0.
\end{align}
\begin{align}
&{\rm SumT_-}\,[\Lambda_c^+,K^0,\Delta^{++}]:\nonumber\\&~~~\,-\mathcal{A}(\Lambda_c^+\to  \pi^-\Delta^{++})+\sqrt{3}\lambda\mathcal{A}(\Lambda_c^+\to K^0\Delta^{+})+\sqrt{3}\lambda\mathcal{A}(\Lambda_c^+\to K^0\Sigma^{*+})=0.
\end{align}
\begin{align}
&{\rm SumT_-}\,[\Lambda_c^+,\overline K^0,\Delta^{++}]:\,-\mathcal{A}(\Lambda_c^+\to  K^-\Delta^{++})+\sqrt{3}\mathcal{A}(\Lambda_c^+\to\overline K^0\Delta^{+})=0.
\end{align}
\begin{align}
&{\rm SumS}\,[\Xi_c^0,\eta_1,\Delta^{0}]:\,2\mathcal{A}(\Xi_c^0\to  \eta_1\Delta^{0})+\sqrt{2}\lambda\mathcal{A}(\Xi_c^0\to \eta_1\Sigma^{*0})=0.
\end{align}
\begin{align}
&{\rm SumS}\,[\Xi_c^0,\eta_1,\Sigma^{*0}]:\,\sqrt{2}\mathcal{A}(\Xi_c^0\to  \eta_1\Delta^{0})-\lambda^2\mathcal{A}(\Xi_c^0\to \eta_1\Xi^{*0})=0.
\end{align}
\begin{align}
&{\rm SumS}\,[\Xi_c^0,\eta_1,\Xi^{*0}]:\,\sqrt{2}\mathcal{A}(\Xi_c^0\to  \eta_1\Sigma^{*0})+2\lambda\mathcal{A}(\Xi_c^0\to \eta_1\Xi^{*0})=0.
\end{align}
\begin{align}
&{\rm SumS}\,[\Xi_c^+,\eta_1,\Delta^{+}]:\,-2\mathcal{A}(\Xi_c^+\to  \eta_1\Delta^{+})+\lambda\mathcal{A}(\Lambda_c^+\to \eta_1\Delta^{+})-\lambda\mathcal{A}(\Xi_c^+\to \eta_1\Sigma^{*+})=0.
\end{align}
\begin{align}
&{\rm SumS}\,[\Xi_c^+,\eta_1,\Sigma^{*+}]:\,\mathcal{A}(\Xi_c^+\to  \eta_1\Delta^{+})+\lambda^2\mathcal{A}(\Lambda_c^+\to \eta_1\Sigma^{*+})=0.
\end{align}
\begin{align}
&{\rm SumS}\,[\Lambda_c^+,\eta_1,\Delta^{+}]:\,\mathcal{A}(\Xi_c^+\to  \eta_1\Delta^{+})+\lambda^2\mathcal{A}(\Lambda_c^+\to \eta_1\Sigma^{*+})=0.
\end{align}
\begin{align}
&{\rm SumS}\,[\Lambda_c^+,\eta_1,\Sigma^{*+}]:\,\mathcal{A}(\Lambda_c^+\to  \eta_1\Delta^{+})-\mathcal{A}(\Xi_c^+\to \eta_1\Sigma^{*+})+2\lambda\mathcal{A}(\Lambda_c^+\to \eta_1\Sigma^{*+})=0.
\end{align}
\begin{align}
&{\rm SumT_-}\,[\Xi_c^0,\eta_1,\Delta^{+}]:\nonumber\\&\,2\mathcal{A}(\Xi_c^0\to  \eta_1\Delta^{0})-\mathcal{A}(\Xi_c^+\to\eta_1\Delta^{+})
+\lambda\mathcal{A}(\Lambda_c^+\to\eta_1\Delta^{+})
+\sqrt{2}\lambda\mathcal{A}(\Xi_c^0\to\eta_1\Sigma^{*0})=0.
\end{align}
\begin{align}
&{\rm SumT_-}\,[\Xi_c^0,\eta_1,\Sigma^{*+}]:\nonumber\\&\,\sqrt{2}\mathcal{A}(\Xi_c^0\to  \eta_1\Sigma^{*0})-\mathcal{A}(\Xi_c^+\to\eta_1\Sigma^{*+})
+\lambda\mathcal{A}(\Lambda_c^+\to\eta_1\Sigma^{*+})
+2\lambda\mathcal{A}(\Xi_c^0\to\eta_1\Xi^{*0})=0.
\end{align}
\begin{align}
&{\rm SumT_-}\,[\Xi_c^+,\eta_1,\Delta^{++}]:\,\mathcal{A}(\Xi_c^+\to  \eta_1\Delta^{+})+\lambda\mathcal{A}(\Xi_c^+\to\eta_1\Sigma^{*+})=0.
\end{align}
\begin{align}
&{\rm SumT_-}\,[\Lambda_c^+,\eta_1,\Delta^{++}]:\,\mathcal{A}(\Lambda_c^+\to  \eta_1\Delta^{+})+\lambda\mathcal{A}(\Lambda_c^+\to\eta_1\Sigma^{*+})=0.
\end{align}

\section{SU(3) sum rules of $\mathcal{B}_{cc}\to M\mathcal{B}_{\overline c3}$ modes}\label{res3}

\begin{align}
&{\rm SumS}\,[\Xi_{cc}^{++},\pi^+,\Xi_c^{+}]:\,-\mathcal{A}(\Xi_{cc}^{++}\to  K^+\Xi_c^{+})+\mathcal{A}(\Xi_{cc}^{++}\to\pi^+\Lambda_c^{+})
+2\lambda\mathcal{A}(\Xi_{cc}^{++}\to\pi^+\Xi_c^{+})=0.
\end{align}
\begin{align}
&{\rm SumS}\,[\Xi_{cc}^{++},\pi^+,\Lambda_c^{+}]:\,\mathcal{A}(\Xi_{cc}^{++}\to  K^+\Lambda_c^{+})+\lambda^2\mathcal{A}(\Xi_{cc}^{++}\to\pi^+\Xi_c^{+})=0.
\end{align}
\begin{align}
&{\rm SumS}\,[\Xi_{cc}^{++},K^+,\Xi_c^{+}]:\,\mathcal{A}(\Xi_{cc}^{++}\to  K^+\Lambda_c^{+})+\lambda^2\mathcal{A}(\Xi_{cc}^{++}\to\pi^+\Xi_c^{+})=0.
\end{align}
\begin{align}
&{\rm SumS}\,[\Xi_{cc}^{++},K^+,\Lambda_c^{+}]:\nonumber\\&~~~\,-2\lambda\mathcal{A}(\Xi_{cc}^{++}\to  K^+\Lambda_c^{+})-\lambda^2\mathcal{A}(\Xi_{cc}^{++}\to K^+\Xi_c^{+})+\lambda^2\mathcal{A}(\Xi_{cc}^{++}\to \pi^+\Lambda_c^{+})=0.
\end{align}
\begin{align}
&{\rm SumS}\,[\Xi_{cc}^{+},\pi^+,\Xi_c^{0}]:\,-\mathcal{A}(\Xi_{cc}^{+}\to  K^+\Xi_c^{0})-\mathcal{A}(\Omega_{cc}^{+}\to \pi^+\Xi_c^{0})+2\lambda\mathcal{A}(\Xi_{cc}^{+}\to \pi^+\Xi_c^{0})=0.
\end{align}
\begin{align}
&{\rm SumS}\,[\Xi_{cc}^{+},K^+,\Xi_c^{0}]:\,-\mathcal{A}(\Omega_{cc}^{+}\to  K^+\Xi_c^{0})+\lambda^2\mathcal{A}(\Xi_{cc}^{+}\to \pi^+\Xi_c^{0})=0.
\end{align}
\begin{align}
&{\rm SumS}\,[\Omega_{cc}^{+},\pi^+,\Xi_c^{0}]:\,-\mathcal{A}(\Omega_{cc}^{+}\to  K^+\Xi_c^{0})+\lambda^2\mathcal{A}(\Xi_{cc}^{+}\to \pi^+\Xi_c^{0})=0.
\end{align}
\begin{align}
&{\rm SumS}\,[\Omega_{cc}^{+},K^+,\Xi_c^{0}]:\,-2\lambda\mathcal{A}(\Omega_{cc}^{+}\to  K^+\Xi_c^{0})+\lambda^2\mathcal{A}(\Xi_{cc}^{+}\to K^+\Xi_c^{0})+\lambda^2\mathcal{A}(\Omega_{cc}^{+}\to \pi^+\Xi_c^{0})=0.
\end{align}
\begin{align}
&{\rm SumS}\,[\Xi_{cc}^{+},\pi^0,\Xi_c^{+}]:\nonumber\\&\,\frac{\mathcal{A}(\Xi_{cc}^{+}\to  K^0\Xi_c^{+})}{\sqrt{2}}+\mathcal{A}(\Xi_{cc}^{+}\to \pi^0\Lambda_c^{+})-\mathcal{A}(\Omega_{cc}^{+}\to \pi^0\Xi_c^{+})+2\lambda\mathcal{A}(\Xi_{cc}^{+}\to \pi^0\Xi_c^{+})=0.
\end{align}
\begin{align}
&{\rm SumS}\,[\Xi_{cc}^{+},\eta_8,\Xi_c^{+}]:\nonumber\\&\,\mathcal{A}(\Xi_{cc}^{+}\to  \eta_8\Lambda_c^{+})-\sqrt{\frac{3}{2}}\mathcal{A}(\Xi_{cc}^{+}\to K^0\Xi_c^{+})-\mathcal{A}(\Omega_{cc}^{+}\to \eta_8\Xi_c^{+})+2\lambda\mathcal{A}(\Xi_{cc}^{+}\to \eta_8\Xi_c^{+})=0.
\end{align}
\begin{align}
&{\rm SumS}\,[\Xi_{cc}^{+},K^0,\Xi_c^{+}]:\nonumber\\&\,\mathcal{A}(\Xi_{cc}^{+}\to  K^0\Lambda_c^{+})-\mathcal{A}(\Omega_{cc}^{+}\to K^0\Xi_c^{+})+\sqrt{\frac{3}{2}}\lambda^2\mathcal{A}(\Xi_{cc}^{+}\to \eta_8\Xi_c^{+})-\frac{\lambda^2\mathcal{A}(\Xi_{cc}^{+}\to \pi^0\Xi_c^{+})}{\sqrt{2}}=0.
\end{align}
\begin{align}
&{\rm SumS}\,[\Xi_{cc}^{+},\overline K^0,\Xi_c^{+}]:\nonumber\\&\,\sqrt{\frac{3}{2}}\mathcal{A}(\Xi_{cc}^{+}\to  \eta_8\Xi_c^{+})+\mathcal{A}(\Xi_{cc}^{+}\to \overline K^0\Lambda_c^{+})-\frac{\mathcal{A}(\Xi_{cc}^{+}\to \pi^0\Xi_c^{+})}{\sqrt{2}}-\mathcal{A}(\Omega_{cc}^{+}\to \overline K^0\Xi_c^{+})=0.
\end{align}
\begin{align}
&{\rm SumS}\,[\Xi_{cc}^{+},\pi^0,\Lambda_c^{+}]:\nonumber\\&\,\frac{\mathcal{A}(\Xi_{cc}^{+}\to  K^0\Lambda_c^{+})}{\sqrt{2}}-\mathcal{A}(\Omega_{cc}^{+}\to \pi^0\Lambda_c^{+})+\frac{\lambda^2\mathcal{A}(\Xi_{cc}^{+}\to \overline K^0\Lambda_c^{+})}{\sqrt{2}}-\lambda^2\mathcal{A}(\Xi_{cc}^{+}\to \pi^0\Xi_c^{+})=0.
\end{align}
\begin{align}
&{\rm SumS}\,[\Xi_{cc}^{+},\eta_8,\Lambda_c^{+}]:\nonumber\\&\,\sqrt{\frac{3}{2}}\mathcal{A}(\Xi_{cc}^{+}\to  K^0\Lambda_c^{+})+\mathcal{A}(\Omega_{cc}^{+}\to \eta_8\Lambda_c^{+})+\lambda^2\mathcal{A}(\Xi_{cc}^{+}\to \eta_8\Xi_c^{+})+\sqrt{\frac{3}{2}}\lambda^2\mathcal{A}(\Xi_{cc}^{+}\to \overline K^0\Lambda_c^{+})=0.
\end{align}
\begin{align}
{\rm SumS}\,[\Xi_{cc}^{+},K^0,\Lambda_c^{+}]:&\,-2\lambda\mathcal{A}(\Xi_{cc}^{+}\to  K^0\Lambda_c^{+})+\sqrt{\frac{3}{2}}\lambda^2\mathcal{A}(\Xi_{cc}^{+}\to \eta_8\Lambda_c^{+})\nonumber\\&~~~-\lambda^2\mathcal{A}(\Xi_{cc}^{+}\to K^0\Xi_c^{+})-\frac{\lambda^2\mathcal{A}(\Xi_{cc}^{+}\to \pi^0\Lambda_c^{+})}{\sqrt{2}}=0.
\end{align}
\begin{align}
&{\rm SumS}\,[\Xi_{cc}^{+},\overline K^0,\Lambda_c^{+}]:\nonumber\\&\,\sqrt{\frac{3}{2}}\mathcal{A}(\Xi_{cc}^{+}\to  \eta_8\Lambda_c^{+})-\frac{\mathcal{A}(\Xi_{cc}^{+}\to \pi^0\Lambda_c^{+})}{\sqrt{2}}-\mathcal{A}(\Omega_{cc}^{+}\to \overline K^0\Lambda_c^{+})+2\lambda\mathcal{A}(\Xi_{cc}^{+}\to \overline K^0\Lambda_c^{+})=0.
\end{align}
\begin{align}
&{\rm SumS}\,[\Omega_{cc}^{+},\pi^0,\Xi_c^{+}]:\nonumber\\&\,\frac{\mathcal{A}(\Omega_{cc}^{+}\to  K^0\Xi_c^{+})}{\sqrt{2}}+\mathcal{A}(\Omega_{cc}^{+}\to \pi^0\Lambda_c^{+})+\lambda^2\mathcal{A}(\Xi_{cc}^{+}\to \pi^0\Xi_c^{+})+\frac{\lambda^2\mathcal{A}(\Omega_{cc}^{+}\to \overline K^0\Xi_c^{+})}{\sqrt{2}}=0.
\end{align}
\begin{align}
{\rm SumS}\,[\Omega_{cc}^{+},\eta_8,\Xi_c^{+}]:&\,\mathcal{A}(\Omega_{cc}^{+}\to  \eta_8\Lambda_c^{+})-\sqrt{\frac{3}{2}}\mathcal{A}(\Omega_{cc}^{+}\to K^0\Xi_c^+)\nonumber\\&~~~+\lambda^2\mathcal{A}(\Xi_{cc}^{+}\to \eta_8\Xi_c^{+})-\sqrt{\frac{3}{2}}\lambda^2\mathcal{A}(\Omega_{cc}^{+}\to \overline K^0\Xi_c^{+})=0.
\end{align}
\begin{align}
{\rm SumS}\,[\Omega_{cc}^{+},K^0,\Xi_c^{+}]:&\,-2\lambda\mathcal{A}(\Omega_{cc}^{+}\to  K^0\Xi_c^{0})+\lambda^2\mathcal{A}(\Xi_{cc}^{+}\to K^0\Xi_c^+)\nonumber\\&~~~+\sqrt{\frac{3}{2}}\lambda^2\mathcal{A}(\Omega_{cc}^{+}\to \eta_8\Xi_c^{+})-\frac{\lambda^2\mathcal{A}(\Omega_{cc}^{+}\to \pi^0\Xi_c^{+})}{\sqrt{2}}=0.
\end{align}
\begin{align}
&{\rm SumS}\,[\Omega_{cc}^{+},\overline K^0,\Xi_c^{+}]:\nonumber\\&\,\sqrt{\frac{3}{2}}\lambda\mathcal{A}(\Omega_{cc}^{+}\to  \eta_8\Xi_c^{+})+\mathcal{A}(\Omega_{cc}^{+}\to \overline K^0\Lambda_c^+)-\frac{\mathcal{A}(\Omega_{cc}^{+}\to \pi^0\Xi_c^{+})}{\sqrt{2}}+2\lambda\mathcal{A}(\Omega_{cc}^{+}\to \overline K^0\Xi_c^{+})=0.
\end{align}
\begin{align}
{\rm SumS}\,[\Omega_{cc}^{+},\pi^0,\Lambda_c^{+}]:&\,-2\lambda\mathcal{A}(\Omega_{cc}^{+}\to  \pi^0\Lambda_c^{+})+\lambda^2\mathcal{A}(\Xi_{cc}^{+}\to  \pi^0\Lambda_c^+)\nonumber\\&~~~+\frac{\lambda^2\mathcal{A}(\Omega_{cc}^{+}\to \overline K^0\Lambda_c^{+})}{\sqrt{2}}-\lambda^2\mathcal{A}(\Omega_{cc}^{+}\to \pi^0\Xi_c^{+})=0.
\end{align}
\begin{align}
{\rm SumS}\,[\Omega_{cc}^{+},\eta_8,\Lambda_c^{+}]:&\,-2\lambda\mathcal{A}(\Omega_{cc}^{+}\to  \eta_8\Lambda_c^{+})+\lambda^2\mathcal{A}(\Xi_{cc}^{+}\to  \eta_8\Lambda_c^+)\nonumber\\&~~~-\lambda^2\mathcal{A}(\Omega_{cc}^{+}\to \eta_8\Xi_c^{+})-\sqrt{\frac{3}{2}}\lambda^2\mathcal{A}(\Omega_{cc}^{+}\to \overline K^0\Lambda_c^{+})=0.
\end{align}
\begin{align}
&{\rm SumS}\,[\Omega_{cc}^{+},K^0,\Lambda_c^{+}]:\nonumber\\&\,\mathcal{A}(\Xi_{cc}^{+}\to  K^0\Lambda_c^{+})+\sqrt{\frac{3}{2}}\mathcal{A}(\Omega_{cc}^{+}\to  \eta_8\Lambda_c^+)-\mathcal{A}(\Omega_{cc}^{+}\to K^0\Xi_c^{+})-\frac{\mathcal{A}(\Omega_{cc}^{+}\to \pi^0\Lambda_c^{+})}{\sqrt{2}}=0.
\end{align}
\begin{align}
&{\rm SumS}\,[\Omega_{cc}^{+},\overline K^0,\Lambda_c^{+}]:\nonumber\\&\,\sqrt{\frac{3}{2}}\mathcal{A}(\Omega_{cc}^{+}\to  \eta_8\Lambda_c^{+})-\frac{\mathcal{A}(\Omega_{cc}^{+}\to \pi^0\Lambda_c^+)}{\sqrt{2}}+\lambda^2\mathcal{A}(\Xi_{cc}^{+}\to \overline K^0\Lambda_c^{+})-\lambda^2\mathcal{A}(\Omega_{cc}^{+}\to \overline K^0\Xi_c^{+})=0.
\end{align}
\begin{align}
&{\rm SumT_-}\,[\Xi_{cc}^{+},\pi^+,\Xi_c^{+}]:\,-\sqrt{2}\mathcal{A}(\Xi_{cc}^{+}\to  \pi^0\Xi_c^{+})-\mathcal{A}(\Xi_{cc}^{++}\to \pi^+\Xi_c^+)+\mathcal{A}(\Xi_{cc}^{+}\to \pi^+\Xi_c^{0})=0.
\end{align}
\begin{align}
&{\rm SumT_-}\,[\Xi_{cc}^{+},\pi^+,\Lambda_c^{+}]:\nonumber\\&\,-\sqrt{2}\mathcal{A}(\Xi_{cc}^{+}\to  \pi^0\Lambda_c^{+})-\mathcal{A}(\Xi_{cc}^{++}\to \pi^+\Lambda_c^+)+\lambda\mathcal{A}(\Xi_{cc}^{+}\to \overline K^0\Lambda_c^{+})-\lambda\mathcal{A}(\Xi_{cc}^{+}\to \pi^+\Xi_c^{0})=0.
\end{align}
\begin{align}
{\rm SumT_-}\,[\Xi_{cc}^{+},K^+,\Xi_c^{+}]:&\,\mathcal{A}(\Xi_{cc}^{+}\to  K^0\Xi_c^{+})-\mathcal{A}(\Xi_{cc}^{++}\to K^+\Xi_c^+)+\mathcal{A}(\Xi_{cc}^{+}\to K^+\Xi_c^{0})\nonumber\\&~~~-\sqrt{\frac{3}{2}}\lambda\mathcal{A}(\Xi_{cc}^{+}\to \eta_8\Xi_c^{+})-\frac{\lambda\mathcal{A}(\Xi_{cc}^{+}\to \pi^0\Xi_c^{+})}{\sqrt{2}}=0.
\end{align}
\begin{align}
{\rm SumT_-}\,[\Xi_{cc}^{+},K^+,\Lambda_c^{+}]:&\,\mathcal{A}(\Xi_{cc}^{+}\to  K^0\Lambda_c^{+})-\mathcal{A}(\Xi_{cc}^{++}\to K^+\Lambda_c^+)-\sqrt{\frac{3}{2}}\lambda\mathcal{A}(\Xi_{cc}^{+}\to \eta_8\Lambda_c^{+})\nonumber\\&~~~-\lambda\mathcal{A}(\Xi_{cc}^{+}\to K^+\Xi_c^{0})-\frac{\lambda\mathcal{A}(\Xi_{cc}^{+}\to \pi^0\Lambda_c^{+})}{\sqrt{2}}=0.
\end{align}
\begin{align}
&{\rm SumT_-}\,[\Omega_{cc}^{+},\pi^+,\Xi_c^{+}]:\nonumber\\&\,-\sqrt{2}\mathcal{A}(\Xi_{cc}^{+}\to  \pi^0\Xi_c^{+})+\mathcal{A}(\Omega_{cc}^{+}\to \pi^+\Xi_c^0)-\lambda\mathcal{A}(\Xi_{cc}^{++}\to \pi^+\Xi_c^+)+\lambda\mathcal{A}(\Omega_{cc}^{+}\to \overline K^0\Xi_c^{+})=0.
\end{align}
\begin{align}
&{\rm SumT_-}\,[\Omega_{cc}^{+},\pi^+,\Lambda_c^{+}]:\nonumber\\&\,-\sqrt{2}\mathcal{A}(\Omega_{cc}^{+}\to  \pi^0\Lambda_c^{+})-\lambda\mathcal{A}(\Xi_{cc}^{++}\to \pi^+\Lambda_c^+)+\lambda\mathcal{A}(\Omega_{cc}^{+}\to \overline K^0\Lambda_c^+)-\lambda\mathcal{A}(\Omega_{cc}^{+}\to  \pi^+\Xi_c^{0})=0.
\end{align}
\begin{align}
{\rm SumT_-}\,[\Omega_{cc}^{+},K^+,\Xi_c^{+}]:&\,\mathcal{A}(\Omega_{cc}^{+}\to  K^0\Xi_c^{+})+\mathcal{A}(\Omega_{cc}^{+}\to K^+\Xi_c^0)-\lambda\mathcal{A}(\Xi_{cc}^{++}\to K^+\Xi_c^+)\nonumber\\&-\sqrt{\frac{3}{2}}\lambda\mathcal{A}(\Omega_{cc}^{+}\to  \eta_8\Xi_c^{+})-\frac{\lambda\mathcal{A}(\Omega_{cc}^{+}\to  \pi^0\Xi_c^{+})}{\sqrt{2}}=0.
\end{align}
\begin{align}
&{\rm SumT_-}\,[\Omega_{cc}^{+},K^+,\Lambda_c^{+}]:\nonumber\\&\,\mathcal{A}(\Xi_{cc}^{++}\to  K^+\Lambda_c^{+})+\sqrt{\frac{3}{2}}\mathcal{A}(\Omega_{cc}^{+}\to \eta_8\Lambda_c^+)+\mathcal{A}(\Omega_{cc}^{+}\to K^+\Xi_c^0)+\frac{\mathcal{A}(\Omega_{cc}^{+}\to  \pi^0\Lambda_c^{+})}{\sqrt{2}}=0.
\end{align}
\begin{align}
&{\rm SumS}\,[\Xi_{cc}^{+},\eta_1,\Xi_c^{+}]:\,\mathcal{A}(\Xi_{cc}^{+}\to  \eta_1\Lambda_c^{+})-\mathcal{A}(\Omega_{cc}^{+}\to \eta_1\Xi_c^+)+2\lambda\mathcal{A}(\Xi_{cc}^{+}\to \eta_1\Xi_c^{+})=0.
\end{align}
\begin{align}
&{\rm SumS}\,[\Xi_{cc}^{+},\eta_1,\Lambda_c^{+}]:\,\mathcal{A}(\Omega_{cc}^{+}\to  \eta_1\Lambda_c^{+})+\lambda^2\mathcal{A}(\Xi_{cc}^{+}\to \eta_1\Xi_c^+)=0.
\end{align}
\begin{align}
&{\rm SumS}\,[\Omega_{cc}^{+},\eta_1,\Xi_c^{+}]:\,\mathcal{A}(\Omega_{cc}^{+}\to  \eta_1\Lambda_c^{+})+\lambda^2\mathcal{A}(\Xi_{cc}^{+}\to \eta_1\Xi_c^+)=0.
\end{align}
\begin{align}
&{\rm SumS}\,[\Omega_{cc}^{+},\eta_1,\Lambda_c^{+}]:\,-2\mathcal{A}(\Omega_{cc}^{+}\to  \eta_1\Lambda_c^{+})+\lambda^2\mathcal{A}(\Xi_{cc}^{+}\to \eta_1\Lambda_c^+)-\lambda^2\mathcal{A}(\Omega_{cc}^{+}\to \eta_1\Xi_c^{+})=0.
\end{align}

\section{SU(3) sum rules of $\mathcal{B}_{cc}\to M\mathcal{B}_{c6}$ modes}\label{res4}

\begin{align}
&{\rm SumS}\,[\Xi_{cc}^{++},\pi^+,\Sigma_c^{+}]:\,\mathcal{A}(\Xi_{cc}^{++}\to  K^+\Sigma_c^{+})+\lambda^2\mathcal{A}(\Xi_{cc}^{++}\to \pi^+\Xi_c^{*+})=0.
\end{align}
\begin{align}
&{\rm SumS}\,[\Xi_{cc}^{++},\pi^+,\Xi_c^{*+}]:\nonumber\\&-\mathcal{A}(\Xi_{cc}^{++}\to  K^+\Xi_c^{*+})+\mathcal{A}(\Xi_{cc}^{++}\to \pi^+\Sigma_c^{+})+2\lambda\mathcal{A}(\Xi_{cc}^{++}\to \pi^+\Xi_c^{*+})=0.
\end{align}
\begin{align}
&{\rm SumS}\,[\Xi_{cc}^{++},K^+,\Sigma_c^{+}]:\nonumber\\&-2\lambda\mathcal{A}(\Xi_{cc}^{++}\to  K^+\Sigma_c^{+})-\lambda^2\mathcal{A}(\Xi_{cc}^{++}\to K^+\Xi_c^{*+})+\lambda^2\mathcal{A}(\Xi_{cc}^{++}\to \pi^+\Sigma_c^{+})=0.
\end{align}
\begin{align}
&{\rm SumS}\,[\Xi_{cc}^{++},K^+,\Xi_c^{*+}]:\,\mathcal{A}(\Xi_{cc}^{++}\to  K^+\Sigma_c^{+})+\lambda^2\mathcal{A}(\Xi_{cc}^{++}\to \pi^+\Xi_c^{*+})=0.
\end{align}
\begin{align}
&{\rm SumS}\,[\Xi_{cc}^{++},\pi^0,\Sigma_c^{++}]:\,\mathcal{A}(\Xi_{cc}^{++}\to  K^0\Sigma_c^{++})+\lambda^2\mathcal{A}(\Xi_{cc}^{++}\to \overline K^0\Sigma_c^{++})=0.
\end{align}
\begin{align}
&{\rm SumS}\,[\Xi_{cc}^{++},\eta_8,\Sigma_c^{++}]:\,\mathcal{A}(\Xi_{cc}^{++}\to  K^0\Sigma_c^{++})+\lambda^2\mathcal{A}(\Xi_{cc}^{++}\to \overline K^0\Sigma_c^{++})=0.
\end{align}
\begin{align}
&{\rm SumS}\,[\Xi_{cc}^{++},K^0,\Sigma_c^{++}]:\nonumber\\&\,-2\lambda\mathcal{A}(\Xi_{cc}^{++}\to  K^0\Sigma_c^{++})+\frac{\sqrt{6}}{2}\lambda^2\mathcal{A}(\Xi_{cc}^{++}\to  \eta_8\Sigma_c^{++})-\frac{\sqrt{2}}{2}\lambda^2\mathcal{A}(\Xi_{cc}^{++}\to  \pi^0\Sigma_c^{++})=0.
\end{align}
\begin{align}
&{\rm SumS}\,[\Xi_{cc}^{++},\overline K^0,\Sigma_c^{++}]:\nonumber\\&\,\sqrt{\frac{3}{2}}\mathcal{A}(\Xi_{cc}^{++}\to  \eta_8\Sigma_c^{++})-\frac{\mathcal{A}(\Xi_{cc}^{++}\to  \pi^0\Sigma_c^{++})}{\sqrt{2}}+2\lambda\mathcal{A}(\Xi_{cc}^{++}\to  \overline K^0\Sigma_c^{++})=0.
\end{align}
\begin{align}
&{\rm SumS}\,[\Xi_{cc}^{+},\pi^+,\Sigma_c^{0}]:\,\mathcal{A}(\Xi_{cc}^{+}\to  K^+\Sigma_c^{0})+\mathcal{A}(\Omega_{cc}^{+}\to  \pi^+\Sigma_c^{0})+\sqrt{2}\lambda^2\mathcal{A}(\Xi_{cc}^{+}\to \pi^+\Xi_c^{*0})=0.
\end{align}
\begin{align}
&{\rm SumS}\,[\Xi_{cc}^{+},\pi^+,\Omega_c^{0}]:\,-\mathcal{A}(\Xi_{cc}^{+}\to  K^+\Omega_c^{0})+\sqrt{2}\mathcal{A}(\Xi_{cc}^{+}\to  \pi^+\Xi_c^{*0})-\mathcal{A}(\Omega_{cc}^{+}\to \pi^+\Omega_c^{0})=0.
\end{align}
\begin{align}
&{\rm SumS}\,[\Xi_{cc}^{+},\pi^+,\Xi_c^{*0}]:\nonumber\\&\,-\mathcal{A}(\Xi_{cc}^{+}\to  K^+\Xi_c^{*0})+\sqrt{2}\mathcal{A}(\Xi_{cc}^{+}\to  \pi^+\Sigma_c^{0})-\mathcal{A}(\Omega_{cc}^{+}\to \pi^+\Xi_c^{*0})+2\lambda\mathcal{A}(\Xi_{cc}^{+}\to \pi^+\Xi_c^{*0})=0.
\end{align}
\begin{align}
&{\rm SumS}\,[\Xi_{cc}^{+},K^+,\Sigma_c^{0}]:\nonumber\\&~~~\,-2\lambda\mathcal{A}(\Xi_{cc}^{+}\to  K^+\Sigma_c^{0})-\sqrt{2}\lambda^2\mathcal{A}(\Xi_{cc}^{+}\to  K^+\Xi_c^{*0})+\lambda^2\mathcal{A}(\Xi_{cc}^{+}\to \pi^+\Sigma_c^{0})=0.
\end{align}
\begin{align}
&{\rm SumS}\,[\Xi_{cc}^{+},K^+,\Omega_c^{0}]:\,\sqrt{2}\lambda\mathcal{A}(\Xi_{cc}^{+}\to  K^+\Xi_c^{*0})-\mathcal{A}(\Omega_{cc}^{+}\to  K^+\Omega_c^{0})+2\lambda\mathcal{A}(\Xi_{cc}^{+}\to K^+\Omega_c^{0})=0.
\end{align}
\begin{align}
{\rm SumS}\,[\Xi_{cc}^{+},K^+,\Xi_c^{*0}]:&\,\sqrt{2}\mathcal{A}(\Xi_{cc}^{+}\to  K^+\Sigma_c^{0})-\mathcal{A}(\Omega_{cc}^{+}\to  K^+\Xi_c^{*0})\nonumber\\&~~~-\sqrt{2}\lambda^2\mathcal{A}(\Xi_{cc}^{+}\to K^+\Omega_c^{0})+\lambda^2\mathcal{A}(\Xi_{cc}^{+}\to \pi^+\Xi_c^{*0})=0.
\end{align}
\begin{align}
&{\rm SumS}\,[\Xi_{cc}^{+},\pi^0,\Sigma_c^{+}]:\nonumber\\&\,\frac{\mathcal{A}(\Xi_{cc}^{+}\to  K^0\Sigma_c^{+})}{\sqrt{2}}-\mathcal{A}(\Omega_{cc}^{+}\to  \pi^0\Sigma_c^{+})+\frac{\lambda^2\mathcal{A}(\Xi_{cc}^{+}\to \overline K^0\Sigma_c^{+})}{\sqrt{2}}-\lambda^2\mathcal{A}(\Xi_{cc}^{+}\to \pi^0\Xi_c^{*+})=0.
\end{align}
\begin{align}
{\rm SumS}\,[\Xi_{cc}^{+},\eta_8,\Sigma_c^{+}]:&\,-\frac{\sqrt{6}}{2}\mathcal{A}(\Xi_{cc}^{+}\to  K^0\Sigma_c^{+})-\mathcal{A}(\Omega_{cc}^{+}\to  \eta_8\Sigma_c^{+})\nonumber\\&~~~-\lambda^2\mathcal{A}(\Xi_{cc}^{+}\to \eta_8\Xi_c^{*+})+\frac{\sqrt{6}}{2}\lambda^2\mathcal{A}(\Xi_{cc}^{+}\to \overline K^0\Sigma_c^{+})=0.
\end{align}
\begin{align}
{\rm SumS}\,[\Xi_{cc}^{+},K^0,\Sigma_c^{+}]:&\,-2\lambda\mathcal{A}(\Xi_{cc}^{+}\to  K^0\Sigma_c^{+})+\frac{\sqrt{6}}{2}\lambda^2\mathcal{A}(\Xi_{cc}^{+}\to  \eta_8\Sigma_c^{+})\nonumber\\&~~~-\lambda^2\mathcal{A}(\Xi_{cc}^{+}\to K^0\Xi_c^{*+})-\frac{\lambda^2\mathcal{A}(\Xi_{cc}^{+}\to \pi^0\Sigma_c^{+})}{\sqrt{2}}=0.
\end{align}
\begin{align}
&{\rm SumS}\,[\Xi_{cc}^{+},\overline K^0,\Sigma_c^{+}]:\nonumber\\&\,\sqrt{\frac{3}{2}}\mathcal{A}(\Xi_{cc}^{+}\to  \eta_8\Sigma_c^{+})-\frac{\mathcal{A}(\Xi_{cc}^{+}\to  \pi^0\Sigma_c^{+})}{\sqrt{2}}-\mathcal{A}(\Omega_{cc}^{+}\to \overline K^0\Sigma_c^{+})-2\lambda\mathcal{A}(\Xi_{cc}^{+}\to \overline K^0\Sigma_c^{+})=0.
\end{align}
\begin{align}
&{\rm SumS}\,[\Xi_{cc}^{+},\pi^0,\Xi_c^{*+}]:\nonumber\\&\,\frac{\mathcal{A}(\Xi_{cc}^{+}\to  K^0\Xi_c^{*+})}{\sqrt{2}}+\mathcal{A}(\Xi_{cc}^{+}\to  \pi^0\Sigma_c^{+})-\mathcal{A}(\Omega_{cc}^{+}\to \pi^0\Xi_c^{*+})+2\lambda\mathcal{A}(\Xi_{cc}^{+}\to \pi^0\Xi_c^{*+})=0.
\end{align}
\begin{align}
&{\rm SumS}\,[\Xi_{cc}^{+},\eta_8,\Xi_c^{*+}]:\nonumber\\&\,\mathcal{A}(\Xi_{cc}^{+}\to  \eta_8\Sigma_c^{+})-\sqrt{\frac{3}{2}}\mathcal{A}(\Xi_{cc}^{+}\to  K^0\Xi_c^{*+})-\mathcal{A}(\Omega_{cc}^{+}\to \eta_8\Xi_c^{*+})+2\lambda\mathcal{A}(\Xi_{cc}^{+}\to \eta_8\Xi_c^{*+})=0.
\end{align}
\begin{align}
&{\rm SumS}\,[\Xi_{cc}^{+},K^0,\Xi_c^{*+}]:\nonumber\\&\,\mathcal{A}(\Xi_{cc}^{+}\to  K^0\Sigma_c^{+})-\mathcal{A}(\Omega_{cc}^{+}\to  K^0\Xi_c^{*+})+\sqrt{\frac{3}{2}}\lambda^2\mathcal{A}(\Xi_{cc}^{+}\to \eta_8\Xi_c^{*+})-\frac{\lambda^2\mathcal{A}(\Xi_{cc}^{+}\to \pi^0\Xi_c^{*+})}{\sqrt{2}}=0.
\end{align}
\begin{align}
&{\rm SumS}\,[\Xi_{cc}^{+},\overline K^0,\Xi_c^{*+}]:\nonumber\\&\,\sqrt{\frac{3}{2}}\mathcal{A}(\Xi_{cc}^{+}\to  \eta_8\Xi_c^{*+})+\mathcal{A}(\Xi_{cc}^{+}\to  \overline K^0\Sigma_c^{+})-\frac{\mathcal{A}(\Xi_{cc}^{+}\to \pi^0\Xi_c^{*+})}{\sqrt{2}}-\mathcal{A}(\Omega_{cc}^{+}\to \overline K^0\Xi_c^{*+})=0.
\end{align}
\begin{align}
&{\rm SumS}\,[\Omega_{cc}^{+},\pi^+,\Sigma_c^{0}]:\,-2\lambda\mathcal{A}(\Omega_{cc}^{+}\to  \pi^+\Sigma_c^{+})+\lambda^2\mathcal{A}(\Xi_{cc}^{+}\to \pi^+\Sigma_c^{0})-\sqrt{2}\mathcal{A}(\Omega_{cc}^{+}\to \pi^+\Xi_c^{*0})=0.
\end{align}
\begin{align}
&{\rm SumS}\,[\Omega_{cc}^{+},\pi^+,\Omega_c^{0}]:\,-\mathcal{A}(\Omega_{cc}^{+}\to  K^+\Omega_c^{+})+\sqrt{2}\mathcal{A}(\Omega_{cc}^{+}\to \pi^+\Xi_c^{*0})+2\lambda\mathcal{A}(\Omega_{cc}^{+}\to \pi^+\Omega_c^{0})=0.
\end{align}
\begin{align}
{\rm SumS}\,[\Omega_{cc}^{+},\pi^+,\Xi_c^{*0}]:&\,-\mathcal{A}(\Omega_{cc}^{+}\to  K^+\Xi_c^{*0})+\sqrt{2}\mathcal{A}(\Omega_{cc}^{+}\to \pi^+\Sigma_c^{0})\nonumber\\&~~~+\lambda^2\mathcal{A}(\Xi_{cc}^{+}\to \pi^+\Xi_c^{*0})-\sqrt{2}\lambda^2\mathcal{A}(\Omega_{cc}^{+}\to \pi^+\Omega_c^{0})=0.
\end{align}
\begin{align}
&{\rm SumS}\,[\Omega_{cc}^{+},K^+,\Sigma_c^{0}]:\,\mathcal{A}(\Xi_{cc}^{+}\to  K^+\Sigma_c^{0})-\sqrt{2}\mathcal{A}(\Omega_{cc}^{+}\to K^+\Xi_c^{*0})+\mathcal{A}(\Omega_{cc}^{+}\to \pi^+\Sigma_c^{0})=0.
\end{align}
\begin{align}
&{\rm SumS}\,[\Omega_{cc}^{+},K^+,\Omega_c^{0}]:\,\sqrt{2}\mathcal{A}(\Omega_{cc}^{+}\to  K^+\Xi_c^{*0})+\lambda^2\mathcal{A}(\Xi_{cc}^{+}\to K^+\Omega_c^{0})+\lambda^2\mathcal{A}(\Omega_{cc}^{+}\to \pi^+\Omega_c^{0})=0.
\end{align}
\begin{align}
{\rm SumS}\,[\Omega_{cc}^{+},K^+,\Xi_c^{*0}]:&\,-2\lambda\mathcal{A}(\Omega_{cc}^{+}\to  K^+\Xi_c^{*0})+\lambda^2\mathcal{A}(\Xi_{cc}^{+}\to K^+\Xi_c^{*0})\nonumber\\&~~~-\sqrt{2}\lambda^2\mathcal{A}(\Omega_{cc}^{+}\to K^+\Omega_c^{0})+\lambda^2\mathcal{A}(\Omega_{cc}^{+}\to \pi^+\Xi_c^{*0})=0.
\end{align}
\begin{align}
{\rm SumS}\,[\Omega_{cc}^{+},\pi^0,\Sigma_c^{+}]:&\,-2\lambda\mathcal{A}(\Omega_{cc}^{+}\to  \pi^0\Sigma_c^{+})+\lambda^2\mathcal{A}(\Xi_{cc}^{+}\to \pi^0\Sigma_c^{+})\nonumber\\&~~~+\frac{\lambda^2\mathcal{A}(\Omega_{cc}^{+}\to \overline K^0\Sigma_c^{+})}{\sqrt{2}}-\lambda^2\mathcal{A}(\Omega_{cc}^{+}\to \pi^0\Xi_c^{*+})=0.
\end{align}
\begin{align}
{\rm SumS}\,[\Omega_{cc}^{+},\eta_8,\Sigma_c^{+}]:&\,-2\lambda\mathcal{A}(\Omega_{cc}^{+}\to  \eta_8\Sigma_c^{+})+\lambda^2\mathcal{A}(\Xi_{cc}^{+}\to \eta_8\Sigma_c^{+})\nonumber\\&~~~-\lambda^2\mathcal{A}(\Omega_{cc}^{+}\to \eta_8\Xi_c^{*+})-\sqrt{\frac{3}{2}}\lambda^2\mathcal{A}(\Omega_{cc}^{+}\to \overline K^0\Sigma_c^{+})=0.
\end{align}
\begin{align}
&{\rm SumS}\,[\Omega_{cc}^{+},K^0,\Sigma_c^{+}]:\nonumber\\&\,\mathcal{A}(\Xi_{cc}^{+}\to  K^0\Sigma_c^{+})+\frac{\sqrt{6}}{2}\mathcal{A}(\Omega_{cc}^{+}\to \eta_8\Sigma_c^{+})-\mathcal{A}(\Omega_{cc}^{+}\to K^0\Xi_c^{*+})-\frac{\mathcal{A}(\Omega_{cc}^{+}\to \pi^0\Sigma_c^{+})}{\sqrt{2}}=0.
\end{align}
\begin{align}
&{\rm SumS}\,[\Omega_{cc}^{+},\overline K^0,\Sigma_c^{+}]:\nonumber\\&\,\sqrt{\frac{3}{2}}\mathcal{A}(\Omega_{cc}^{+}\to  \eta_8\Sigma_c^{+})-\frac{\mathcal{A}(\Omega_{cc}^{+}\to \pi^0\Sigma_c^{+})}{\sqrt{2}}+\lambda^2\mathcal{A}(\Xi_{cc}^{+}\to \overline K^0\Sigma_c^{+})-\lambda^2\mathcal{A}(\Omega_{cc}^{+}\to \overline K^0\Xi_c^{*+})=0.
\end{align}
\begin{align}
&{\rm SumS}\,[\Omega_{cc}^{+},\pi^0,\Xi_c^{*+}]:\nonumber\\&\,\frac{\mathcal{A}(\Omega_{cc}^{+}\to  K^0\Xi_c^{*+})}{\sqrt{2}}+\mathcal{A}(\Omega_{cc}^{+}\to \pi^0\Sigma_c^{+})+\lambda^2\mathcal{A}(\Xi_{cc}^{+}\to \pi^0\Xi_c^{*+})+\frac{\lambda^2\mathcal{A}(\Omega_{cc}^{+}\to \overline K^0\Xi_c^{*+})}{\sqrt{2}}=0.
\end{align}
\begin{align}
{\rm SumS}\,[\Omega_{cc}^{+},\eta_8,\Xi_c^{*+}]:&\,\mathcal{A}(\Omega_{cc}^{+}\to  \eta_8\Sigma_c^{+})-\sqrt{\frac{3}{2}}\mathcal{A}(\Omega_{cc}^{+}\to K^0\Xi_c^{*+})\nonumber\\&~~~+\lambda^2\mathcal{A}(\Xi_{cc}^{+}\to \eta_8\Xi_c^{*+})-\sqrt{\frac{3}{2}}\lambda^2\mathcal{A}(\Omega_{cc}^{+}\to \overline K^0\Xi_c^{*+})=0.
\end{align}
\begin{align}
{\rm SumS}\,[\Omega_{cc}^{+},K^0,\Xi_c^{*+}]:&\,-2\lambda\mathcal{A}(\Omega_{cc}^{+}\to  K^0\Xi_c^{*+})+\lambda^2\mathcal{A}(\Xi_{cc}^{+}\to K^0\Xi_c^{*+})\nonumber\\&~~~+\sqrt{\frac{3}{2}}\lambda^2\mathcal{A}(\Omega_{cc}^{+}\to \eta_8\Xi_c^{*+})-\frac{\lambda^2\mathcal{A}(\Omega_{cc}^{+}\to \pi^0\Xi_c^{*+})}{\sqrt{2}}=0.
\end{align}
\begin{align}
&{\rm SumS}\,[\Omega_{cc}^{+},\overline K^0,\Xi_c^{*+}]:\nonumber\\&\,\sqrt{\frac{3}{2}}\mathcal{A}(\Omega_{cc}^{+}\to  \eta_8\Xi_c^{*+})+\mathcal{A}(\Omega_{cc}^{+}\to \overline K^0\Sigma_c^{+})-\frac{\mathcal{A}(\Omega_{cc}^{+}\to \pi^0\Xi_c^{*+})}{\sqrt{2}}+2\lambda\mathcal{A}(\Omega_{cc}^{+}\to \overline K^0\Xi_c^{*+})=0.
\end{align}
\begin{align}
{\rm SumT_-}\,[\Xi_{cc}^{++},\pi^+,\Sigma_c^{++}]:&\,-\sqrt{2}\mathcal{A}(\Xi_{cc}^{++}\to  \pi^0\Sigma_c^{+})+\sqrt{2}\mathcal{A}(\Xi_{cc}^{++}\to \pi^+\Sigma_c^{+})\nonumber\\&~~~+\lambda\mathcal{A}(\Xi_{cc}^{++}\to \overline K^0\Sigma_c^{+})+\sqrt{2}\lambda\mathcal{A}(\Xi_{cc}^{++}\to \pi^+\Xi_c^{*+})=0.
\end{align}
\begin{align}
{\rm SumT_-}\,[\Xi_{cc}^{++},K^+,\Sigma_c^{++}]:&\,\mathcal{A}(\Xi_{cc}^{++}\to  K^0\Sigma_c^{+})+\sqrt{2}\mathcal{A}(\Xi_{cc}^{++}\to K^+\Sigma_c^{+})-\sqrt{\frac{3}{2}}\lambda\mathcal{A}(\Xi_{cc}^{++}\to \eta_8\Sigma_c^{++})\nonumber\\&~~~+\sqrt{2}\lambda\mathcal{A}(\Xi_{cc}^{++}\to K^+\Xi_c^{*+})-\frac{\lambda\mathcal{A}(\Xi_{cc}^{++}\to \pi^0\Sigma_c^{++})}{\sqrt{2}}=0.
\end{align}
\begin{align}
{\rm SumT_-}\,[\Xi_{cc}^{+},\pi^+,\Sigma_c^{+}]:&\,-\sqrt{2}\mathcal{A}(\Xi_{cc}^{+}\to  \pi^0\Sigma_c^{+})-\mathcal{A}(\Xi_{cc}^{++}\to \pi^+\Sigma_c^{+})+\sqrt{2}\mathcal{A}(\Xi_{cc}^{+}\to \pi^+\Sigma_c^{0})\nonumber\\&~~~+\lambda\mathcal{A}(\Xi_{cc}^{+}\to \overline K^0\Sigma_c^{+})+\lambda\mathcal{A}(\Xi_{cc}^{+}\to \pi^+\Xi_c^{*0})=0.
\end{align}
\begin{align}
&{\rm SumT_-}\,[\Xi_{cc}^{+},\pi^+,\Xi_c^{*+}]:\,-\sqrt{2}\mathcal{A}(\Xi_{cc}^{+}\to  \pi^0\Xi_c^{*+})-\mathcal{A}(\Xi_{cc}^{++}\to \pi^+\Xi_c^{*+})+\sqrt{2}\mathcal{A}(\Xi_{cc}^{+}\to \pi^+\Xi_c^{*0})=0.
\end{align}
\begin{align}
{\rm SumT_-}\,[\Xi_{cc}^{+},K^+,\Sigma_c^{+}]:&\,\mathcal{A}(\Xi_{cc}^{+}\to  K^0\Sigma_c^{+})-\mathcal{A}(\Xi_{cc}^{++}\to K^+\Sigma_c^{+})+\sqrt{2}\mathcal{A}(\Xi_{cc}^{+}\to K^+\Sigma_c^{0})\nonumber\\&~~~-\frac{\sqrt{6}}{2}\lambda\mathcal{A}(\Xi_{cc}^{+}\to \eta_8\Sigma_c^{+})+\lambda\mathcal{A}(\Xi_{cc}^{+}\to K^+\Xi_c^{*0})\nonumber\\&~~~~~-\frac{\lambda\mathcal{A}(\Xi_{cc}^{+}\to \pi^0\Sigma_c^{+})}{\sqrt{2}}=0.
\end{align}
\begin{align}
{\rm SumT_-}\,[\Xi_{cc}^{+},K^+,\Xi_c^{*+}]:&\,\mathcal{A}(\Xi_{cc}^{+}\to  K^0\Xi_c^{*+})-\mathcal{A}(\Xi_{cc}^{++}\to K^+\Xi_c^{*+})+\mathcal{A}(\Xi_{cc}^{+}\to K^+\Xi_c^{*0})\nonumber\\&~~~-\sqrt{\frac{3}{2}}\lambda\mathcal{A}(\Xi_{cc}^{+}\to \eta_8\Xi_c^{*+})+\sqrt{2}\lambda\mathcal{A}(\Xi_{cc}^{+}\to K^+\Omega_c^{0})\nonumber\\&~~~~~-\frac{\lambda\mathcal{A}(\Xi_{cc}^{+}\to \pi^0\Xi_c^{*+})}{\sqrt{2}}=0.
\end{align}
\begin{align}
{\rm SumT_-}\,[\Xi_{cc}^{+},\pi^0,\Sigma_c^{++}]:&\,-\mathcal{A}(\Xi_{cc}^{++}\to  \pi^0\Sigma_c^{++})+\frac{\lambda\mathcal{A}(\Xi_{cc}^{++}\to K^-\Sigma_c^{++})}{\sqrt{2}}+\sqrt{2}\lambda\mathcal{A}(\Xi_{cc}^{+}\to \pi^0\Sigma_c^{+})\nonumber\\&~~~+\sqrt{2}\lambda\mathcal{A}(\Xi_{cc}^{+}\to \pi^-\Sigma_c^{++})+\sqrt{2}\lambda\mathcal{A}(\Xi_{cc}^{+}\to \pi^0\Xi_c^{*0})=0.
\end{align}
\begin{align}
{\rm SumT_-}\,[\Xi_{cc}^{+},\eta_8,\Sigma_c^{++}]:&\,-\mathcal{A}(\Xi_{cc}^{++}\to  \eta_8\Sigma_c^{++})+\sqrt{2}\mathcal{A}(\Xi_{cc}^{+}\to \eta_8\Sigma_c^{+})\nonumber\\&~~~+\sqrt{2}\lambda\mathcal{A}(\Xi_{cc}^{+}\to \eta_8\Xi_c^{*+})+\sqrt{\frac{3}{2}}\lambda\mathcal{A}(\Xi_{cc}^{+}\to K^-\Sigma_c^{++})=0.
\end{align}
\begin{align}
{\rm SumT_-}\,[\Xi_{cc}^{+},K^0,\Sigma_c^{++}]:&\,-\mathcal{A}(\Xi_{cc}^{++}\to  K^0\Sigma_c^{++})-\lambda\mathcal{A}(\Xi_{cc}^{+}\to \pi^-\Sigma_c^{++})\nonumber\\&~~~+\sqrt{2}\lambda\mathcal{A}(\Xi_{cc}^{+}\to K^0\Sigma_c^{+})+\sqrt{2}\lambda\mathcal{A}(\Xi_{cc}^{+}\to K^0\Xi_c^{*+})=0.
\end{align}
\begin{align}
&{\rm SumT_-}\,[\Xi_{cc}^{+},\overline K^0,\Sigma_c^{++}]:\nonumber\\&~~~\,-\mathcal{A}(\Xi_{cc}^{+}\to  K^-\Sigma_c^{++})-\mathcal{A}(\Xi_{cc}^{++}\to \overline K^0\Sigma_c^{++})+\sqrt{2}\mathcal{A}(\Xi_{cc}^{+}\to \overline K^0\Sigma_c^{+})=0.
\end{align}
\begin{align}
{\rm SumT_-}\,[\Omega_{cc}^{+},\pi^+,\Sigma_c^{+}]:&\,-\sqrt{2}\mathcal{A}(\Omega_{cc}^{+}\to  \pi^0\Sigma_c^{+})+\sqrt{2}\mathcal{A}(\Omega_{cc}^{+}\to \pi^+\Sigma_c^{0})-\lambda\mathcal{A}(\Xi_{cc}^{++}\to  \pi^+\Sigma_c^{+})\nonumber\\&~~~+\lambda\mathcal{A}(\Omega_{cc}^{+}\to \overline K^0\Sigma_c^{+})+\lambda\mathcal{A}(\Omega_{cc}^{+}\to  \pi^+\Xi_c^{*0})=0.
\end{align}
\begin{align}
{\rm SumT_-}\,[\Omega_{cc}^{+},\pi^+,\Xi_c^{*+}]:&\,-\sqrt{2}\mathcal{A}(\Omega_{cc}^{+}\to  \pi^0\Xi_c^{*+})+\mathcal{A}(\Omega_{cc}^{+}\to \pi^+\Xi_c^{*0})-\lambda\mathcal{A}(\Xi_{cc}^{++}\to  \pi^+\Xi_c^{*+})\nonumber\\&~~~+\lambda\mathcal{A}(\Omega_{cc}^{+}\to \overline K^0\Xi_c^{*+})+\sqrt{2}\lambda\mathcal{A}(\Omega_{cc}^{+}\to  \pi^+\Omega_c^{0})=0.
\end{align}
\begin{align}
&{\rm SumT_-}\,[\Omega_{cc}^{+},K^+,\Sigma_c^{+}]:\nonumber\\&\,-\mathcal{A}(\Xi_{cc}^{++}\to  K^+\Sigma_c^{+})-\frac{\sqrt{6}}{2}\mathcal{A}(\Omega_{cc}^{+}\to \eta_8\Sigma_c^{+})+\mathcal{A}(\Omega_{cc}^{+}\to  K^+\Xi_c^{*0})-\frac{\mathcal{A}(\Omega_{cc}^{+}\to \pi^0\Sigma_c^{+})}{\sqrt{2}}=0.
\end{align}
\begin{align}
{\rm SumT_-}\,[\Omega_{cc}^{+},K^+,\Xi_c^{*+}]:&\,\mathcal{A}(\Omega_{cc}^{+}\to  K^0\Xi_c^{*+})+\mathcal{A}(\Omega_{cc}^{+}\to K^+\Xi_c^{*0})-\lambda\mathcal{A}(\Xi_{cc}^{++}\to  K^+\Xi_c^{*+})\nonumber\\&~-\sqrt{\frac{3}{2}}\lambda\mathcal{A}(\Omega_{cc}^{+}\to \eta_8\Xi_c^{*+})+\sqrt{2}\lambda\mathcal{A}(\Omega_{cc}^{+}\to K^+\Omega_c^{0})\nonumber\\&~~~~~-\frac{\lambda\mathcal{A}(\Omega_{cc}^{+}\to \pi^0\Xi_c^{*+})}{\sqrt{2}}=0.
\end{align}
\begin{align}
{\rm SumT_-}\,[\Omega_{cc}^{+},\pi^0,\Sigma_c^{++}]:&\,-\mathcal{A}(\Xi_{cc}^{++}\to  \pi^0\Sigma_c^{++})+\frac{\mathcal{A}(\Omega_{cc}^{+}\to K^-\Sigma_c^{++})}{\sqrt{2}}+\sqrt{2}\mathcal{A}(\Omega_{cc}^{+}\to \pi^0\Sigma_c^{+})\nonumber\\&~~~+\sqrt{2}\mathcal{A}(\Omega_{cc}^{+}\to \pi^-\Sigma_c^{++})+\sqrt{2}\mathcal{A}(\Omega_{cc}^{+}\to \pi^0\Xi_c^{*+})=0.
\end{align}
\begin{align}
{\rm SumT_-}\,[\Omega_{cc}^{+},\eta_8,\Sigma_c^{++}]:&\,\sqrt{2}\mathcal{A}(\Omega_{cc}^{+}\to  \eta_8\Sigma_c^{+})-\lambda\mathcal{A}(\Xi_{cc}^{++}\to \eta_8\Sigma_c^{++})\nonumber\\&~~~+\sqrt{2}\lambda\mathcal{A}(\Omega_{cc}^{+}\to \eta_8\Xi_c^{*+})+\sqrt{\frac{3}{2}}\mathcal{A}(\Omega_{cc}^{+}\to K^-\Sigma_c^{++})=0.
\end{align}
\begin{align}
&{\rm SumT_-}\,[\Omega_{cc}^{+},K^0,\Sigma_c^{++}]:\nonumber\\&~~~\,\mathcal{A}(\Xi_{cc}^{++}\to  K^0\Sigma_c^{++})-\sqrt{2}\mathcal{A}(\Omega_{cc}^{+}\to K^0\Xi_c^{*+})+\mathcal{A}(\Omega_{cc}^{+}\to \pi^-\Sigma_c^{++})=0.
\end{align}
\begin{align}
{\rm SumT_-}\,[\Omega_{cc}^{+},\overline K^0,\Sigma_c^{++}]:&\,-\mathcal{A}(\Omega_{cc}^{+}\to  \overline K^0\Sigma_c^{++})+\sqrt{2}\mathcal{A}(\Omega_{cc}^{+}\to \overline K^0\Sigma_c^{+})\nonumber\\&~~~-\lambda\mathcal{A}(\Xi_{cc}^{++}\to \overline K^0\Sigma_c^{++})+\sqrt{2}\lambda\mathcal{A}(\Omega_{cc}^{+}\to \overline K^0\Xi_c^{*+})=0.
\end{align}
\begin{align}
&{\rm SumS}\,[\Xi_{cc}^{+},\eta_1,\Sigma_c^{+}]:\,\mathcal{A}(\Omega_{cc}^{+}\to  \eta_1\Sigma_c^{+})+\lambda^2\mathcal{A}(\Xi_{cc}^{+}\to \eta_1\Xi_c^{*+})=0.
\end{align}
\begin{align}
&{\rm SumS}\,[\Xi_{cc}^{+},\eta_1,\Xi_c^{*+}]:\,\mathcal{A}(\Xi_{cc}^{+}\to  \eta_1\Sigma_c^{+})-\mathcal{A}(\Omega_{cc}^{+}\to \eta_1\Xi_c^{*+})+2\lambda\mathcal{A}(\Xi_{cc}^{+}\to \eta_1\Xi_c^{*+})=0.
\end{align}
\begin{align}
&{\rm SumS}\,[\Omega_{cc}^{+},\eta_1,\Sigma_c^{+}]:\,-2\lambda\mathcal{A}(\Omega_{cc}^{+}\to  \eta_1\Sigma_c^{+})+\lambda^2\mathcal{A}(\Xi_{cc}^{+}\to \eta_1\Sigma_c^{+})-\lambda^2\mathcal{A}(\Omega_{cc}^{+}\to \eta_1\Xi_c^{*+})=0.
\end{align}
\begin{align}
&{\rm SumS}\,[\Omega_{cc}^{+},\eta_1,\Xi_c^{*+}]:\,\mathcal{A}(\Omega_{cc}^{+}\to  \eta_1\Sigma_c^{+})+\lambda^2\mathcal{A}(\Xi_{cc}^{+}\to \eta_1\Xi_c^{*+})=0.
\end{align}
\begin{align}
&{\rm SumT_-}\,[\Xi_{cc}^{+},\eta_1,\Sigma_c^{++}]:\nonumber\\&\,-\mathcal{A}(\Xi_{cc}^{++}\to  \eta_1\Sigma_c^{++})+\sqrt{2}\mathcal{A}(\Xi_{cc}^{+}\to \eta_1\Sigma_c^{+})+\sqrt{2}\lambda\mathcal{A}(\Xi_{cc}^{+}\to \eta_1\Xi_c^{*+})=0.
\end{align}
\begin{align}
&{\rm SumT_-}\,[\Omega_{cc}^{+},\eta_1,\Sigma_c^{++}]:\nonumber\\&~~~\,-\lambda\mathcal{A}(\Xi_{cc}^{++}\to  \eta_1\Sigma_c^{++})+\sqrt{2}\mathcal{A}(\Omega_{cc}^{+}\to \eta_1\Sigma_c^{+})+\sqrt{2}\lambda\mathcal{A}(\Omega_{cc}^{+}\to \eta_1\Xi_c^{*+})=0.
\end{align}

\end{document}